\documentclass[aps,amsmath,amssymb,showpacs,showkeys]{revtex4-2}
\usepackage[dvips]{graphicx}
\usepackage{times}
\usepackage{braket}
\usepackage{xcolor}
\usepackage{adjustbox}
\usepackage{orcidlink}
\usepackage{hyperref}
\usepackage{booktabs}
\hypersetup{
  colorlinks=true,
  urlcolor=blue,
  linkcolor=blue,
  citecolor=blue
}

\usepackage{multirow}

\begin{document}
\title{Strong Lensing Effect and Quasinormal Modes of Oscillations of Black 
Holes in $\boldsymbol{f(R,T)}$ Gravity Theory}

\author{Gayatri Mohan \orcidlink{0009-0008-0654-5227}}
\email[Email: ]{gayatrimohan704@gmail.com}

\author{Ronit Karmakar \orcidlink{0000-0002-9531-7435}}
\email[Email: ]{ronit.karmakar622@gmail.com}

\author{Rupam Jyoti Borah \orcidlink{0009-0005-5134-0421}}
\email[Email: ]{rupamjyotiborah856@gmail.com}

\author{Umananda Dev Goswami \orcidlink{0000-0003-0012-7549}}
\email[Email: ]{umananda@dibru.ac.in (Corresponding author)}

\affiliation{Department of Physics, Dibrugarh University, Dibrugarh 786004, 
Assam, India}

\begin{abstract}
In this work, we analyze the strong lensing phenomenon and quasinormal modes 
(QNMs) in the case of black holes (BHs) surrounded by fluids within the 
framework of $f(R,T)$ gravity, adopting a minimally coupled model of the 
theory. Our analysis is conducted for three surrounding fields corresponding 
to three different values of the parameter $\omega$ of the equations of state, 
each representing a unique class of BH solutions. A universal method developed 
by V.~Bozza is employed for strong lensing analysis and the WKB approximation 
method to compute the QNMs of oscillation of the BHs. The influences of the 
model parameters $\beta$ and $c_2$ on the deflection angle and associated 
lensing coefficients are analyzed. Our findings on lensing reveal that smaller 
values of $\beta$ and $c_2$ cause photon divergence at larger impact 
parameters as well as the lensing results converge to the 
Schwarzschild limit. Extending the analysis to 
the supermassive BH Sgr A*, we examine the outermost Einstein rings, estimate 
three lensing observables: angular position $\vartheta_{\infty}$, angular 
separation $s$ and relative magnification $r_\text{mag}$ for the BHs. For a 
specific values of $\beta$ and $c_2$, BHs with different field configurations exhibit 
substantial variations in their observable properties. The variation of 
amplitude and damping of QNMs with respect to the model parameter $\beta$ and $c_2$ is 
analyzed for the BHs. We found that the $\beta$ parameter has a direct 
correlation with the amplitude and an inverse relation with the damping of the 
QNMs, while $c_2$ has direct correlation with amplitude as well as damping. 
Further, we use the time domain analysis to verify the results and found 
a good match between the two methods.
\end{abstract}
	
	
\maketitle

\section{Introduction}
\label{sec.1}

General Relativity (GR) is the most fundamental and robust theory of 
gravity thus far. GR has been backed by two most fascinating and crucial 
observations of recent times in addition to earlier supporting observations 
\cite{r15-1}. One is the detection of gravitational waves (GWs) by the LIGO 
and Virgo team in 2015 \cite{r1,r2,r3,r4,r5,r6} and the other is the snap of 
the direct image of the black hole (BH) at the center of galaxy M87 
\cite{r16} by the Event Horizon Telescope (EHT) group in 2019 
\cite{r7,r8,r9,r10,r11,r12}. However, there are some issues with GR that 
need to be addressed so that a complete theory of gravity can be obtained. 
GR faces the normalization issue in the high energy regime \cite{r16-1}. 
GR can not explain the dark components in the Universe, especially it fails 
in the IR regime as it cannot explain the accelerated expansion of the 
Universe as recently confirmed \cite{r16-2,r16-3} and can not predict anything 
regarding the existence of dark matter \cite{r16-4}. It has compatibility 
issues with quantum mechanical phenomena \cite{r16-5}. These issues have 
motivated researchers to look for modified as well as alternative theories of 
gravity (ATGs) \cite{r16,r17,r18,r19}. Physicists modify either the 
matter-energy sector or the geometry part of the Einstein field equations in 
their efforts to mitigate the issues faced by GR. The former way of 
modification leads to the theories of dark energy and the latter one leads to 
the development of modified theories of gravity (MTGs). Some of the most 
pursued MTGs are $f(R)$ gravity \cite{r20,r21,r22}, $f(R,T$) gravity 
\cite{r23,r24,r24-1,r24-2}, Rastall gravity \cite{r27,r28}, $f(R,L_m)$ 
gravity \cite{r29} and so on. On the other hand ATGs are based on 
fundamentally different geometrical approaches than the approach of GR. One 
of the most studied ATGs is the $f(Q)$ gravity \cite{r25,r26} theory. 

BHs have gained significant attention in recent times. The two 
major milestones of modern physics that have propelled the growing interest in 
the field of BH physics are the already mentioned detection of GWs and the 
photograph of the first ever image of the black hole M87$^*$ at the center 
of the M87 galaxy. Since then a lot of efforts have been directed towards a 
better understanding of this aspect of BH physics \cite{r12-1,r12-2,r12-3}. 
Though a lot remains to be uncovered about BHs, some insights have been 
achieved and it is expected that advanced detectors of the future will shed 
more light regarding these mysterious objects. BHs came out as a mathematical 
solution of Einstein field equations in the GR framework in 1916, obtained by 
Karl Schwarzschild for the vacuum case. Ever since a number of papers have 
derived BH solutions in different spacetimes and different gravity theories 
(see Ref.~\cite{r13,r14,r15} for a review).

Gravitational lensing is a gravitational effect on light, which occurs due to 
the bending of light when passing the curved spacetime produced by massive 
objects such as galaxies, galaxy clusters, BHs, etc. It was predicted by 
Einstein in 1915 in his theory of GR \cite{Einstein_1915} and now become an 
indispensable tool in modern astrophysics and cosmology \cite{Bartelmann_2010, 
Wambsganss_1998, Blandford_1992, Congdon_2018}. This phenomenon was first 
observed around the sun by Eddington, Dyson and Davidson during the time of 
the solar eclipse in 1919, when starlight from the Hyades cluster passed near 
the sun. It was recorded as the first experimental verification of GR 
\cite{Dyson_1920}. The lensing effect produces magnified, distorted images and 
delays the time taken by light to arrive at the observer. Specifically, 
lensing in the strong field limit exhibits distinct and notable features, and 
the deflection angle offers a critical pathway for exploring the optical 
characteristics of lensing objects. Consequently, as a lensing object, BHs 
serve as unique platforms for studying the effects of strong gravitational 
fields. Within this limit, a ray of light passing very close to the objects 
becomes highly nonlinear. The trajectories followed by light in such 
propagation undergo multiple circular loops around the object before 
leaving it \cite{2002_Bozza,2022_kumar}. As a 
result, a series of discrete images, referred to as relativistic images are 
formed from a single object which cannot be explained by classical weak-field 
lensing theories \cite{Zhao_2017,Darwin_1959}. Indeed, gravitational lensing 
functions as a reliable astrophysical technique for investigating the 
gravitational fields of massive objects and shedding light on the elusive 
nature of dark matter \cite{Clowe_2006,Massey_2010,Fu_2022}.

In the strong field regime, Darwin's pioneering work on the interpretation  
of Einstein's equations for orbits around the sun to determine whether 
particles or light rays captured by the sun is served as the groundwork 
for subsequent developments in this area \cite{Darwin_1959}. Over the years, 
investigations in the strong field limit have gained significant attention due 
to their potential to extract information about BHs \cite{2002_Bozza,
Bozza_2003,Whisker_2005,Chen_2009,Liu_2010,Ding_2011,Tsukamoto_2022,
Sotani_2015,Tsukamoto_2023,Choudhuri_2023,Mustafa_2024,Soares_2023a,
2007_bozza}. A wide range of work has been performed on the strong deflection 
lensing extensively for Schwarzschild BHs \cite{hans_1987,Nemiro_1993,
Virbhadra_1998,2001_Bozza,Virbhadra_2000}. Virbhadra and his co-workers have 
made a significant contribution to the study of gravitational lensing by 
introducing the lens equation in the investigations of the strong field regime 
\cite{Virbhadra_2000,Virbhadra_1998}. Later they analyzed the formation 
and properties of relativistic images for a Schwarzschild BH in this  
context \cite{Virbhadra_2009,Virbhadra_2022}. Further investigations expanding 
to other static and symmetric BH spacetimes have been performed by many 
researchers \cite{Eiroa_2002,2011_Eiroa,Whisker_2005,Perlick_2004,Wang_2016,
Chen_2009,Keeton_2005,Olmo_2023}. In fact, the past decade has witnessed a 
significant progress in studies on gravitational lensing including 
investigations for naked singularities \cite{Atamurotov_2022, Chen_2024, 
Hossain_2024}, wormholes \cite{Shaikh_2019, Saurav_2024, Godani_2023,
Shaikh_2017}, and other exotic entities \cite{Surajit_2023, Davies_2020}. A 
plethora of parallel efforts have been made on strong field deflection in GR 
as well as in context of MTGs, revealing remarkable phenomena such as 
relativistic images and shadow formation \cite{EHT_2022,Konoplya_2019,Olmo_2023,2025_Maryam,
Pantig_2022, Javed_2022,Soares_2023,Nashiba_2023b,Badia_2017,Zhao_2016,
Islam_2021,Gibbons_2008,Kuang_2022,2021_Gao,Gyulchev_2019,Dong_2024,
ghosh_2021}. However, when one considers the MTGs, the deflection angle 
encapsulates additional information about deviations from the Schwarzschild or 
Kerr metrics. By analyzing the deflection in such scenarios, one can constrain 
the parameters of MTGs and assess their viability in describing astrophysical 
observations. 

Detection of GWs has opened up a new window to look at the Universe and 
develop our understanding on how gravity works in strong gravity regimes. 
When a BH is perturbed by some 
field, it oscillates and releases the excess energy to go back to its steady 
state. This release of energy is in the form of spacetime ripples with very 
minute amplitudes and decaying as it travels through spacetime. These are 
nothing but quasinormal modes (QNMs) of oscillations \cite{r30,r31}. The 
concept of QNMs theoretically emerged for the first time via the work of 
Vishveshwara in 1970 \cite{r31-1}. Later, in 1975, Chandrasekhar and 
Detweiler computed the QNMs for the first time \cite{r32}. Schrutz and Will 
computed the QNMs for the first time using the WKB approximation method 
\cite{r33}. Since then many numerical as well as analytical methods have been 
employed to compute the QNMs. Konoplya developed the WKB method up to the 13th 
order along with the Pad\'e averaging technique \cite{r34}. Other methods such 
as the asymptotic iteration method (AIM), Bernstein spectral method, Leaver’s 
technique, etc.~have been widely used in the literature to compute the QNMs 
(see Ref.~\cite{r30} for a review). In Ref.~\cite{r35}, the authors computed 
QNMs of BHs in scalar-vector-tensor (SVT) gravity with a surrounding 
quintessence field. They also studied the thermodynamic and optical properties 
of the obtained BH solution. Ref.~\cite{r36} discusses BHs in MTGs in both 
static and rotating configurations and studies their optical properties. 
Recently, the authors in \cite{Santos_2023} computed a class of Kiselev BH 
solutions in the framework of $f(R,T)$ gravity. In another work 
\cite{Bidyut_2024}, the authors obtained different solutions of BHs 
considering various models of $f(R,T)$ gravity framework and studied its 
properties. QNMs of black hole solutions in the framework of Hu-Sawicki model 
of $f(R)$ gravity has been studied in Ref. \cite{r16}.

An important observational as well as theoretical aspect with regard
to QNMs is that they can be used to test the cosmic censorship conjecture, 
which basically states that a singularity must exist within an event horizon. 
The abnormal behaviour of QNMs like growing modes and the absence of 
a well-defined QNM spectrum can point towards the presence of naked 
singularities \cite{n1}. Another such aspect that is relevant to the QNMs is 
the testing of the No-Hair theorem which states that the BH has only three 
properties: mass, charge and spin (angular momentum). Thus, QNMs should 
depend only on these three conserved quantities. Any deviation of QNMs from 
this dependency can be an indication of the presence of `hairs' \cite{n2}. 
It is observed that the QNMs for some values of model parameters are found 
to be indeterminate, meaning that QNMs do not exist. It may signal presence 
of singularities in the BH spacetime.
However, to the best of our knowledge, QNMs have not been detected yet 
but it is believed that future detectors of GWs will be able to observe them. 
In this respect, it should be mentioned that Ref.~\cite{r31-01} claims to 
have detected QNMs for $l=m=2$ and $l=m=3$ modes. It should be noted that such 
claims are subject to criticism and signal-to-noise (SNR) issues. Having 
said that, some works \cite{r31-01,ref} have claimed to detect some modes of QNMs but a 
more robust analysis is anticipated in the near future with advanced future 
detectors for more clarity. Readers can refer to Ref.~\cite{ref} for an 
overview of this topic.

Motivated by the factors mentioned above, our study aims to explore the 
strong limit lensing characteristics and QNMs of BHs within the framework of 
the $f(R,T)$ gravity. The recent formulation of a BH spacetime in this 
theoretical framework \cite{Santos_2023,Bidyut_2024} has sparked our interest 
in exploring gravitational bending effects as well as QNMs of the BH. In the 
strong field lensing investigation, the technique proposed by V.~Bozza 
\cite{2002_Bozza} is adopted, whereas for QNMs calculations 3rd order WKB 
approximation is employed.

The rest of the paper is structured as follows: In Section \ref{sec.2}, 
we outline the theoretical framework of the study. In Section~\ref{sec.3}, we 
focus on the derivation of the gravitational bending angle in the strong 
field limit in the BH spacetime. Additionally, this section includes the 
computation and analysis of lensing observables by considering the 
supermassive BH SgrA* as a $f(R,T)$ gravity BH. Constraining of the
model parameters using the EHT observational data of the Einstein ring is 
presented in this section. Then we compute QNMs of the BH in 
Section~\ref{sec.4}.  Section~\ref{sec.5} presents the time profile analysis 
of the BH. Finally, we present our concluding remark in Section~\ref{sec.6}. 

\section{$\boldsymbol{f(R,T)}$ Gravity Black Holes}
\label{sec.2}
 
The $f(R,T)$ gravity is a modification of Einstein’s GR where the 
gravitational Lagrangian $\mathcal{L}$ in the Einstein-Hilbert action is 
generalized as a function of the Ricci scalar $R$ and the trace $T$ of the 
energy-momentum tensor $T_{\mu\nu}$. The core concept of this theory is the 
coupling between the geometry and matter. This can significantly influence 
the BH’s spacetime by introducing an extra term in the field equations. Thus
the action in this theory is given as \cite{Harko_2011} 
\begin{equation} 
S = \frac{1}{2\kappa^2}\int\!\sqrt{-\, g}\,{f(R,T)}\, d ^4x + \int\!\sqrt{- \, g}\, \mathcal{L}_m\, d^4x,
	\label{eq1}
\end{equation} 
where $\mathcal{L}_{m}$ is the Lagrangian density of matter field, $g$ is 
determinant of the metric $g_{\mu\nu}$ and $\kappa^2 = 8 \pi G = 1/M^2_{pl}$. 
$G$ and $M_{pl}$ are the gravitational constant and reduced Planck mass 
respectively. The energy-momentum tensor of the matter field can be written as 
\begin{equation} 	
	T_{\mu\nu} = -\,\frac{2\, \delta(\sqrt{- \, g}\mathcal{L}_{m})}{\sqrt{- \, g}\ \delta g^{\mu\nu}} = -\, 2\ \frac{\partial \mathcal{L}_{m}}{\partial g^{\mu\nu}} + g_{\mu\nu}\, \mathcal{L}_{m}.
	\label{eq2}
\end{equation}     
This equation gives us the trace of $T_{\mu\nu}$ as
\begin{equation}
	T  =  -\, 2\, g^{\mu\nu}\frac{\partial \mathcal{L}_{m}}{\partial g^{\mu\nu}} + 4\,\mathcal{L}_m.
	\label{eq3}
\end{equation}
The variation of the action \eqref{eq1} with respect to the metric 
$g_{\mu\nu}$ yields the field equations of this gravity theory as 
\begin{equation}
	F_R\, R_{\mu\nu} - \frac{1}{2}\, f(R,T)\, g_{\mu\nu} + \left(g_{\mu\nu} \square - \nabla_\mu \nabla_ \nu\right)F_R = \kappa^2 T_{\mu\nu} -  F_T\left(T_{\mu\nu} + \theta_{\mu\nu}\right),
	\label{eq4}	
\end{equation}
where $ F_R = df(R,T)/dR$, $F_T = df(R,T)/dT$ and the energy tensor 
$\theta_{\mu\nu}$ defined as \cite{Harko_2011,Alvarenga_2013}
\begin{equation}
\theta_{\mu\nu}= g^{ab}\,\frac{\delta{T}_{ab}}{\delta{g^{\mu\nu}}}= -\,2\,T_{\mu\nu} + g_{\mu\nu}\, \mathcal{L}_{m} - 2\,g^{ab}\frac{\partial^2{\mathcal{L}_m}}{\partial{g^{\mu\nu}}\partial{g^{ab}}}.
\label{eq4a}	
\end{equation}
Importantly, the inclusion of the trace $T$ of the energy-momentum 
tensor $T_{\mu\nu}$ in the gravitational action of $f(R,T)$ gravity theory 
results in some extra terms in the field equations of the theory. Because of 
this, it can be treated as a two fluid theory with the effective 
energy-momentum tensor $T_{\mu\nu}^{eff} = T_{\mu\nu} + T_{\mu\nu}^I$. Here 
$T_{\mu\nu}$ represents the energy-momentum tensor for ordinary matter 
field and $T_{\mu\nu}^I$ is the matter-curvature interaction term which may 
act as a new matter source and can be defined for a geometrically originated 
anisotropic matter distribution.

It is noteworthy that depending on the different forms of the function 
$f(R,T)$, the models of this gravity may be placed into three categories, 
viz., the minimally, non-minimally and purely non-minimally coupling models. 
The forms of these three types are proposed respectively as 
\cite{Shabani_2014,Shabani_2013}:
$ f(R,T) = f_1(R) + f_2(T)$,
$ f(R,T) = f_1(R)[1 +  f_2(T)]$ and 
$ f(R,T) = f_1(R)\, f_2(T)$. 
In our work, we assume the minimally coupled matter and geometry sector in 
$f(R,T)$ gravity and so we consider the first kind of model for the analysis 
of the BHs spacetime in this theory. However, investigations of BHs 
characteristics in such models of $f(R,T)$ modified gravity result in some 
remarkable features that are attributed to these compact objects surrounded 
by some fluids~\cite{Kiselev_2003,Santos_2023}.  Recently, as mentioned 
earlier a study by Ref.~\cite{Bidyut_2024} has derived the BH solutions in 
the context of this gravity theory and explored the thermodynamic properties 
of these BHs from their thermodynamic topology and thermodynamic geometry. 
Following this reference, we consider a static spherically symmetric metric 
ansatz in the form:   
\begin{equation}
ds^2 =  A(r) dt^2 - B(r) dr^2 - r^2(d\theta^2 + \sin^2\!\theta\, d\phi^2).
\label{eq5}
\end{equation}
As mentioned above, we are focusing on a minimally coupled $f(R,T)$ model 
characterized by the parameters $\alpha$ and $\beta$, expressed in the 
following form: 
\begin{equation}
	f(R,T) = \alpha f_1(R) +  \beta f_2(T),
	\label{eq6}
\end{equation}	
where, $f_1(R) = R$\, and $f_2(T) = T$. In this study, $T$ is chosen 
as the trace of the energy-momentum tensor connected to the spherically 
symmetric Kiselev BHs. The energy-momentum tensor governing Kiselev
BHs is effectively associated with an anisotropic fluid, whose components 
satisfy the following correlations~\cite{Kiselev_2003,Santos_2023}:
\begin{align} 
	T^t_t & = T^r_r = \rho,
	\label{eq7a}\\[5pt] 
	T^\theta_\theta & = T^\phi_\phi = -\frac{1}{2}\rho\,(3\, \omega + 1),
	\label{eq7b}
\end{align}
where $\rho$ represents energy density of the fluid and $\omega$ is the 
parameter of equation of state. Choosing the matter Lagrangian density
$L_m = P = (-1/3)(p_r + 2\,p_t)$,\, with the radial and tangential pressures $p_r= -\rho$ 
and $p_t=\frac{1}{2}(3\,\omega +1)$ respectively, calculated from the 
general form of the energy-momentum tensor
$T_{\mu\nu} = - p_t g_{\mu\nu}+(\rho + p_t)u_\mu u_\nu + (p_r - p_t)v_\mu v_\nu $ of 
an anisotropic fluid, where $u_\mu$ and $v_\mu$ represent the radial-four 
vectors and the four velocity vectors respectively~\cite{Deb_2018}, equation \eqref{eq4a} can be rewritten as
\begin{equation}
	\theta_{\mu\nu}= -2 T_{\mu\nu} - \frac{1}{3}(p_r+2p_t)g_{\mu\nu}
	\label{eq7c}
\end{equation}
Using these results the solution obtained for metric functions of spacetime
in $f(R,T)$ gravity framework for the model \eqref{eq6} is as follows~\cite{Bidyut_2024}
\begin{equation}
A(r) = \frac{1}{B(r)} = 1 -\frac{2 M}{r} + c_2\, r^{-\frac{8 (\beta  \omega + \pi (6 \omega +2))}{16 \pi-\beta  (\omega -3)}},
\label{eq7}
\end{equation}
where $M$ represents the total mass of the BH enclosed by the surrounding 
fluid, $\omega$ is the equation of state parameter of the fluid and $c_2$ is a 
constant. 
It is evident from equation \eqref{eq7} that this BH solution is 
independent of the parameter $\alpha$. Notably, in our 
\begin{figure}[!h]
	\centerline{
		\includegraphics[scale = 0.6]{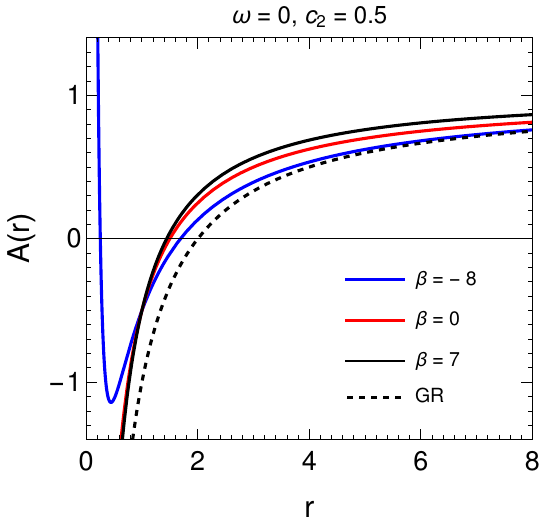}\hspace{0.4cm}
		\includegraphics[scale = 0.6]{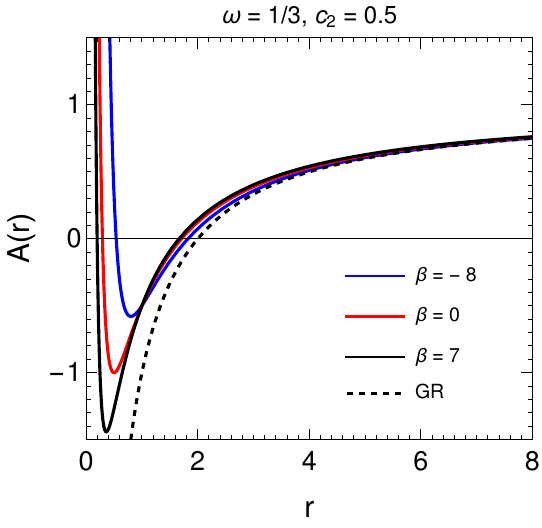}\hspace{0.3cm}
		\includegraphics[scale = 0.62]{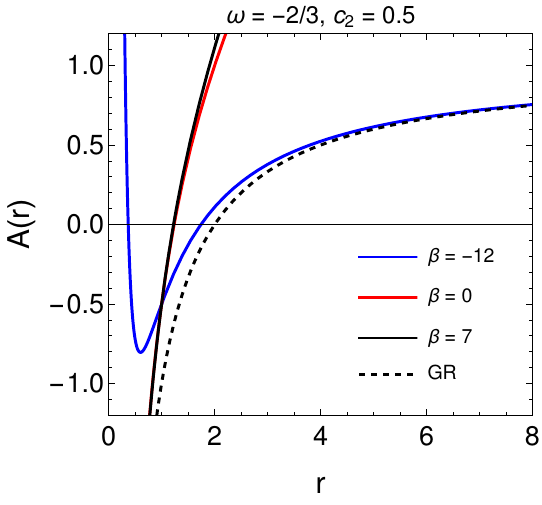}}\vspace{0.2cm}
	\centerline{
		\includegraphics[scale = 0.6]{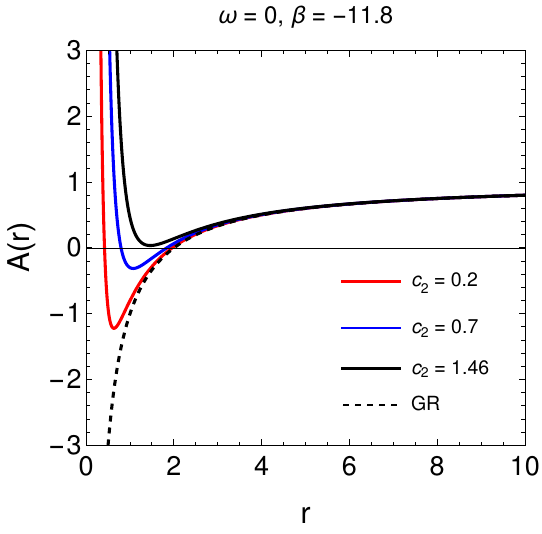}\hspace{0.4cm}
		\includegraphics[scale = 0.6]{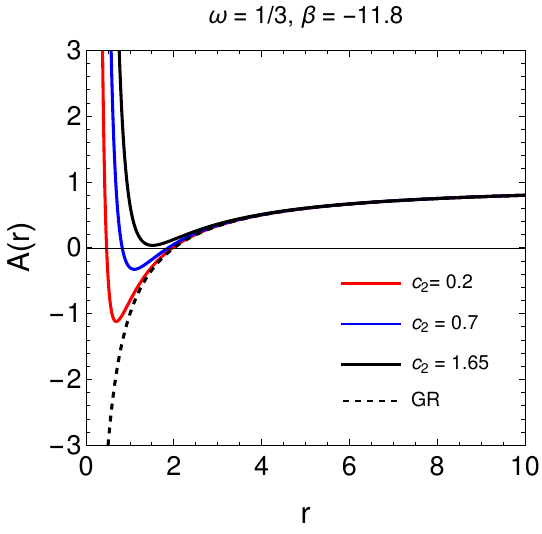}\hspace{0.4cm}
		\includegraphics[scale = 0.6]{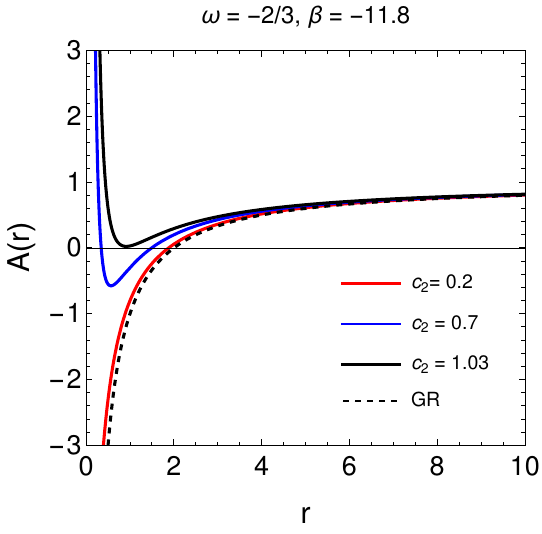}}\vspace{-0.2cm}
	\caption{The horizon structure of minimally coupled $f(R,T)$ 
gravity BHs with different values of $\omega$ as well as for different values 
of model parameters $\beta$ and $c_2$, considering $M = 1$ for all the cases.}
	\label{fig1}
\end{figure}
analysis, we are concerned with three choices of the parameter $\omega$ 
corresponding to dust field ($\omega = 0 $), radiation field ($\omega = 1/3$) 
and quintessence field ($\omega = -2/3$) around the BH which yield three 
different BH solutions. In the absence of fluid, the solution \eqref{eq7} 
simplifies to the 
Schwarzschild BH solution, but simply $\omega \to 0$, $\omega \to 1/3$ or 
$\omega \to -2/3$ do not reduce the solution to a BH surrounded by dust, 
radiation or a quintessence field in the framework of GR due to the presence 
of parameter $\beta$. When $\beta\to 0$, $\omega\to 0$, the solution 
corresponds to the metric of a Schwarzschild BH with an effective mass 
$M_{eff}=2M-c_2$~\cite{Santos_2023}. Similarly, for $\beta \to 0$, 
$\omega \to 1/3$, the solution behaves as the Reissner-Nordstr\"om BH with 
effective charge $q^2_{eff}=c_2$ \cite{Santos_2023,Bidyut_2024}. 

It is also crucial to outline some other key aspects of the 
considered BHs' solution to ensure the completeness of the subsequent 
analyses. As a part, the metric function $A(r)$ is plotted with respect to 
radial distance $r$ as shown in Fig.~\ref{fig1} to analyze the horizon 
structure of the BHs under consideration. It is done by the root analysis of 
equation $A(r) = 0$ for different values of model parameters $\beta$ and $c_2$ 
respectively, as the positive roots of the equation $A(r) = 0$ correspond to 
the locations of horizons of the BHs. In Fig.~\ref{fig1}, we have presented 
three plots in the upper row for $c_2=0.5$, and $\omega=0,1/3,-2/3$ 
respectively, each showcasing different values of $\beta$ along with the GR 
case. It is observed that the BHs can possess at most two horizons for all 
three cases while the number may be less depending on the values of the 
parameters. Similarly, in the lower row, we have plotted three corresponding 
graphs by changing $c_2$ for $\beta = -11.8$ and three aforementioned values of $\omega$ 
respectively along with the GR case as well. Significantly, here also it is 
seen that the 
maximum number of horizons that the BHs can possess does not exceed two at a
fixed value of $\beta$. It is observed that corresponding to each surrounding 
field, for both $\beta$ and $c_2$ varying scenarios, the outer horizon moves 
away from the GR case with increasing values of $\beta$ and $c_2$ respectively.  Remarkably, the BHs exhibit a horizonless nature for  $c_2$ values higher than $1.46, 1.65$ and $1.03$ corresponding to $\omega=0,1/3,-2/3$ respectively with respect to $\beta = -11.8$
as depicted in lower row of Fig~\ref{fig1}.

\begin{figure}[!h]
	\centerline{
		\includegraphics[scale = 0.39]{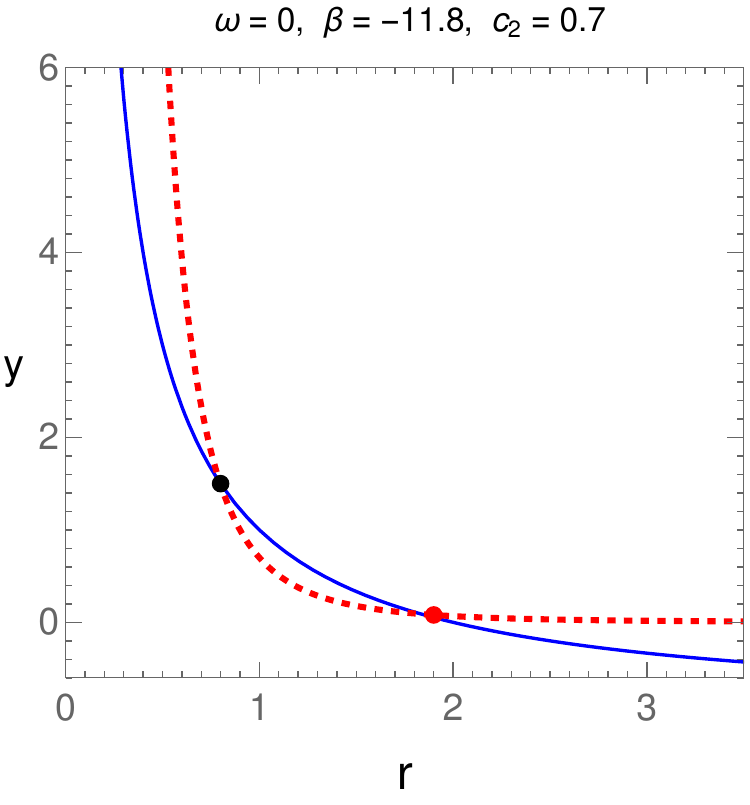}\hspace{1cm}
		\includegraphics[scale = 0.385]{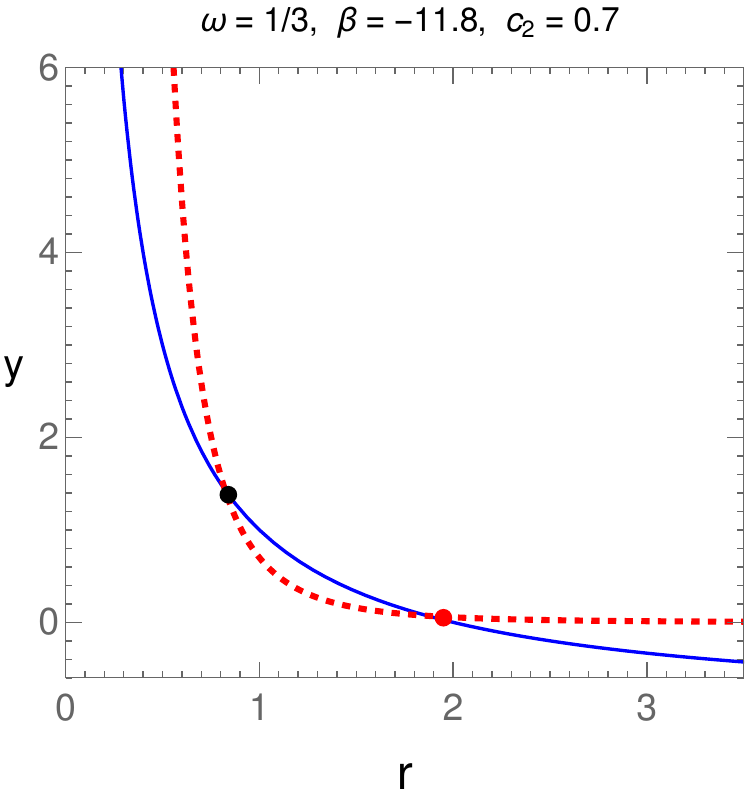}\hspace{.8cm}
		\includegraphics[scale = 0.39]{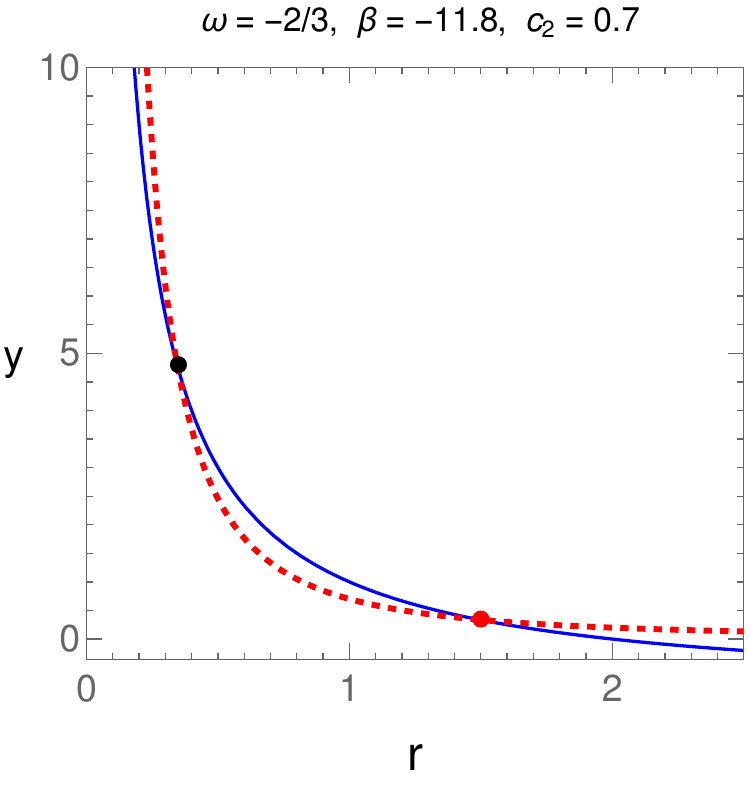}}\vspace{-0.2cm}
\caption{Behavior of $y = 2 M/r - 1 $  (solid line) and 
$y=c_2\, r^{-n}$ 
(red dashed line) as functions of $r$. These plots are presented to examine 
the maximum number of positive real roots of equation $A(r) =0$ of $f(R,T)$ 
BHs for $\omega = 0,\, 1/3, \, -2/3$ cases considering $M = 1$. The black and 
red dots are two intersecting points of the solid curve with the red dashed 
line providing the locations of the inner horizon $r_{ih}$ and the event 
horizon $r_{eh}$ respectively.}
\label{fig1a}
\end{figure}
Furthermore, we adopt a standard graphical method (Fig.~\ref{fig1a}) 
to analyze the number of positive real roots of the equation $A(r)=0$ 
\cite{Strogatz_2015,2017_Azreg}, in which the behavior of both sides of the 
equation $2 M/r - 1 = c_2\, r^{-n}$ obtained from equation $A(r)=0$, where 
$n = [8 (\beta  \omega + \pi (6 \omega +2))]/[16 \pi-\beta  (\omega -3)]$, 
are examined as functions of the radial coordinate $r$. The left-hand side 
(LHS), $2M/r - 1 $, is a monotonically decreasing 
function for $r > 0$, diverging to positive infinity as $r \to 0^+$ and 
asymptotically approaching $-1$ as $r \to \infty$. In contrast, although the 
right-hand side (RHS), $c_2\, r^{-n}$ is also a monotonically decreasing 
function, continues asymptotically to zero as $r \to \infty$ for 
$c_2 > 0$, $n>0$, and diverges to infinity without bound as $r \to 0^+ $. The 
points of intersection of these two graphs correspond to the real positive 
solutions of the equation $A(r)=0$. Depending on the values of $c_2 $ and $n$, 
the system can exhibit the following three distinct scenarios:
\begin{itemize}
	\vspace{-0.1cm}
	\item[(i)] Two positive real roots, when the LHS and RHS curves 
		intersect at two distinct points, indicating two physically relevant solutions.
	\vspace{-0.2cm}
	\item[(ii)] A single positive root, when the curves meet at a single point, marking a critical case.
	\vspace{-0.2cm}
	\item[(iii)] No real roots, when the LHS lies entirely below the RHS for all $r > 0$, indicating the absence of any intersection.
\end{itemize}
Fig.~\ref{fig1a} demonstrates a case that the equation $A(r)=0$ can 
admit a maximum of two real positive solutions, though the actual number may 
reduce to one or zero depending on the parameter values as shown in 
Fig.~\ref{fig1}. Hence, our BHs can have a maximum two number of horizons 
only. It can be inferred that the spacetime geometry of the considered BHs 
differs fundamentally from that of the Schwarzschild BH due to the presence of 
the Cauchy horizon. Thus observing such a horizon structure scenario, we can 
remark that the structure of a minimally coupled $f(R,T)$ gravity BHs changes 
typically and reflects a significant change in the spacetime's causal 
structure as the model parameters vary. 
 
\section{Bending of Light in the Strong Field Limit}
\label{sec.3}
In this section, we explore the bending of light in the strong field limit of 
the $f(R,T)$ gravity BHs spacetime. In the limit of strong gravitational 
lensing, a photon coming from a far distant source passes extremely close to 
a massive object and encounters its intense gravitational field. As a result, 
it is deviated at a closest approach distance $r = r_0$ before moving to the 
observer at infinity. The path of the photon undergoes sharp bending as it 
approaches closer to the object leading to an increased deflection angle. At 
a certain value of $r_0$ the deflection angle becomes $2 \pi$ causing the 
photon to make a complete loop around the BH. A further reduction of $r_0$, 
the photon orbits the BH multiple times, thereby resulting in the deflection 
angle greater than $2\pi$. When $r_0$ becomes equal to the radius of the 
photon sphere $r_p$, the deflection angle diverges \cite{1972_stefan,
2002_Bozza,2017_Naoki,2011_Eiroa,2024_gayatri1}. Here our objective is to 
analyze the deflection angle and related lensing observables in the strong 
field limit for a static, spherically symmetric BH spacetime described by 
the line element \eqref{eq5}. For this, we assume photon trajectories 
confined in the equatorial plane $(\theta= \pi/2)$ of the object and adopt 
the strong lensing effect investigation method developed by V.~Bozza, 
suitable for the motion of photons described by a standard geodesic equation 
in any spacetime and within any gravitational framework \cite{2002_Bozza}. 
Notably, the entire trajectory of the photon is governed by the equation, 
$g_{\mu\nu} k^{\mu} k^{\nu} = 0,$ where $k^{\mu} = dx^\mu/d\tau$ 
represents the wave number of the photon and $\tau$ is the affine parameter 
along its trajectory \cite{2017_Naoki,2024_islum}. Thus, considering 
$C(r) = r^2$ in equation \eqref{eq5}, the equation of photon trajectories can 
be written as \cite{1985_chandra,2020_islum}
\begin{equation}
	 A(r)\,\dot{t}^2 - B(r)\,\dot{r}^2 - C(r)\, \dot{\phi}^2 = 0.
	\label{eq8}
\end{equation}

As the spherical symmetry of the metric \eqref{eq5} ensures the conservation 
of energy $E = A(r)\dot{t}$ and angular momentum $L = r^2 \dot{\phi}$ 
respectively during the journey, hence specified by these two constant 
quantities and defining impact parameter $\zeta$ as $\zeta = L/E$, the above 
equation of trajectories \eqref{eq8} can be rewritten as 
\cite{1985_chandra,2017_Naoki}
\begin{equation}
	\dot{r}^2 = V_{eff}(r),
	\label{eq9}
\end{equation}
where $V_{eff}(r)$ is the effective radial potential for the motion of a 
photon and is defined as
 \begin{equation}
 V_{eff}(r) = E^2\left[1 - \zeta^2 \frac{A(r)}{r^2}\right].
 \label{eq10}
 \end{equation}
\begin{figure}[!h]
        \centerline{
                \includegraphics[scale = 0.42]{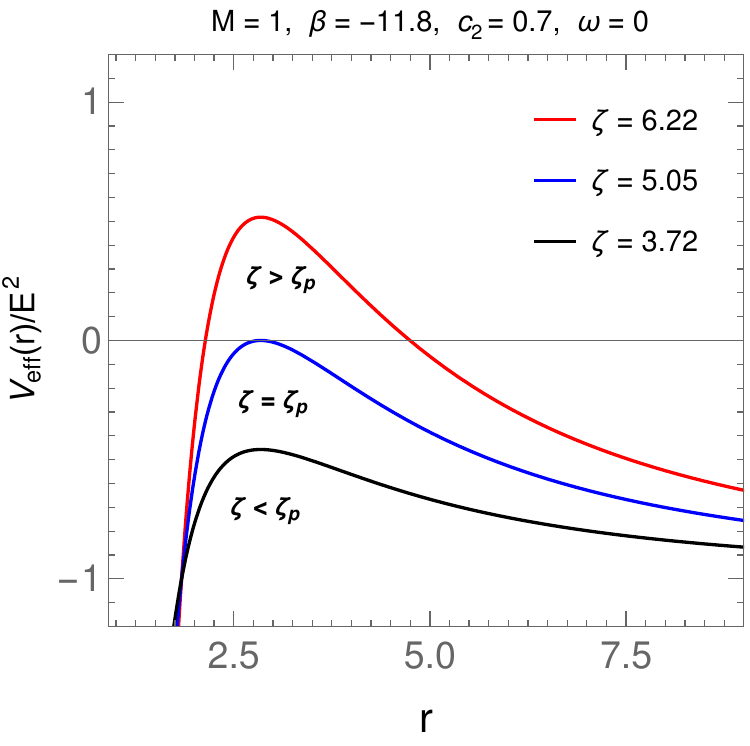}\hspace{0.4cm}
                \includegraphics[scale = 0.42]{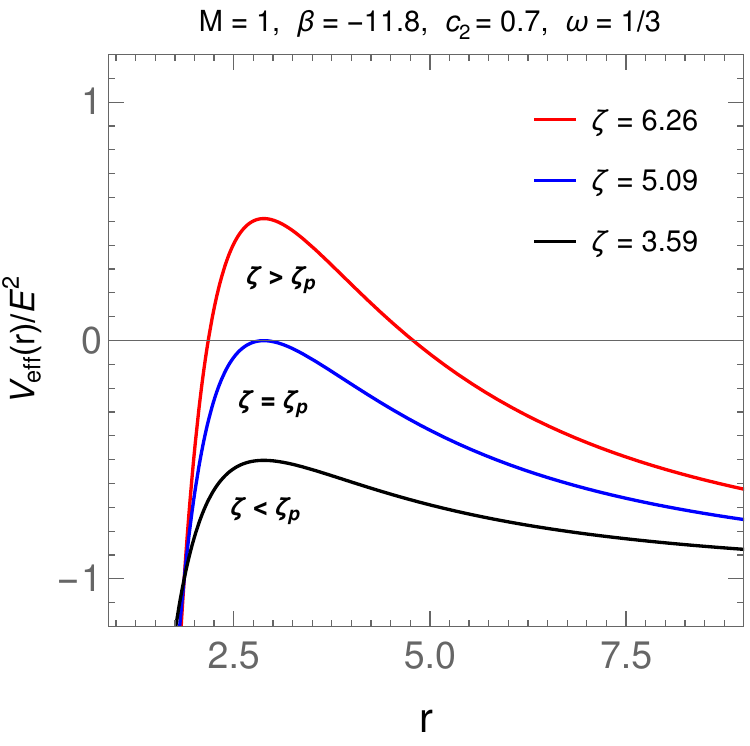}\hspace{0.4cm}
                \includegraphics[scale = 0.42]{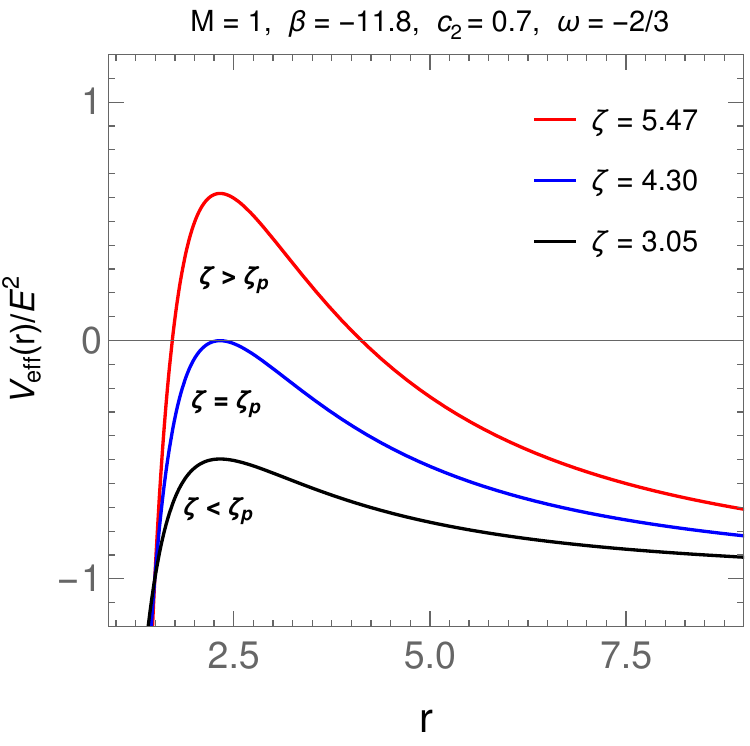}}\vspace{-0.2cm}
        \caption{Variation of the effective potential $V_{eff}(r)$ for 
different values of $\omega$ as a function of radial coordinate $r$ for 
different values of impact parameter $\zeta$.}
\label{fig2}
\end{figure}
The above equation characterizes various possible photon orbits, which are 
graphically represented in Fig.~\ref{fig2} for $\omega = 0,$ $1/3$ and $ -2/3$. 
The figure depicts the effective radial potential corresponding to different 
impact parameter values such as $\zeta<\zeta_p$,\, $\zeta=\zeta_p$ and 
$\zeta >\zeta_p$, which describe three different orbits of a photon. Here, 
$\zeta_p$ is the critical or minimum impact parameter defined at the photon 
sphere radius $r_p$ \cite{2024_islum,2024_gayatri1}, at a point 
where $dV_{eff}(r/dr = 0$. A photon with 
impact parameter $\zeta<\zeta_p$ is captured by the strong field of the BHs, 
while one with $\zeta >\zeta_p$, is deflected by the BHs when reaches 
$r=r_0$ and with $\zeta=\zeta_p$, whirls the BHs for sometimes at photon sphere 
radius $r_p$ before leaving it~\cite{2024_wang}. The photon sphere radius is  
the largest positive root of the equation \cite{2000_virbhadra,2002_Bozza,Virbhadra_2000}, 
\begin{equation}
	\frac{C^{\prime}(r)}{A^{\prime}(r)} = \frac{C(r)}{A(r)}.
	\label{eq11}
\end{equation}
For our BHs' solution \eqref{eq7} this equation takes the form:
\begin{equation}
\frac{3 M}{r}-\frac{4 c_2 (\beta  \omega +\pi  (6 \omega +2)) r^{-\frac{8 (\beta  \omega +\pi  (6 \omega +2))}{16 \pi -\beta  (\omega -3)}}}{16 \pi -\beta  (\omega -3)}- c_2\, r^{-\frac{8 (\beta  \omega +\pi  (6 \omega +2))}{16 \pi -\beta  (\omega -3)}}-1=0.
\label{eq12}
\end{equation}
Due to the complexity of this equation, the photon sphere radius $r_p$ for 
different classes of BHs are computed numerically for a range of values
of $\beta$ and $c_2$, and graphical representations of their dependence on 
the parameters are shown in Fig.~\ref{fig3}. Now, as an example, to specify the 
location of the photon sphere for specific values of $\beta$ and $c_2$, we 
illustrate the behavior of both sides of 
the equation $2M/r -1 =  (n/2 + 1)c_2\, r^{-n} - M/r$, obtained from 
Eq.~\eqref{eq12} in Fig.~\ref{fig1b} as functions of $r$. It is seen from the
figure that the curve for the RHS of the above equation intersects the solid 
curve 
for $2M/r -1$ at two points, one of them (red dot marked point in  Fig.~\ref{fig1b}) corresponds to the peak of the effective potential $V_{eff}$ (see Fig.~\ref{fig2}), giving 
the
position of the photon sphere in the region beyond event horizon
$r_{eh}$ (see Fig.~\ref{fig1a}).
\begin{figure}[!h]
	\centering{
	    \includegraphics[scale=0.53]{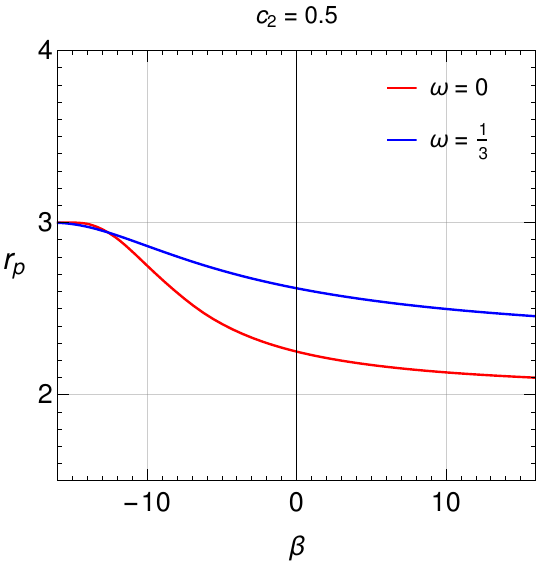}\hspace{1cm}
		\includegraphics[scale=0.54]{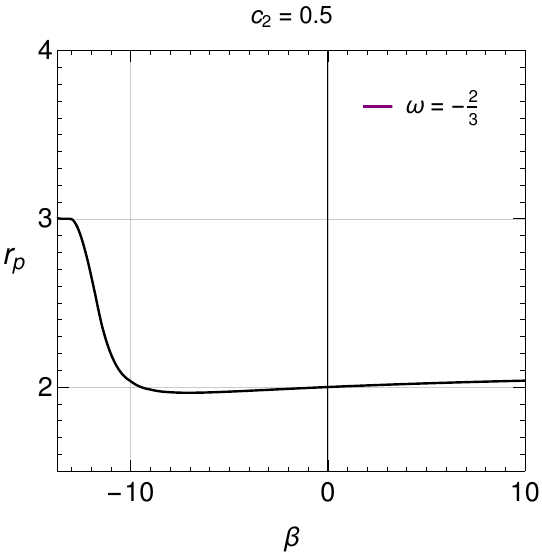}\hspace{1cm}
		\includegraphics[scale=0.53]{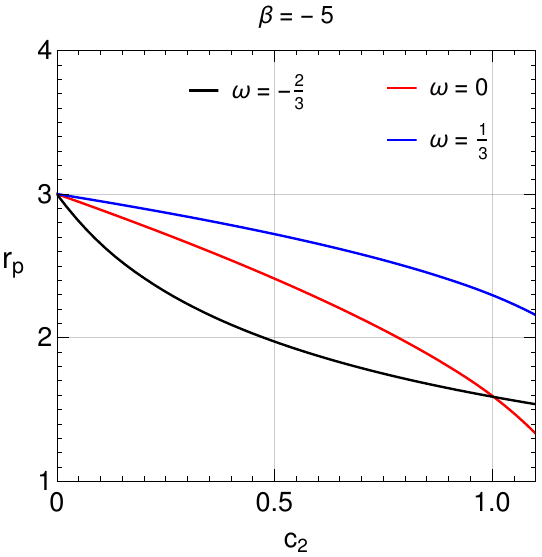}}\vspace{0.2cm}
		\centering{
        \includegraphics[scale=0.53]{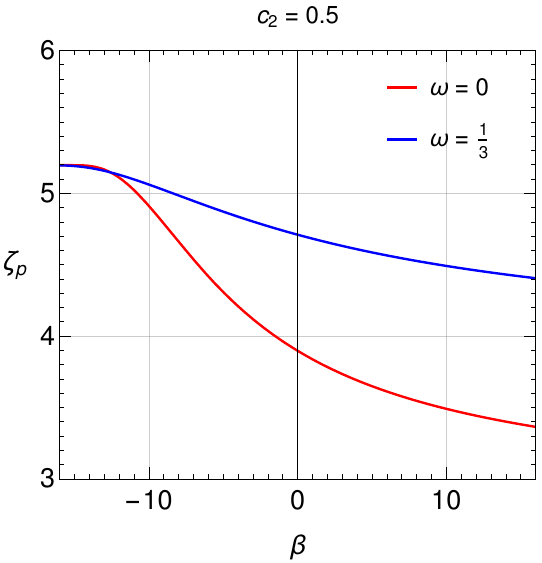}\hspace{1cm}
		\includegraphics[scale=0.54]{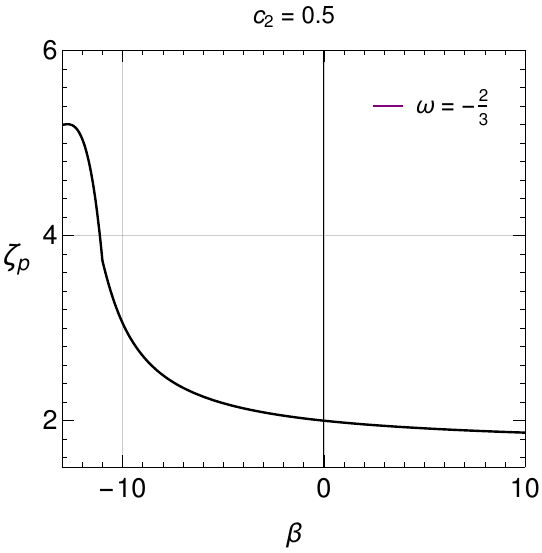}\hspace{1cm}
		\includegraphics[scale=0.53]{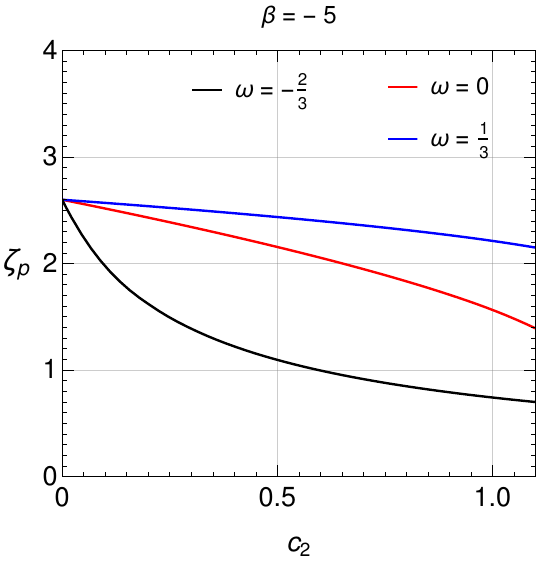}}\vspace{-0.2cm}
	\caption{Behaviors of photon sphere radius $r_p$ and critical impact 
parameter $\zeta_p$ for minimally coupled $f(R,T)$ gravity BHs as a function 
of model parameter $\beta$ and $c_2$ for $\omega = 0,$ $1/3$ and $ -2/3$ respectively.}
\label{fig3}
\end{figure}
\begin{figure}[!h]
        \centerline{
                \includegraphics[scale = 0.39]{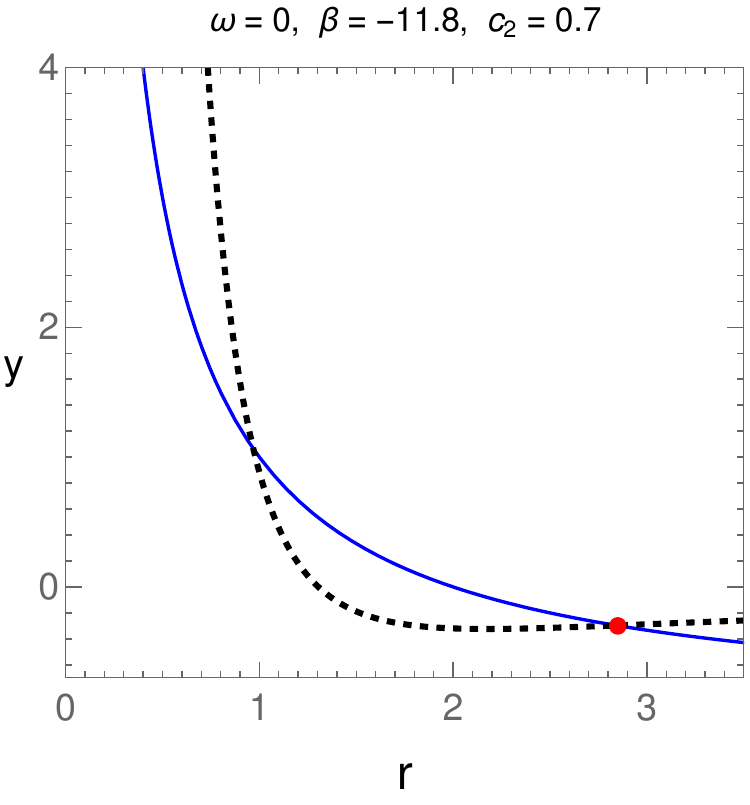}\hspace{1cm}
                \includegraphics[scale = 0.39]{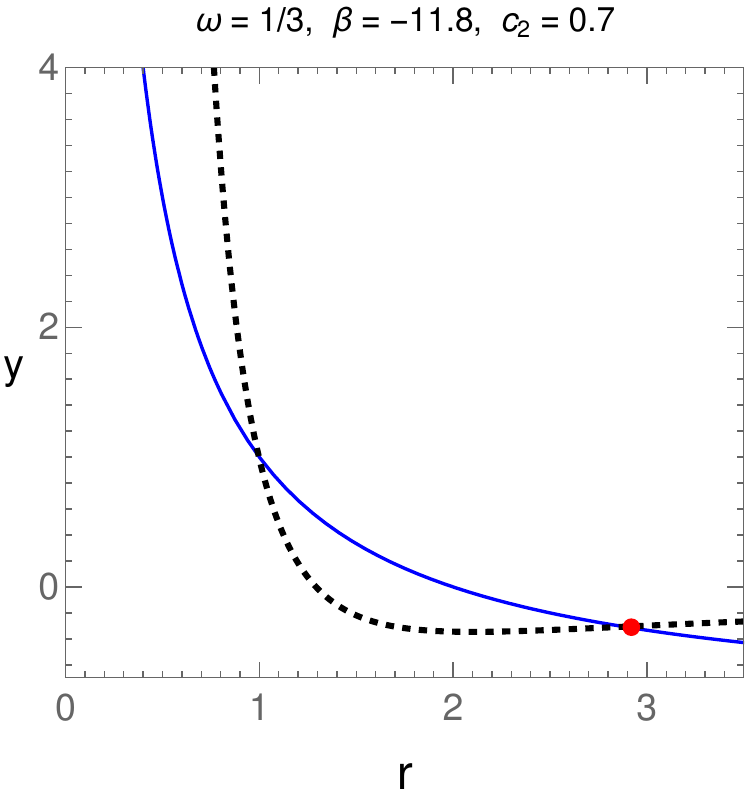}\hspace{1cm}
                \includegraphics[scale = 0.39]{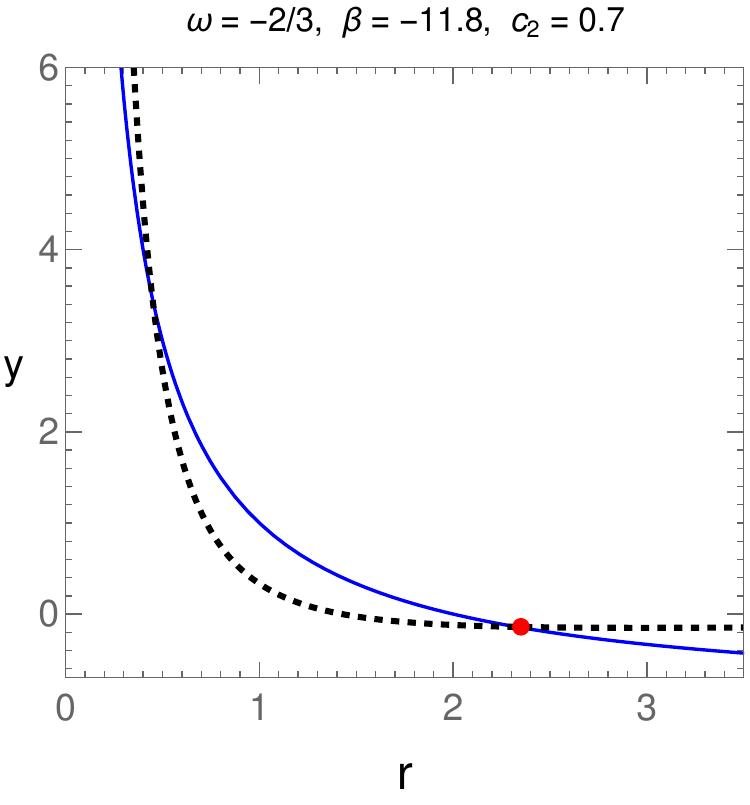}}\vspace{-0.2cm}
        \caption{Behavior of $y = 2 M/r - 1 $  (solid line) 
and $y = (n/2 + 1)c_2\, r^{-n} - M/r$ (black dashed line) as functions of 
$r$. These plots are presented to locate the photon sphere radius $r_p$ of 
$f(R,T)$  BHs for $\omega = 0,\, 1/3, \, -2/3$ considering $M = 1$. The red 
dot mark in each plot is the intersecting point of the solid curve and the black dashed 
line corresponding to the effective potential peak specifying the location of the photon sphere radius.}
\label{fig1b}
\end{figure}
Furthermore, at the point of closest approach $r = r_0$ 
the impact parameter $\zeta_0$ can be expressed from the orbit equation 
\eqref{eq9} as $\zeta_0=\sqrt{C(r_0)/A(r_0)}$. This expression allows us to 
determine the critical impact parameter at $r_0 = r_p$ as given below:
 \begin{equation} 
	\zeta_p=\sqrt{\frac{C(r_p)}{A(r_p)}} = \frac{r_p}{\sqrt{1 -\frac{2 M}{r_p} + \, c_2 r_p^{-\frac{8 (\beta  \omega + \pi (6 \omega +2))}{16 \pi-\beta  (\omega -3)}}}}.
	\label{eq13}
\end{equation}
We depict its behavior as a function of $\beta$ in Fig.~\ref{fig3} along with 
the same for $r_p$ for all three cases. The figure demonstrates that both 
$r_p$ and $\zeta_p$ for $\omega = -2/3$ show steep initial fall in comparison 
to $\omega = 0$ and $1/3$ cases.

\subsection{Deflection angle in minimally coupled $f(R,T)$ gravity spacetime}
The deflection angle $\hat{\alpha}$ which is formed between the asymptotic 
paths of the incoming and outgoing photons can be expressed using orbit 
equation \eqref{eq9} as \cite{1972_weinberg,1998_virbhadra,2002_Bozza}
\begin{equation} 
	\hat{\alpha}(r_0)=I(r_0) - \pi,  
	\label{eq14}
\end{equation}
where 
\begin{equation} 
	I(r_0)= 2\!\int_{r_0}^{\infty}\!\frac{1}{\sqrt{A(r)C(r)}\sqrt{\frac{C(r)A(r_0)}{C(r_0)A(r)}-1}}\,  dr.
	\label{eq15}
\end{equation}  
As mentioned earlier, we focus on the investigation of the deflection angle in 
the strong field limit where photon trajectories approach closely to the 
photon sphere of the lensing BHs. For completeness of our analysis following 
the Bozza's method we define a new variable $z$ as
\cite{2017_Naoki,2022_kumar,Chen_2009}
\begin{equation} 
	z = 1 - \frac{r_0}{r},  
	\label{eq16}
\end{equation}
and rewrite the integral \eqref{eq15} as \cite{2002_Bozza,Chen_2009}
\begin{equation} 
	I(r_0)=\int_{0}^1\! N(z,r_0)\, g(z,r_0)\, dz,
	\label{eq17}
\end{equation}
where the functions $N(z,r_0)$ and $g(z,r_0)$ can be written as follows \cite{zhang_2017,Chen_2009,2022_kumar}: 
\begin{equation} 
	N(z,r_0)= \frac{2 r^2\sqrt{A(r)B(r)C(r_0)}}{r_0\,C(r)},
	\label{eq18}
\end{equation}
and
\begin{equation} 
	g(z,r_0)= \frac{1}{\sqrt{A(r_0)- \frac{A(r)}{C(r)}\,C(r_0)}}.
	\label{eq19}
\end{equation}
For any value of $z$ and $r_0$ the function $N(z,r_0)$ is regular, whereas 
$g(z,r_0)$ diverges as the photon approaches the photon sphere, i.e.~when 
$z\to0$. Therefore, the integral \eqref{eq17} can be decomposed into two 
parts, the regular part $I_R(r_0)$ and the divergence part $I_D(r_0)$ as 
given below \cite{2002_Bozza}:  
\begin{equation} 
	I(r_0)= I_D(r_0) + I_R (r_0),
	\label{eq20}
\end{equation}
where
\begin{equation} 
	I_D(r_0)=\int_{0}^1\!N(0,r_p)\, g_0(z,r_0)\,dz,
	\label{eq21}
\end{equation}
\begin{equation} 
	I_R(r_0)=\int_{0}^1\!\big[N(z,r_0)\, g(z,r_0)- N(0,r_p)\, g_0(z,r_0)\big]\,dz.
	\label{eq22}
\end{equation}

Next, to approximate the rate of divergence, the argument of the square root 
of $g(z,r_0)$ is expanded to second-order in $z$ which results 
\cite{2002_Bozza,Ding_2011}:
\begin{equation} 
	 g(z,r_0)\sim g_0(z,r_0)= \frac{1}{\sqrt{m(r_0)\, z+ n(r_0)\, z^2}},
	\label{eq23}
\end{equation}
where the coefficients $m(r_0)$ and $n(r_0)$ are obtained as 
\cite{Wang_2016,Chen_2012} 
\begin{align} 
	m(r_0) & = \frac{r_0}{C(r_0)}\left[A(r_0)\, C^\prime (r_0) - A^\prime(r_0)\, C(r_0) \right]\notag \\[5pt]
	& = 2 - \frac{6 M}{r_0} + \left[1 + \frac{4 (\beta  \omega + \pi (6 \omega +2))}{(16 \pi-\beta  (\omega -3))}\right]2\,c_2\,r_0^{\!-\frac{8 (\beta  \omega + \pi (6\, \omega +2))}{16 \pi-\beta  (\omega -3)}},
	\label{eq24}\\[8pt]	
	n(r_0) & = \frac{r_0}{2\, C^2 (r_0)} \left[2\Big(C(r_0) - r_0\, C^\prime (r_0)\Big) \Big(A(r_0) \,C^\prime(r_0) - A^\prime(r_0)\, C(r_0)\Big) + r_0 C(r_0) \Big(A(r_0)\, C^{\prime\prime}(r_0) - A^{\prime\prime}(r_0) C(r_0)\Big)\right] \notag \\[5pt]
	&= \frac{6 M}{r_0}  - \left[1 + \frac{8 \,(\beta  \omega +\pi  (6 \omega +2))}{16 \pi -\beta  (\omega -3)} + \frac{4 \,(\beta  \omega +\pi  (6 \omega +2)) \left(\frac{8 (\beta  \omega +\pi  (6 \omega +2))}{16 \pi -\beta  (\omega -3)}+1\right)}{16 \pi -\beta  (\omega -3)} \right]c_2\,r_0^{-\frac{8 (\beta  \omega +\pi  (6 \omega +2))}{16 \pi -\beta  (\omega -3)}} - 1.
	\label{eq25}
\end{align}
Eventually, when $r_0$ coincides with the radius of the photon sphere $r_p$, 
the coefficient $m(r_0)$ becomes zero, thereby leaving $z^{-1}$ as the leading 
divergence term in the function $g_0(z,r_0)$. It results in the logarithmic 
divergence of the integral \eqref{eq17}. Accordingly, in the strong field limit 
an expression of the deflection angle in terms of impact parameter $\zeta$ can 
be approximated as \cite{2002_Bozza,Kumar_2020}
\begin{equation}
\hat{\alpha}(\zeta)=-\,\bar{a}\log \left(\frac{\zeta}{\zeta_p}-1\right) + \bar{b} + \mathcal{O} (\zeta-\zeta_p),
	\label{eq26}
\end{equation}
where $\bar {a}$, $\bar{b}$ represent the strong field limit lensing 
coefficients which are dependent on spacetime geometry. These coefficients 
are evaluated at $r_0 = r_p$ and can be given by the following 
expressions \cite{2002_Bozza,Chen_2012,Chen_2009}
\begin{align} 
	\bar{a} & = \frac{N(0,r_p)}{2\sqrt{n(r_p)}},
	\label{eq27}\\[8pt]
	\bar{b} & = -\,\pi + b_R + b_D, 
	\label{eq28}
\end{align}
where $b_D$ and $b_R$ are defined as \cite{Chen_2009,zhang_2017}
\begin{align}
	b_D & =\bar{a}\, \log \Bigg[\frac{r_p^2[C^{\prime\prime}(r_p) \,A(r_p) - C(r_p)\,A^{\prime\prime}(r_p)]}{\zeta_p\,\sqrt{A^3(r_p)\,C(r_p)}}\Bigg] \notag\\[5pt]
	& =\bar{a}\, \log \Bigg[2 + \frac{[4 M\,(16 \pi -\beta  (\omega -3))-8 c_2\,(\beta  \omega +\pi  (6 \omega +2)) \left(\frac{8 (\beta  \omega +\pi  (6 \omega +2))}{16 \pi -\beta  (\omega -3)}+1\right) r_p^{-\frac{8 (\beta  \omega +\pi  (6 \omega +2))}{16 \pi -\beta  (\omega -3)}+1}]}{(-\frac{2 M}{r_p}+ c_2\,r_p^{-\frac{8 (\beta  \omega +\pi  (6 \omega +2))}{16 \pi -\beta  (\omega -3)}}+1)(16 \pi -\beta  (\omega -3))r_p}\Bigg], 
	\label{eq29}\\[8pt]
	b_R & = I_R(r_p) = \int_{0}^1\!\big[N(z,r_p)\, g(z,r_p) -N(0,r_p)\, g_0(z,r_p)\big]\,dz.
	\label{eq30}
\end{align}
\begin{figure}[!h]
	\centering{
	\includegraphics[scale=0.55]{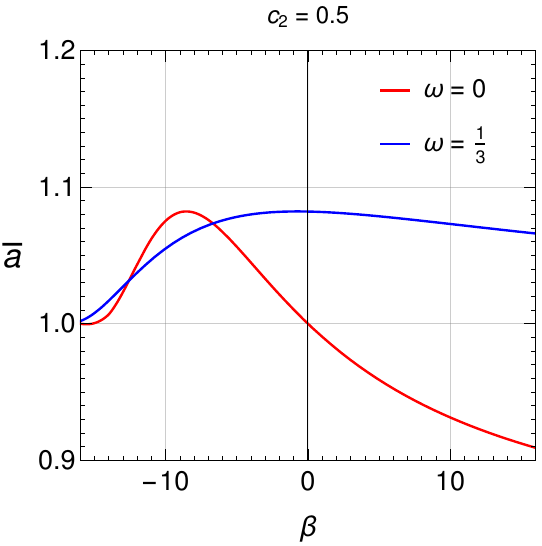}\hspace{1cm}
	\includegraphics[scale=0.55]{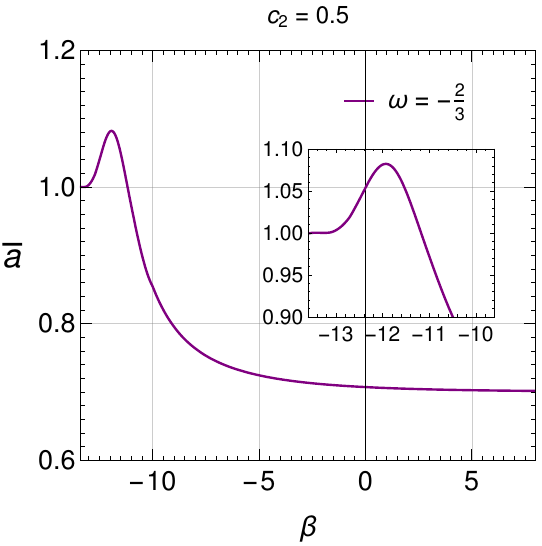}\hspace{1cm}
    \includegraphics[scale=0.55]{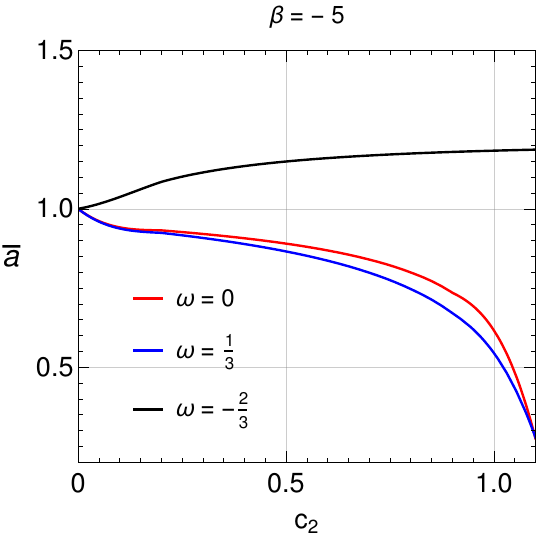}}\vspace{0.2cm}
		\centering{
		\includegraphics[scale=0.57]{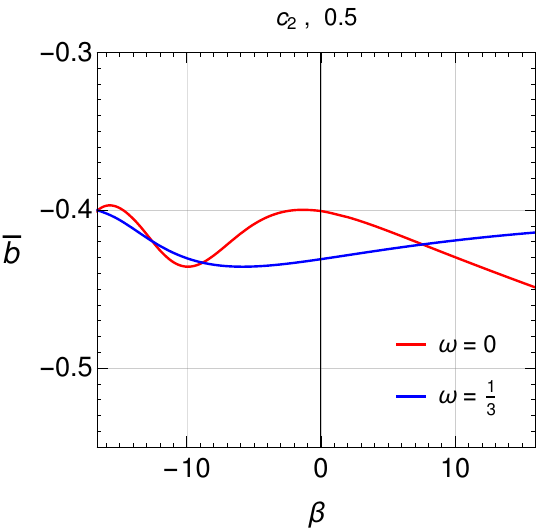}\hspace{.9cm}
		\includegraphics[scale=0.57]{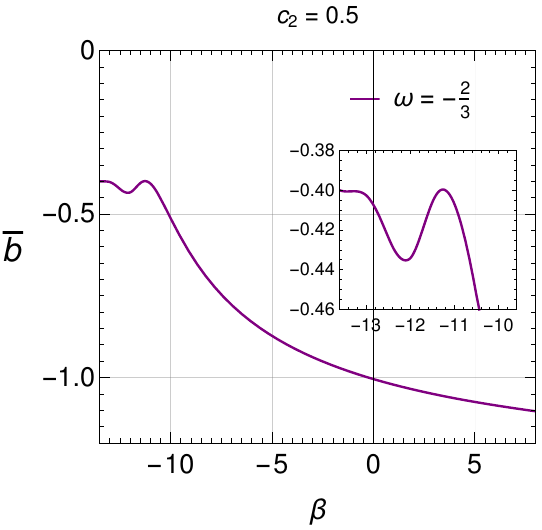}\hspace{.9cm}
		\includegraphics[scale=0.57]{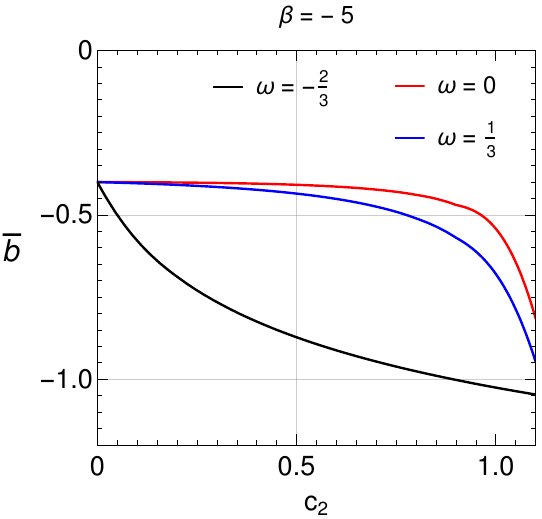}}\vspace{-0.2cm}
	\caption{Behaviors of lensing coefficients $\bar{a}$ (upper 
row) and $\bar{b}$ (lower row) in minimally coupled $f(R,T)$ gravity BHs 
spacetime as functions of model parameters $\beta$ and $c_2$ for 
$\omega = 0$, $1/3$ and $-2/3$ cases.}
	\label{fig4}
\end{figure}
To examine the behavior of the lensing coefficients in our modified BHs 
spacetime, we analyze their variation with model parameters $\beta$ and $c_2$ 
in Fig.~\ref{fig4} and list their values in Tables \ref{table1}, \ref{table2} 
and \ref{table2a}. For the $\omega = 0$ case, the coefficient $\bar{a}$ 
initially increases moderately, reaches a peak around $\beta \approx -8$, then 
decreases as $\beta$ grows. When $\omega = 1/3$, $\bar{a}$ exhibits a 
continuous but slow growth up to $\beta \approx- 4.7$ then decreases smoothly 
as $\beta$ increases.  In contrast, the coefficient $\bar{b}$ shows opposite 
trends. For the $\omega = 0$ case, $\bar{b}$ decreases 
initially, reaches a minimum near $\beta \approx -10$, then increases as 
$\beta$ approaches to $-2$. After that it follows a smooth declination for 
$\beta>0$, while for the $\omega = 1/3$ case, $\bar{b}$ falls gradually up to 
$\beta \approx -7.5$ from the negative end and shows a slow growth beyond this 
value. On the other hand, for the $\omega = -2/3$ case, we observe significant 
variations in the lensing coefficients within the range of 
$-13.5 \leq\beta \leq-7.5$. Similar to the cases of $\omega = 0$ and 
$\omega = 1/3$, coefficients $\bar{a}$ and $\bar{b}$ exhibit analogous 
behavior in this case also. However, for $\beta > -7.5$, $\bar{a}$ goes 
almost as flat and $\bar{b}$ declines rapidly than in the previous two 
cases beyond $\beta=-11$. In fact, we observe contrasting behaviors of the 
lensing coefficients as a function of $\beta$ across these scenarios. 
Moreover, these coefficients show nearly identical behavior with 
respect to parameter $c_2$ for both $\omega = 0$ and $\omega = 1/3$ cases. 
It is observed that initially they decrease slowly and then undergo a sudden 
fall beyond $c_2=1$. In contrast, the coefficients exhibit opposite trends 
with the parameter $c_2$ in the $\omega = -2/3$ case as presented in 
Fig.~\ref{fig4}.  

\begin{figure}[!h]
	\centerline{
		\includegraphics[scale = 0.6]{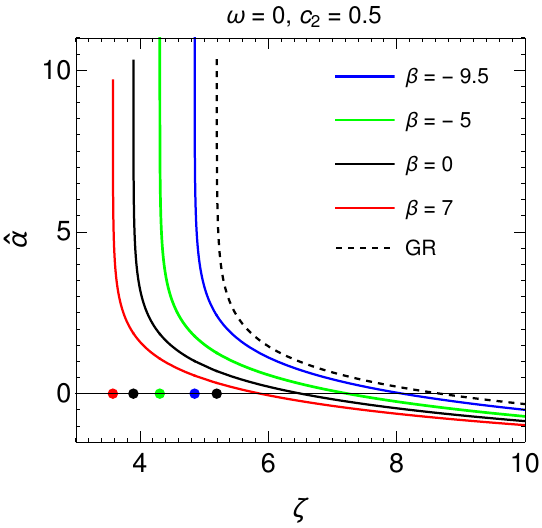}\hspace{.4cm}
		\includegraphics[scale = 0.6]{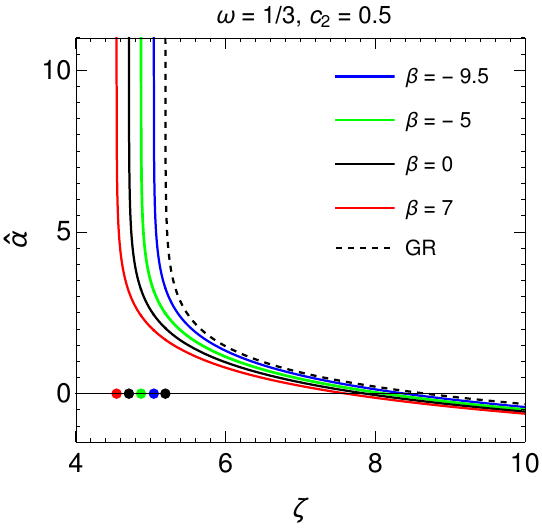}\hspace{.4cm}
		\includegraphics[scale = 0.6]{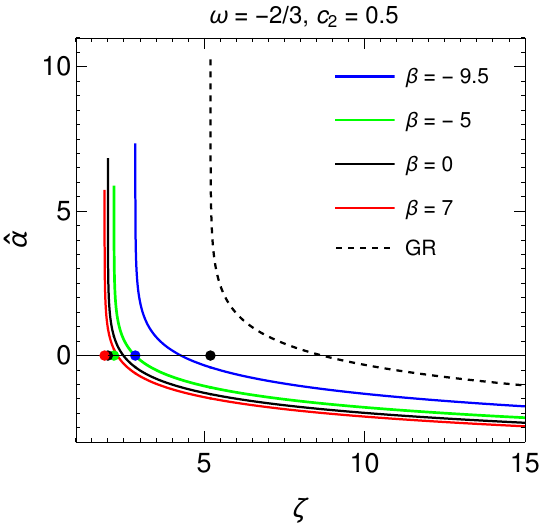}}\vspace{0.2cm}
	\centerline{
		\includegraphics[scale = 0.6]{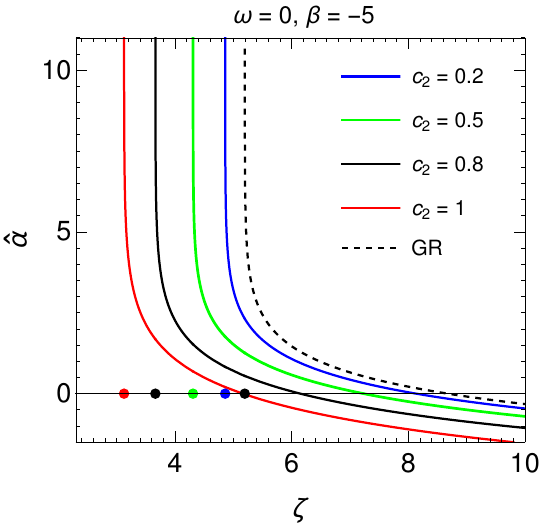}\hspace{.4cm}
		\includegraphics[scale = 0.6]{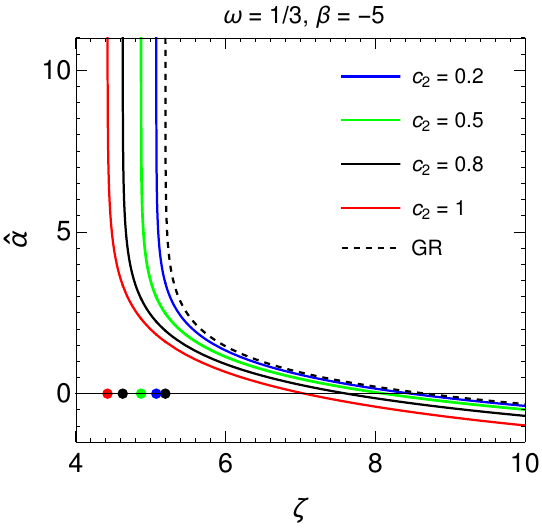}\hspace{.4cm}
		\includegraphics[scale = 0.61]{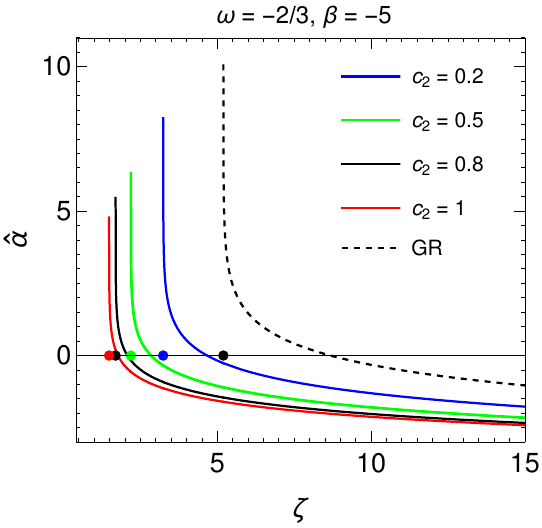}}\vspace{-0.05cm}
	\caption{Variations of the deflection angle with the impact parameter 
$\zeta$ for different values of the model parameters $\beta$ and $c_2$ for 
$\omega = 0,$ $1/3$ and $-2/3$ cases respectively, where the black dashed 
curves represent the Schwarzschild BH. The points on the horizontal axes 
represent the values of the impact parameter $\zeta = \zeta_p$ at which the 
deflection angle diverges. $M=1$ is taken for all the cases.}
	\label{fig5}
\end{figure}

Also we illustrate the variation in the deflection angle resulting from the 
bending of light in the strong field limit of minimally coupled $f(R,T)$ 
gravity BHs as given by equation \eqref{eq26}. This variation is analyzed as 
a function of the impact parameter $\zeta$ for different values of model 
parameters $\beta$ and $c_2$, employing the values of $\zeta_p$ and lensing coefficients 
$\bar{a}$ and $\bar{b}$ determined from equations \eqref{eq13}, \eqref{eq27} 
and \eqref{eq28} respectively. To compute $\bar{a}$ we rely on equations 
\eqref{eq18} and \eqref{eq25}, evaluated at $r_0 = r_p$. Meanwhile, the 
coefficient $\bar{b}$ is calculated by estimating $b_R$ numerically and 
determining $b_D$ using equations \eqref{eq27} in \eqref{eq29}. Thus, the 
behaviors of the strong limit deflection angle $\hat{\alpha}(\zeta)$ are 
depicted for varying values of the model parameters $\beta$ and $c_2$, 
separately for the cases of $\omega = 0,$  $1/3$ and $-2/3$ respectively 
in Fig.~\ref{fig5}.
\begin{table}[!h]
	\centering
	\caption{Estimation of numerical values of lensing coefficients and 
		corresponding impact parameter $\zeta_p$ with respect 
to different values of parameters $\beta$ and $c_2$ for $f(R,T)$ BHs with 
$\omega = 0$ case. The impact parameter is listed as $\zeta_p/R_s$, 
where $R_s$ is the Schwarzschild radius.}
	\vspace{0.2cm}
	\setlength{\tabcolsep}{7pt}
	\scalebox{0.99}{
		\begin{tabular}{c c c c  || c c c c|| c c c c} \hline\\[-12pt]
			$\beta$&~$\bar{a}$~&$\bar{b}$ & $\zeta_p/R_s$&$\beta$ &$\bar{a}$~&$\bar{b}$ &$\zeta_p/R_s$ &$c_2$ &$\bar{a}$~&$\bar{b}$ &$\zeta_p/R_s$ 
			\\[2pt] \hline \hline\\[-12pt]  		  	 
			- 16.731    & 1.00000   &  -0.40064   &  2.59808  & SBH    & 1.00000 & -0.40063  &  2.59808 & 0   & 1       &-0.40064 & 1.29904\\ [5pt] 
			- 14        & 1.00660   &  -0.40543   &  2.59560  & 2      & 0.98156 & -0.40447  &  1.89158 & 0.2 & 0.93188 &-0.40192 & 1.21509\\ [5pt]
			- 12        & 1.04324   &  -0.42545   &  2.55485  & 5      & 0.95891 & -0.41319  &  1.82442 & 0.4 & 0.90693 &-0.40549 & 1.12510\\ [5pt]
			- 9.5       & 1.07920   &  -0.43546   &  2.42575  & 7      & 0.94658 & -0.41979  &  1.78873 & 0.5 & 0.89005 &-0.40895 & 1.07718\\ [5pt]  
			- 8         & 1.08128   &  -0.42885   &  2.33060  & 10     & 0.93135 & -0.42981  &  1.74508 & 0.8 & 0.79887 &-0.43947 & 0.91595\\ [5pt]  
			0         & 0.99998   &  -0.40061   &  1.94855  & 12     & 0.92291 & -0.43624  &  1.72106 & 0.9 & 0.73501 &-0.47014 & 0.85301 \\[2pt]\hline

	\end{tabular}}
	\label{table1}
\end{table}

\begin{table}[!h]
	\centering
	\caption{Estimation of numerical values of lensing coefficients and 
		corresponding impact parameter $\zeta_p$ with respect to different values of parameters $\beta$ and $c_2$ for $f(R,T)$ BHs with 
		$\omega =1/3$ case. The impact parameter is listed as $\zeta_p/R_s$, 
		where $R_s$ is the Schwarzschild radius.}
	\vspace{0.2cm}
	\setlength{\tabcolsep}{7pt}
	\scalebox{0.99}{
		\begin{tabular}{ c c c  c ||   c c c c||c c c c} \hline\\[-12pt]
			$\beta$&$\bar{a}$&$\bar{b}$ &$\zeta_p/R_s$&$\beta$ &$\bar{a}$&$\bar{b}$ &$\zeta_p/R_s$&$c_2$ &$\bar{a}$&$\bar{b}$ &$\zeta_p/R_s$ 
			\\[2pt] \hline \hline\\[-12pt]   
			- 18.8495     & 1        &  -0.40064   &  2.59808 & SBH     & 1       & -0.40063  &  2.59808& 0   & 1       &-0.40064  & 1.29904  \\ [5pt]
			- 14       & 1.01666  &  -0.41168   &  2.58877 & 2       & 1.08114 & -0.42854  &  2.32791& 0.2 & 0.92310 &-0.40918  & 1.26873  \\ [5pt]
			- 12       & 1.03728  &  -0.42266   &  2.56476 & 5       & 1.07858 & -0.42476  &  2.29274& 0.4 & 0.88874 &-0.42410  & 1.23564  \\ [5pt]
			- 9.5      & 1.05859  &  -0.43179   &  2.52108 & 7       & 1.07644 & -0.42241  &  2.27229& 0.5 & 0.86539 &-0.43570  & 1.21778  \\ [5pt]
			- 8        & 1.06756  &  -0.43458   &  2.49219 & 9       & 1.07413 & -0.42024  &  2.25391& 0.8 & 0.74655 &-0.51098  & 1.15675  \\ [5pt]
			0        & 1.08204  &  -0.43109   &  2.35480 & 12      & 1.07059 & -0.41736  &  2.22966& 0.9 & 0.67062 &-0.569633 & 1.13271  \\[2pt]\hline   
	\end{tabular}}
	\label{table2}
\end{table}
\begin{table}[!h]
	\centering
	\caption{Estimation of numerical values of lensing coefficients and 
corresponding impact parameter $\zeta_p$ \textbf{with respect to different 
values of parameters $\beta$ and $c_2$ for $f(R,T)$ BHs} with $\omega =-2/3$ case. The impact parameter is listed as $\zeta_p/R_s$, where $R_s$ is the 
Schwarzschild radius.}
	\vspace{0.2cm}
	\setlength{\tabcolsep}{7pt}
	\scalebox{0.99}{
		\begin{tabular}{ c c c  c || c c c c ||c c c c} \hline\\[-12pt]
			$\beta$&$\bar{a}$&$\bar{b}$ &$\zeta_p/R_s$&$\beta$&$\bar{a}$&$\bar{b}$ &$\zeta_p/R_s$ &$c_2$ &$\bar{a}$&$\bar{b}$ &$\zeta_p/R_s$ 
			\\[2pt] \hline \hline\\[-12pt]   
			- 13.7088  & 1.00004  &  -0.40064   &  2.59808 & SBH     & 1       & -0.40063  &  2.59808 & 0   & 1       & -0.40064 & 2.59808 \\ [5pt]
			- 13       & 1.00239  &  -0.40252   &  2.59744 & -9.5    & 0.82067 & -0.57138  &  1.42679 & 0.2 & 1.08472 & -0.68839 & 1.61999 \\ [5pt]
			- 12       & 1.08116  &  -0.43441   &  2.40080 & -9      & 0.79594 & -0.62411  &  1.35108 & 0.4 & 1.13480 & -0.82506 & 1.22075 \\ [5pt]
			- 11.5     & 1.05033  &  -0.40639   &  2.12771 &  0      & 0.76393 & -1.00503  &  1.00021 & 0.5 & 1.14925 & -0.87285 & 1.09440 \\ [5pt]
			- 11       & 0.97273  &  -0.40731   &  1.86521 &  7      & 0.70173 & -1.09473  &  0.94797 & 0.8 & 1.17446 & -0.97636 & 0.84669 \\ [5pt] 
			- 10       & 0.85517  &  -0.51272   &  1.52817 &  10     & 0.70107 & -1.11889  &  0.93542 & 0.9 & 1.17947 & -1.00255 & 0.74103 \\[2pt]\hline  
	\end{tabular}}
	\label{table2a}
\end{table}

From each graphical representation presented in this figure, it is evident 
that $\hat{\alpha}(\zeta)$ increases as impact parameter $\zeta$ decreases and shows 
divergence at $\zeta = \zeta_p$ as expected. When compared with the 
Schwarzschild BH, the deflection angle of a minimally coupled $f(R,T)$ 
gravity BH is found smaller than the Schwarzschild case in all the 
scenarios. The deflection angle decreases with increasing model parameters 
$\beta$ and $c_2$, eventually becoming negative for larger impact parameter values. This negativity is more pronounced for $\beta > 0$ and the higher 
values of $c_2$, whereas for smaller values of $\beta$ and  $c_2$ the 
negativity reduces and the deflection angle approaches to resemble the 
Schwarzschild behavior beyond some specific impact 
parameter values. Regarding the negative deflection angle, it is noteworthy 
that at high impact parameter values photons experience a repulsive 
interaction with the BH's gravitational field. Additionally, in the 
$\omega = -2/3$ case, the deflection angle attains a more negative value 
compared to the $\omega = 0$ and $\omega = 1/3$ scenarios. Moreover, it is 
seen that for a specific value of $\beta$, and also $c_2$, a photon exhibits 
divergence at a larger $\zeta_p$ in the 
$\omega = 1/3$ case relative to others. These observations suggest that the 
gravitational field of a minimally coupled $f(R,T)$ gravity BH surrounded by 
the radiation field is relatively stronger compared to when it is surrounded 
by a dust or quintessence field. In fact, the appearance of a negative 
deflection angle suggests valuable insights into the gravitational behavior 
of BHs. Several pioneering studies have also highlighted the phenomenon of 
negative deflection angles in BH spacetime \cite{Nashiba_2023c, 
Panpanich_2019, Nakashi_2019, Kitamura_2013,2024_gayatri1}.

\subsection{Observables}
\begin{figure}[!h]
	\centerline{
		\includegraphics[scale = 0.71]{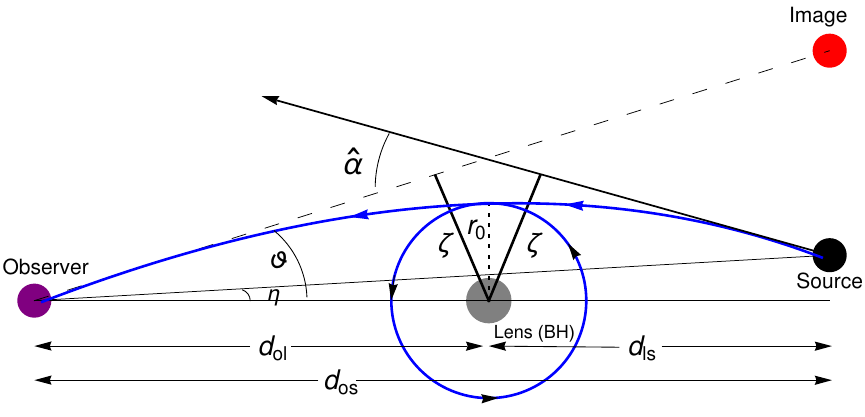}}	
                \vspace{-0.2cm}
	\caption{Schematic diagram of the strong field limit gravitational 
lensing by a BH as a lens.}
	\label{fig6}
\end{figure}

In this subsection we estimate the strong field limit three observables viz., 
the angular position of the asymptotic relativistic images $\vartheta_\infty$, 
the angular separation between the outermost and asymptotic relativistic 
images $s$ and the relative magnification of the outermost relativistic image 
with other relativistic images $r_{mag}$ in the BHs spacetime under 
consideration. To proceed further, the source and the observer are assumed to 
be at infinity relative to the BH such that the spacetime curvature impacts 
the deflection angle close to the lens (BH) only \cite{Bozza_2010,2024_islum}. 
It is also assumed that the observer, the lens and the source are in perfect 
alignment along the line of sight (see Fig.~\ref{fig6}). This assumption is 
made because the images appear significantly diminished in size unless the 
source, lens, and observer are nearly perfectly aligned \cite{Virbhadra_2000}. 
The lens equation which relates the positions of the source, observer and lens 
to the positions of the image can be written as \cite{Virbhadra_1998,
Virbhadra_2000,2001_Bozza}  
\begin{equation} 
	\eta = \vartheta - \frac{d_{ls}}{d_{os}}\, \Delta \alpha_k,
	\label{eq31}
\end{equation}
where $\eta$ and $\vartheta$ are the angular positions of the source and the 
image respectively from the optic axis, 
$\Delta \alpha_k = \hat{\alpha} - 2k\pi$ is an extra deflection angle, $k$ 
is a positive integer that denotes the total 
number of complete loops made by a photon around the BHs, and $d_{ls}$ and 
$d_{os}$ represent the distances of the source and the observer from the lens 
respectively. 
The position of the $k$th relativistic image $\vartheta_k$ in terms of the 
coefficients $\bar{a}$, $\bar{b}$ and critical impact parameter $\zeta_p$ can 
be derived from equations \eqref{eq26} and \eqref{eq31}. Thus, use of the 
relation $\zeta \approx \vartheta d_{ol}$ in equation \eqref{eq26} yields 
$\vartheta_k$ in the following form \cite{2002_Bozza,2022_kumar}:
\begin{equation} 
	\vartheta_k = \vartheta_k^0 +\frac{\zeta_p e_k(\eta - \vartheta_k^0)d_{os}}{\bar{a}\,d_{ls}d_{ol}},
	\label{eq32}
\end{equation}	
where $\vartheta_k^0 = \zeta_p(1 + e_k)/d_{ol}$ is the relativistic 
image position associated with deflection angle $\hat{\alpha}=2 k \pi$ with 
$e_k  = \exp\left({\bar{b} - 2k\pi}/\bar{a}\right)$. It is seen that 
$\vartheta_k^0 $ decreases rapidly with $k$ and in the limit $k\to \infty$, 
one finds $e_k\to 0$, which leads $\vartheta_{\infty}$ to the form as given 
below \cite{2002_Bozza,2020_Nascimento,Chen_2009}:
\begin{equation} 
    \vartheta_{\infty}=\frac{\zeta_p}{d_{ol}}.
	\label{eq33}
\end{equation}
Under the assumption of perfect alignment i.e., for $\eta=0$, visibility of 
relativistic images increases. In this scenario, the geometry of the system 
is symmetric about the line of sight. Thus, the light emitted from the source 
can reach the observer after being bent by the gravitational field of the BH 
along multiple paths in all possible directions \cite{2023_Walia}. As a 
result, the distorted image of the source is observed as a circular ring, 
known as Einstein ring \cite{2023_Walia,Bozza_2004}. Taking $d_{os}=2d_{ol}$, 
the radius of the $k$th relativistic Einstein ring can be obtained from 
equation \eqref{eq32} as
\begin{equation} 
	\vartheta_k^E = \vartheta_k^0(1 -\frac{2 \zeta_p e_k}{\bar{a}\,d_{ol}})=\frac{\zeta_p}{d_{ol}}\left(1 + e_k\right)\left(1 -\frac{2 \zeta_p e_k}{\bar{a}\,d_{ol}}\right).
	\label{eq34}
\end{equation}	
As $d_{ol}\gg\zeta_p$, \eqref{eq34} takes the form:
\begin{equation} 
	\vartheta_k^E = \frac{\zeta_p}{d_{ol}}\left(1 + e_k\right).
	\label{eq35}
\end{equation}
\begin{figure}[!h]
        \centerline{
        \includegraphics[scale = 0.50]{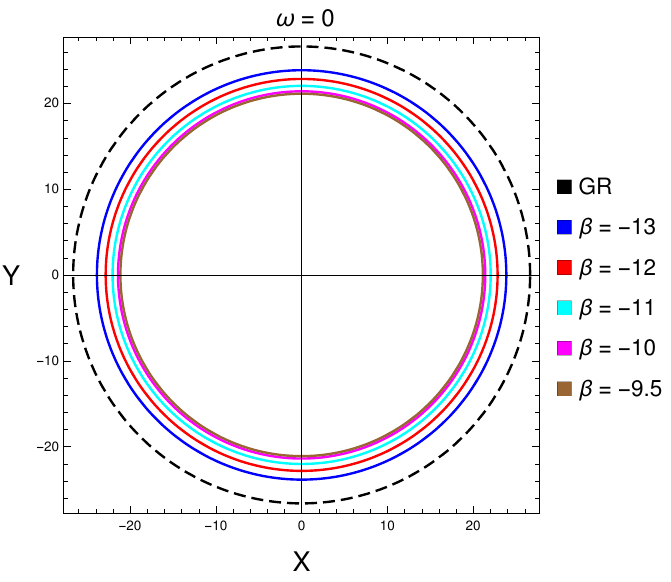}\hspace{.01cm}
        \includegraphics[scale = 0.50]{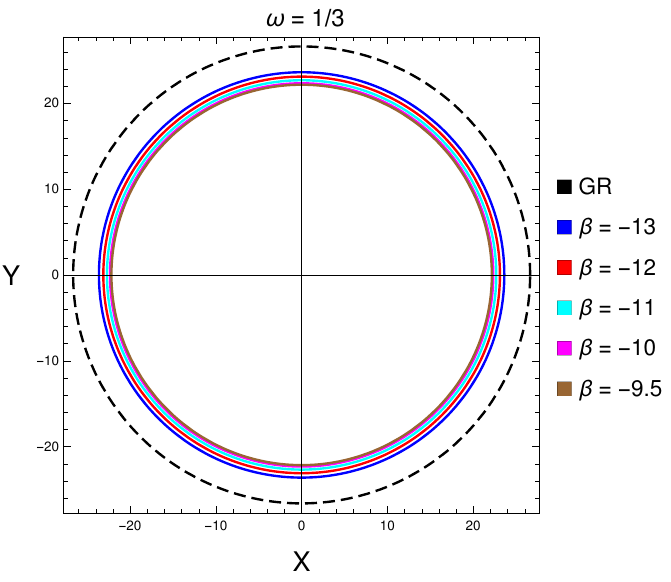}\hspace{.02cm}
        \includegraphics[scale = 0.50]{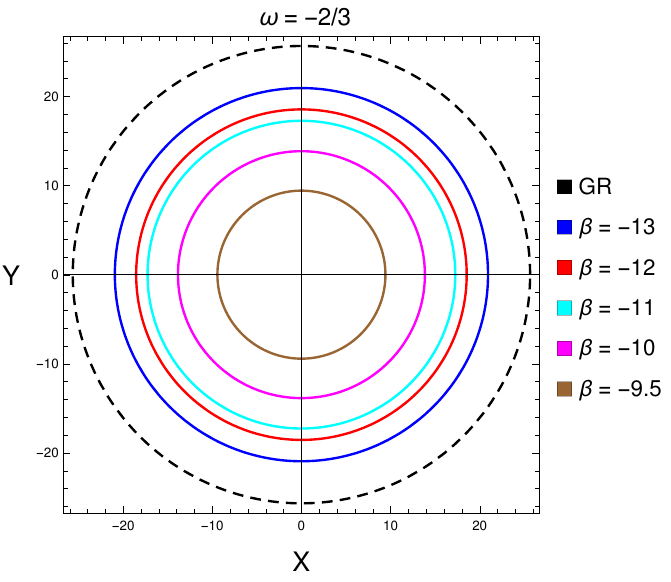}}\vspace{0.2cm}
        \centerline{
        \includegraphics[scale = 0.50]{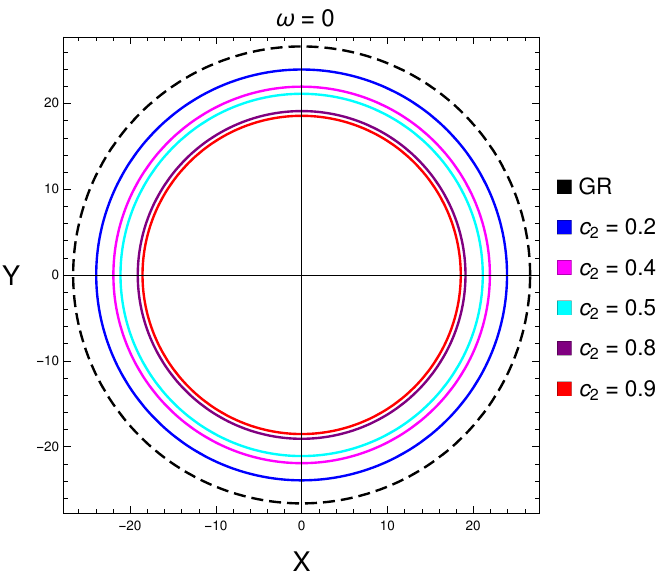}\hspace{.5cm}
        \includegraphics[scale = 0.50]{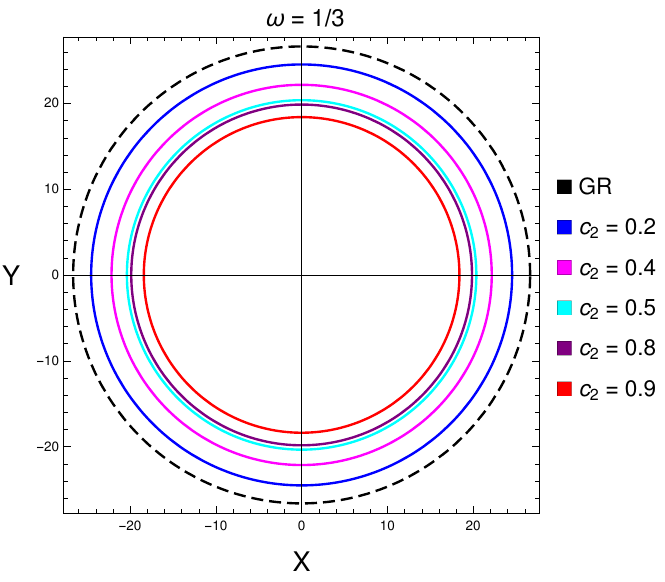}\hspace{.35cm}
        \includegraphics[scale = 0.50]{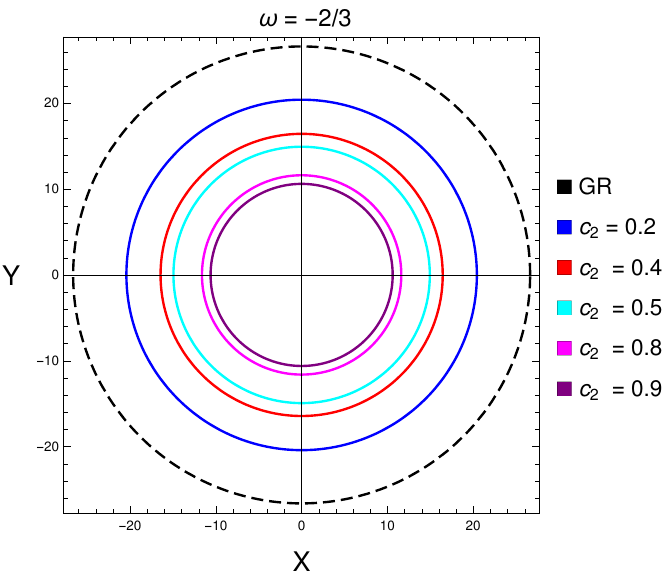}}\vspace{-0.2cm}
        \caption{The outermost relativistic Einstein rings for Sgr A* by
considering it as a minimally coupled $f(R,T)$ BH, for different values of
model parameters $\beta$ and $c_2$ respectively. The upper row is for 
changing $\beta$ with $c_2 = 0.5$, and the lower row is for varying $c_2$ at 
$\beta = -9.5$. The left plots are for $\omega=0$, the middle plots are for 
$\omega=1/3$ and the right two are for $\omega = -2/3$. The black dashed rings
correspond to Schwarzschild BHs.}
       \label{fig7}
\end{figure}
It should be noted that $k=1$ corresponds to the outermost Einstein ring. We 
depict the outermost Einstein ring of the supermassive BH SgrA* located at 
the center of the Milky Way, by modeling it as a minimally coupled $f(R,T)$ 
gravity BH for different values of $\beta$ and $c_2$ individually. The mass $M$ and the distance 
$d_{ol}$ from the Earth of Sgr A* are taken as $4.3 \times 10^6  M_\odot$ and 
$ 8.35$ kpc respectively \cite{EHT_2022,Kuang_2022,Dong_2024,2008_Ghez,
2019_Do,2009_Gill}. The rings are portrayed for $\omega=0$, 
$\omega=1/3$ and $\omega=-2/3$ separately in Fig.~\ref{fig7}, for different 
values of parameters $\beta$ and $c_2$. It is observed that relative 
to each respective field, the radii of rings decrease as $\beta$ increases. 
Similar trend is noticed with increase of $c_2$ also.

The magnification of the images serves as another essential source of 
information. This can be examined from the expression of the magnification 
of the $k$th relativistic image as given by \cite{2001_Bozza,2002_Bozza,
Whisker_2005} 
\begin{equation} 
	\mu_k =\left. \left(\frac{\eta}{\vartheta}\,\frac{\partial \eta}{\partial \vartheta}\right)^{-1} \right|_{ \vartheta_k^0} \simeq \frac{e_k(1+e_k)d_{os}}{\bar{a}\, \eta\ d_{ls}\ d^2_{ol}}\, \zeta^2_p.
	\label{eq36}
\end{equation}
This hints that the magnification decreases rapidly as $k$ increases, thus 
resulting in the outermost image as the brightest of all the other relativistic 
images. Also, the presence of the significantly large factor $d^2_{ol}$ 
ensures that the overall luminosity of all images remains relatively weak. 
Moreover, one would observe a diverging magnification as $\beta \to 0$. 

Thus, we can infer that the outermost single loop image $\vartheta_1$ is 
distinguishable from the group of remaining inner images packed as 
$\vartheta_\infty$. Based on this separation the observable 
$\vartheta_\infty$ is obtained as given by equation \eqref{eq33}. The other 
two observables, $s$ and $r_{mag}$, are defined as follows
\cite{2002_Bozza,2020_islum,2020_Nascimento,Zhao_2017}:   
\begin{align} 
 	s  & = \vartheta_1 - \vartheta_\infty \approx \vartheta_\infty \exp\left(\frac{\bar{b} - 2\pi}{\bar{a}}\right),
 	\label{eq37}\\[8pt]
	r_{\text{mag}} & =\frac{\mu_1}{\sum_{k=2}^{\infty}\mu_k} = \exp\left(\frac{2\pi}{\bar{a}}\right).
	\label{eq38}
\end{align} 

\begin{figure}[!h]
	\centerline{
		\includegraphics[scale = 0.58]{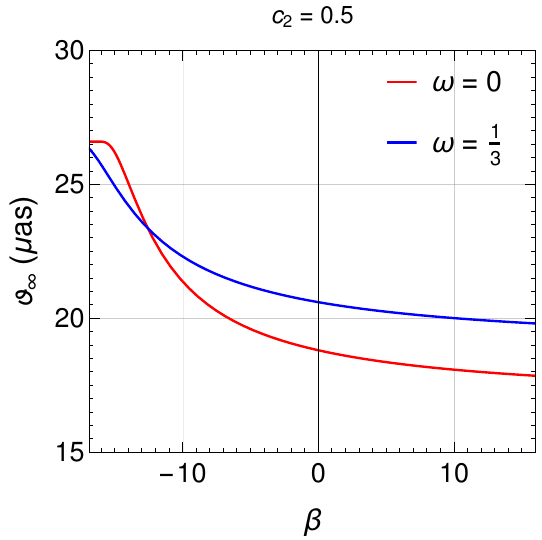}\hspace{.3cm}
		\includegraphics[scale = 0.61]{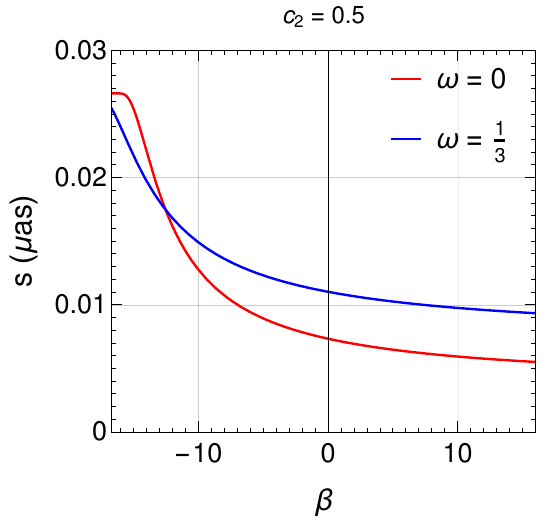}\hspace{.3cm}
		\includegraphics[scale = 0.55]{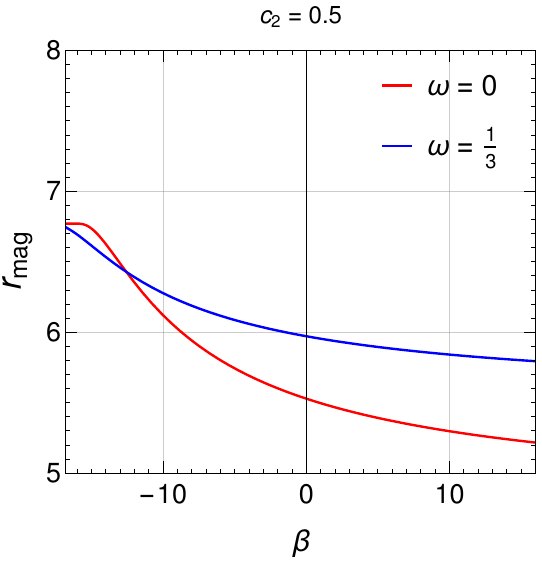}}\vspace{0.2cm}
	\centerline{
		\includegraphics[scale = 0.59]{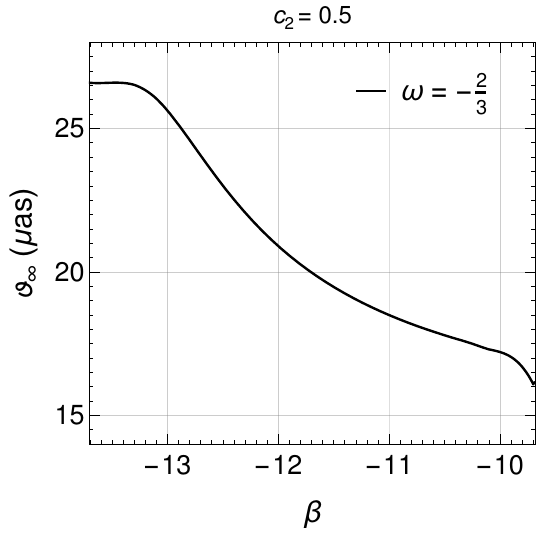}\hspace{.3cm}
		\includegraphics[scale = 0.61]{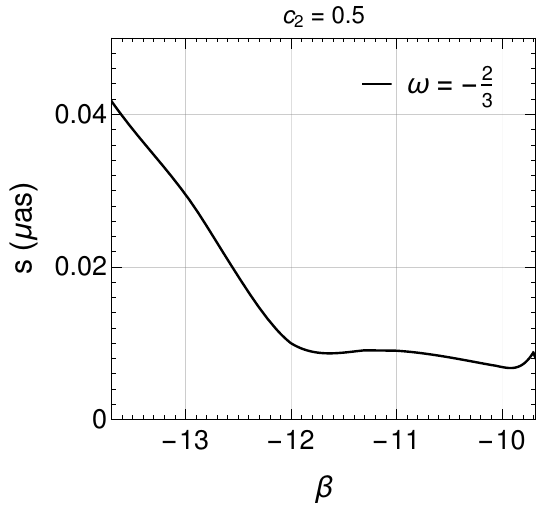}\hspace{.3cm}
		\includegraphics[scale = 0.55]{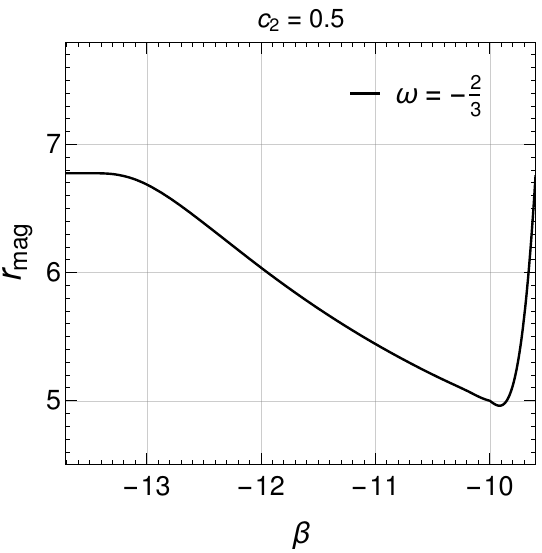}}\vspace{0.2cm}
	\centerline{
		\includegraphics[scale = 0.57]{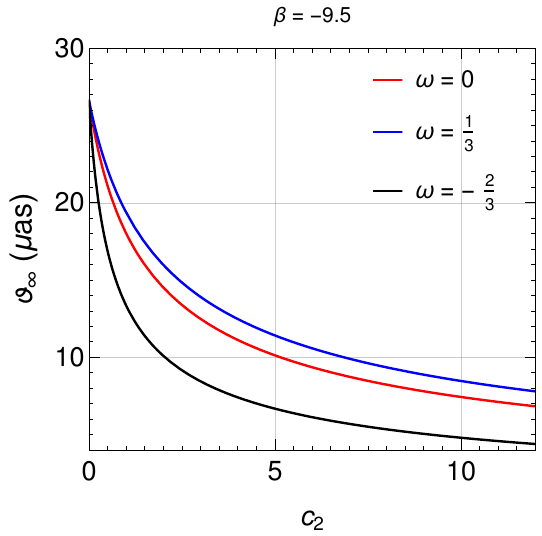}\hspace{.3cm}
		\includegraphics[scale = 0.60]{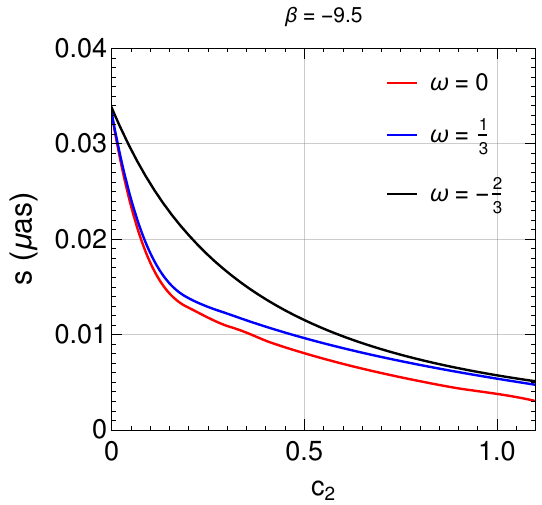}\hspace{.3cm}
		\includegraphics[scale = 0.54]{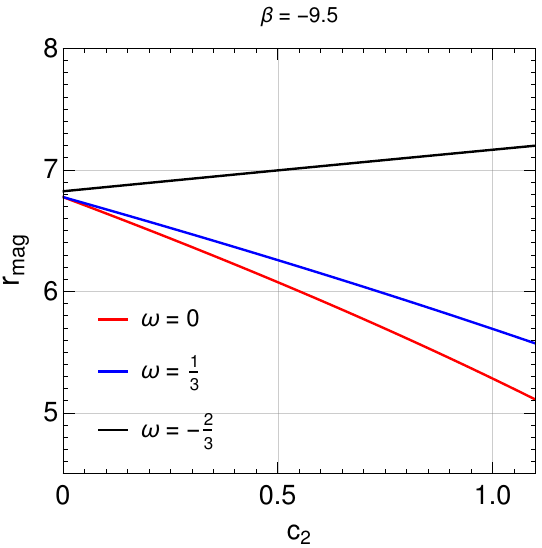}}\vspace{-0.2cm}
	\caption{Behaviors of three lensing observables with $\beta$ and 
$c_2$ respectively. The left plots show the variations of the angular 
position of the asymptotic relativistic images $\vartheta_\infty$, the middle 
plots are for the variations of the angular separation between the outermost 
and asymptotic relativistic images $s$ and the right plots depict the 
variations of relative magnification $r_{mag}$. The top two rows show the 
variations with respect to $\beta$, while the bottom plots show the 
corresponding variations with $c_2 $.}
	\label{fig8}
\end{figure}

Subsequently, we proceed to estimate these three key lensing observables in 
the strong-field regime, as their evaluation reveals the nature of lensing BHs.
To do this, considering the aforementioned mass and distance of SgrA*, the 
numerical values of the lensing observables $\vartheta_\infty$, $s$ and 
$r_{mag}$ for the BH are computed and summarized in Table \ref{table3}. To 
analyze their behaviors in the $f(R,T)$ BHs spacetime, the variations of these 
observables with respect to the modification parameters $\beta$ 
fixing $c_2$ at $0.5$, and with $c_2$ taking $\beta = -9.5$ are depicted in 
Fig.~\ref{fig8}. It is evident that all three observables decrease with 
increasing $\beta$ for all three scenarios, $\omega = 0$, $\omega = 1/3$ and 
$\omega = -2/3$. For a given $\beta$, the observables attain smaller values 
in the presence of a dust field compared to a radiation field around it 
when $\beta \geq -12.23$. Below this critical limit of $\beta$, the values of 
all three observables exceed those corresponding to the radiation field. 
Further, $s$ exhibits a relatively steep decline with increasing $\beta$ for 
$\omega = 0$, followed by a very slower variation, whereas for $\omega = 1/3$, 
the decrease is more gradual before showing nearly constant behavior at higher 
values of $\beta$. On the other hand, for $\omega = -2/3$, although all three 
observables initially follow a similar trend as in the cases of $\omega = 0$ 
and $\omega = 1/3$ from about $\beta = -13.5$, their behavior deviates beyond 
$\beta \approx -9.7$. Specifically, the observable $\vartheta_{\infty}$ 
exhibits a sudden decline, while $s$ and $r_{\text{mag}}$ undergo steep growth 
beyond $\beta= -9.7$. Notably, good results for these observables are obtained 
only within this range of $\beta$. The observables approach their 
corresponding GR values at $\beta = -16.731$, $\beta =-18.8495$ and 
$\beta =-13.7088$ for the three respective fields and show deviations from 
GR as $\beta$ increases. Again, $\vartheta_{\infty}$ shows a rapid initial 
decline with respect to $c_2$, followed by a slow variation for all three 
surrounding fields. However, in the case of $s$ the change related to 
$\omega = -2/3$ is moderate from the initial point, while for $\omega = 0$ 
and $\omega = 1/3$, it exhibits a steep decline before decreasing slowly 
with $c_2$. Further, $r_{mag}$ decreases gradually in an even manner for the 
dust and radiation fields respectively, whereas it grows slowly as a function 
of $c_2$ in the case of the quintessence field.   
  \begin{table}[!h]
 	\centering
 	\caption{Estimation of numerical values of characteristic strong 
lensing observables \textbf{for different values of parameters $\beta$ and 
$c_2$} by considering supermassive BH Sgr A* as a minimally coupled $f(R,T)$ 
gravity BH for $\omega = 0$, $\omega = 1/3$ and $\omega = -2/3$ respectively.}
 	\vspace{8pt}
 	\setlength{\tabcolsep}{10pt}
 	\scalebox{0.99}{
 		\begin{tabular}{cc c c c |c c c c c } \hline 
 			&$\beta$~~ & $\vartheta_{\infty}(\mu \text{as})$& s $(\mu \text{as})$ &~~$r_{\text{mag}}$~~&$c_2$~~ &~~~ $\vartheta_{\infty}(\mu\text{as})$& s $(\mu \text{as})$ & ~~$r_{\text{mag}}$~~\\[2pt]\hline \hline  \\[-12pt]     		
 			& SBH     & 26.5972  & 0.033239   & 6.77155 &  0  &26.5972&0.033239&6.77592 \\ [5pt]
 			&-16.731  & 26.5972  & 0.033245   & 6.77155 & 0.2 &23.9244&0.012753&6.50475 \\ [5pt]			
 			&-13.7088 & 24.6790  & 0.025693   & 6.58678 & 0.4 &21.9233&0.009338&6.22178 \\ [5pt]
 			$\omega = 0$&-12      & 22.8102  & 0.017886   & 6.35276 & 0.5 &21.0943&0.008035&6.07535 \\ [5pt]
 			&-9.4     & 21.0634  & 0.009731   & 6.06947 & 0.7 &19.6843&0.005951&5.77136 \\ [5pt]
 			&0        & 18.8071  & 0.009157   & 5.52895 & 0.8 &19.0777&0.005100&5.61320 \\ [5pt]
 			&7        & 18.2374  & 0.008667   & 5.35067 & 0.9 &18.5238&0.004469&5.45045 \\ [5pt]\hline   			     
 			& SBH     & 26.5972  & 0.033239   & 6.77155 & 0   &26.5972&0.033292&6.77592 \\ [5pt] 			 		
 			&-18.8495 & 26.5972  & 0.033245   & 6.77155 & 0.2 &24.5137&0.013801&6.57327 \\ [5pt]
 			&-13.7088 & 24.0187  & 0.022997   & 6.51123 & 0.4 &22.8538&0.010815&6.36418 \\ [5pt]
 			$\omega = 1/3$	      &-12       & 23.0824    & 0.019064   & 6.39098 & 0.5 &22.1411&0.009619&6.25701 \\ [5pt]
 			&-9.4     & 22.1228  & 0.008827   & 5.68459 & 0.7 &20.8952&0.007637&6.03697 \\ [5pt]
 			& 0       & 20.6021  & 0.007679   & 5.97195 & 0.8 &20.3462&0.006804&5.92389 \\ [5pt]
 			& 7       & 20.1400  & 0.006877   & 5.87211 & 0.9 &19.8384&0.006052&5.80861 \\[5pt]\hline 
 			&SBH      & 26.5972  & 0.033239   & 6.77155 &   0   & 26.5995  & 0.033810  & 6.82302 \\ [5pt]
 			&-13.7088 & 26.5972  & 0.041883   & 6.77592&   0.2 & 20.4061  & 0.020423  & 6.89216 \\ [5pt]			
 			&-12      & 20.9018  & 0.010009   & 6.03828 &   0.4 & 16.4323  & 0.013695  & 6.96094 \\ [5pt]
 			$\omega = -2/3$ 	  &-11.5     & 19.47890    & 0.00828 & 5.71995 &   0.5  & 14.9315   & 0.011512  & 6.99518 \\ [5pt]
 			&-11      & 18.4999  & 0.009013   & 5.44296 &   0.7 & 12.5612  & 0.004787  & 7.06336 \\ [5pt]
 			&-10      & 17.2128  & 0.006880   & 5.04514 &   0.8 & 11.6097  & 0.007357  & 7.09729 \\ [5pt]
 			&-9.4    & 16.9101  & 0.031534   & 10.3246 &   0.9 & 10.7762  & 0.006459  & 7.13111 \\[2mm]\hline 			
 	\end{tabular}}
 	\label{table3}     
 \end{table}  

\subsection{Constraints on the model parameters $\beta$ and $c_2$ from 
EHT observations of Sgr A*} 

\begin{figure}[!h]
	\centerline{
\includegraphics[scale = 0.36]{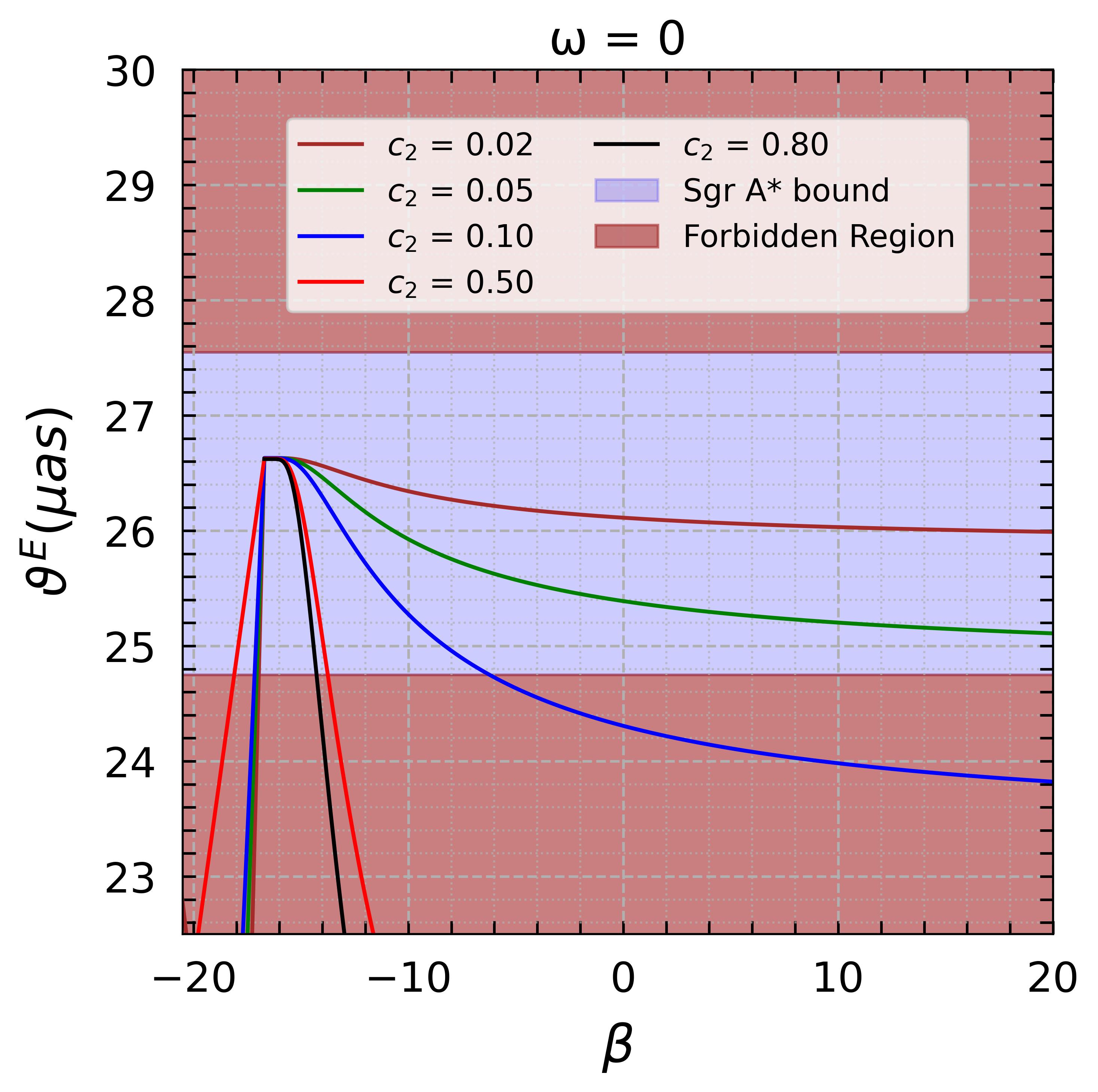}\hspace{.3cm}
\includegraphics[scale = 0.36]{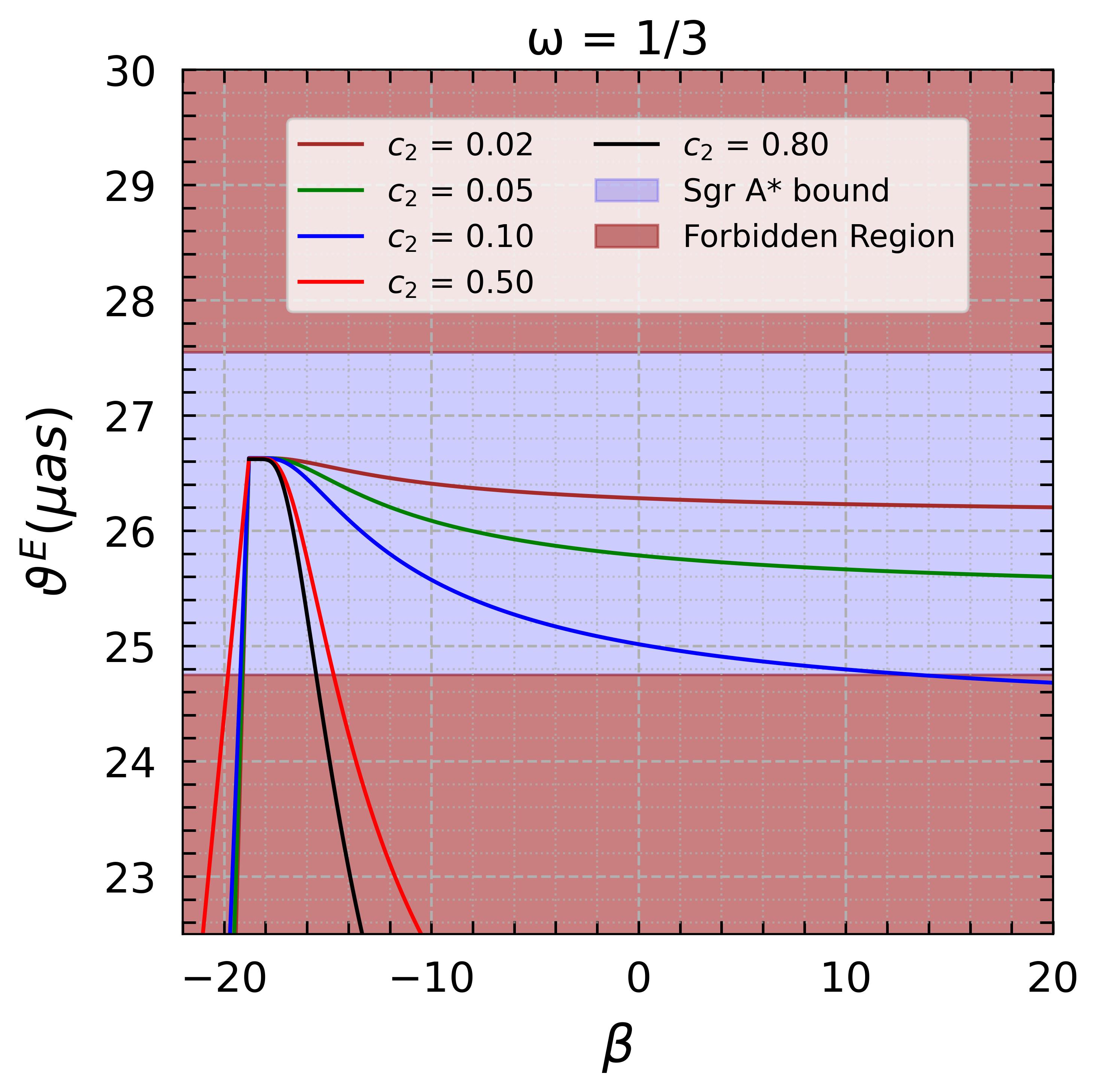}\hspace{.3cm}
\includegraphics[scale = 0.36]{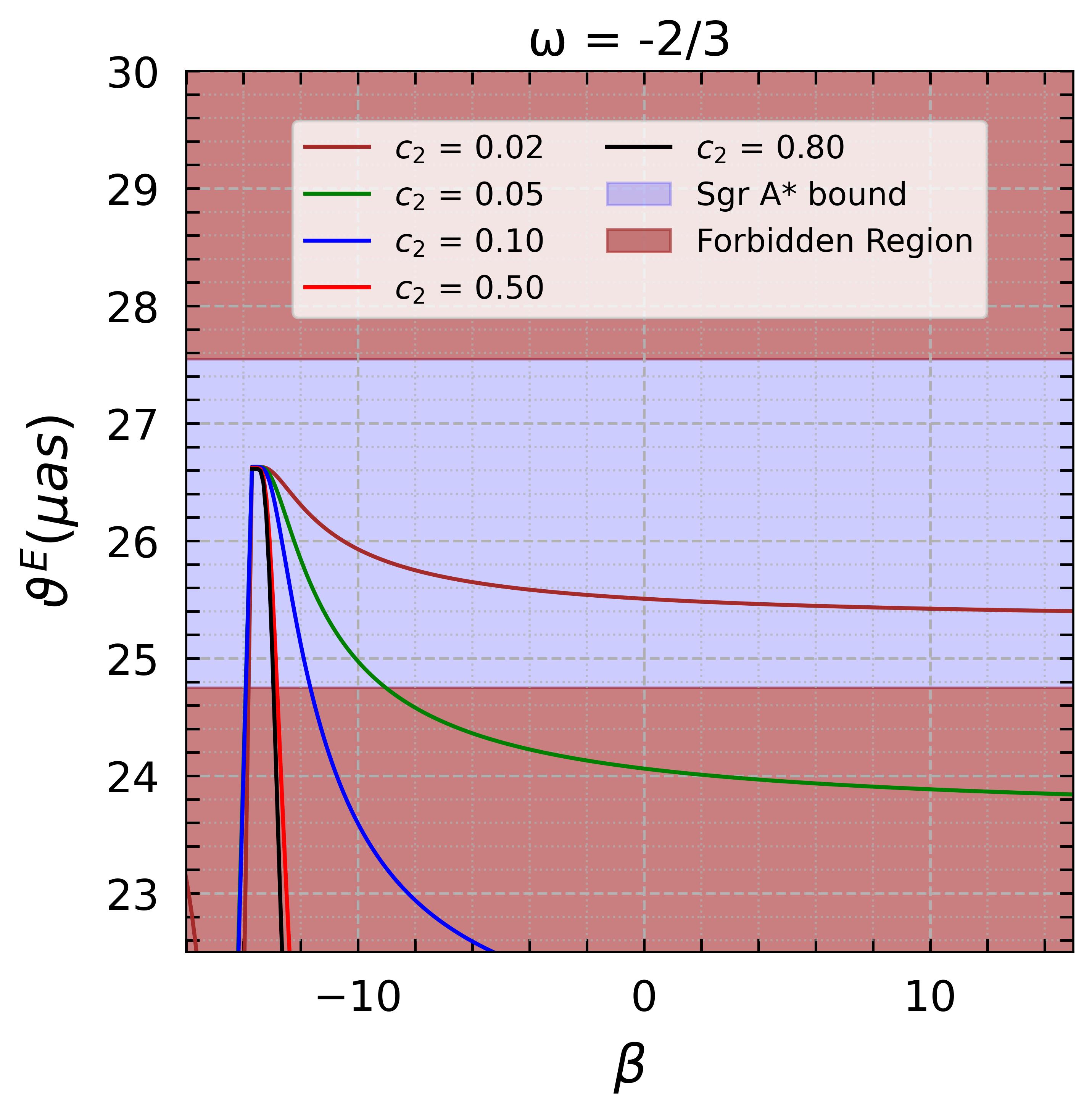}}\vspace{-0.2cm}
\caption{\textbf{Constraining of free model parameters $\beta$ and $c_2$ of 
$f(R,T)$ gravity BHs spacetime from the observed EHT Einstein radius 
$\vartheta^E$ of Sgr A*, by plotting $\vartheta^E$ as a function $\beta$ with 
respect to varying values of $c_2$ for three respective surrounding fields 
$\omega =0, 1/3$ and $-2/3$.}}
\label{fig8a}
\end{figure}

In 2022, the EHT unveiled the groundbreaking observations of the 
supermassive BH Sgr A* at the center of our galaxy by providing detailed direct 
measurements such as its emission ring diameter with the surrounding emission 
features \cite{EHT_2022}. Such observations serve as powerful means for 
exploring MTGs by constraining their free parameters, and understanding the 
strong gravitational field effects \cite{2025_Rahman, 2022_Kuang,
2023_Kumar,Dong_2024}. For the Sgr A* BH, the Einstein radius is 
$\vartheta^E = (25.90 \pm 1.15)\mu as$ as reported by the EHT 
collaboration \cite{EHT_2022,2025_Maryam}. Using these bounds, we constrain 
the associated free parameters $\beta$ and $c_2$ of our BHs for three 
considered surrounding fields. For this, we depict $\vartheta^E$ as a 
function of $\beta$ by varying $c_2$ over a broad range of $\beta$ as shown 
in Fig.~\ref{fig8a}. It is found that the observed EHT constraints on 
$\vartheta^E$ for Sgr A* significantly restrict the parameters of the 
$f(R,T)$ BHs spacetime as indicated by the blue band in the figure. 
Accordingly, the EHT observations impose bounds to $\beta$, narrowing its 
allowable range as $c_2$ increases across all three cases. The figure further 
illustrates that for each value of $c_2$, only a limited set of $\beta$ values 
lie within the blue band.

\begin{figure}[!h]
\centerline{
	\includegraphics[scale = 0.4]{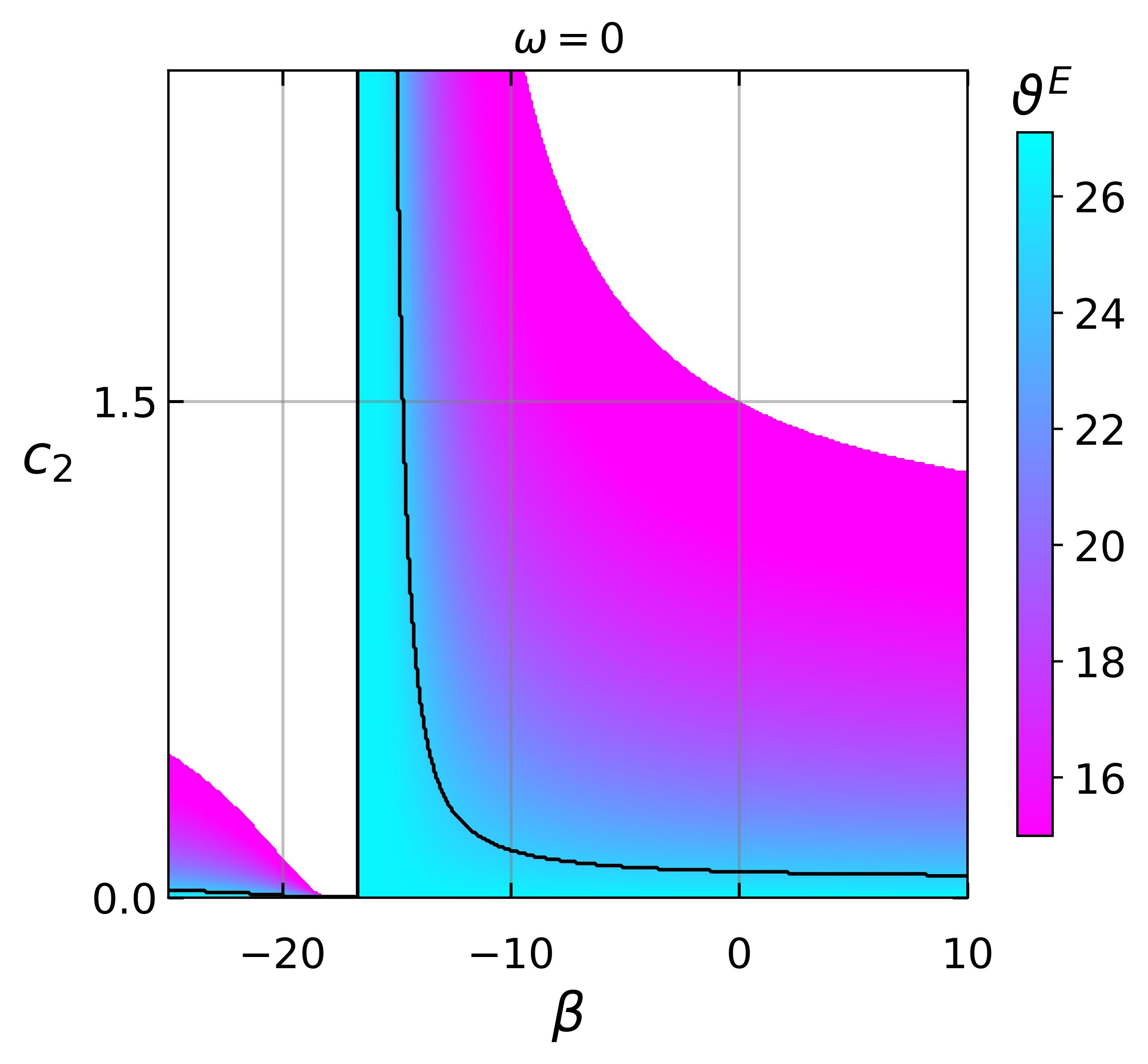}\hspace{.4cm}
	\includegraphics[scale = 0.4]{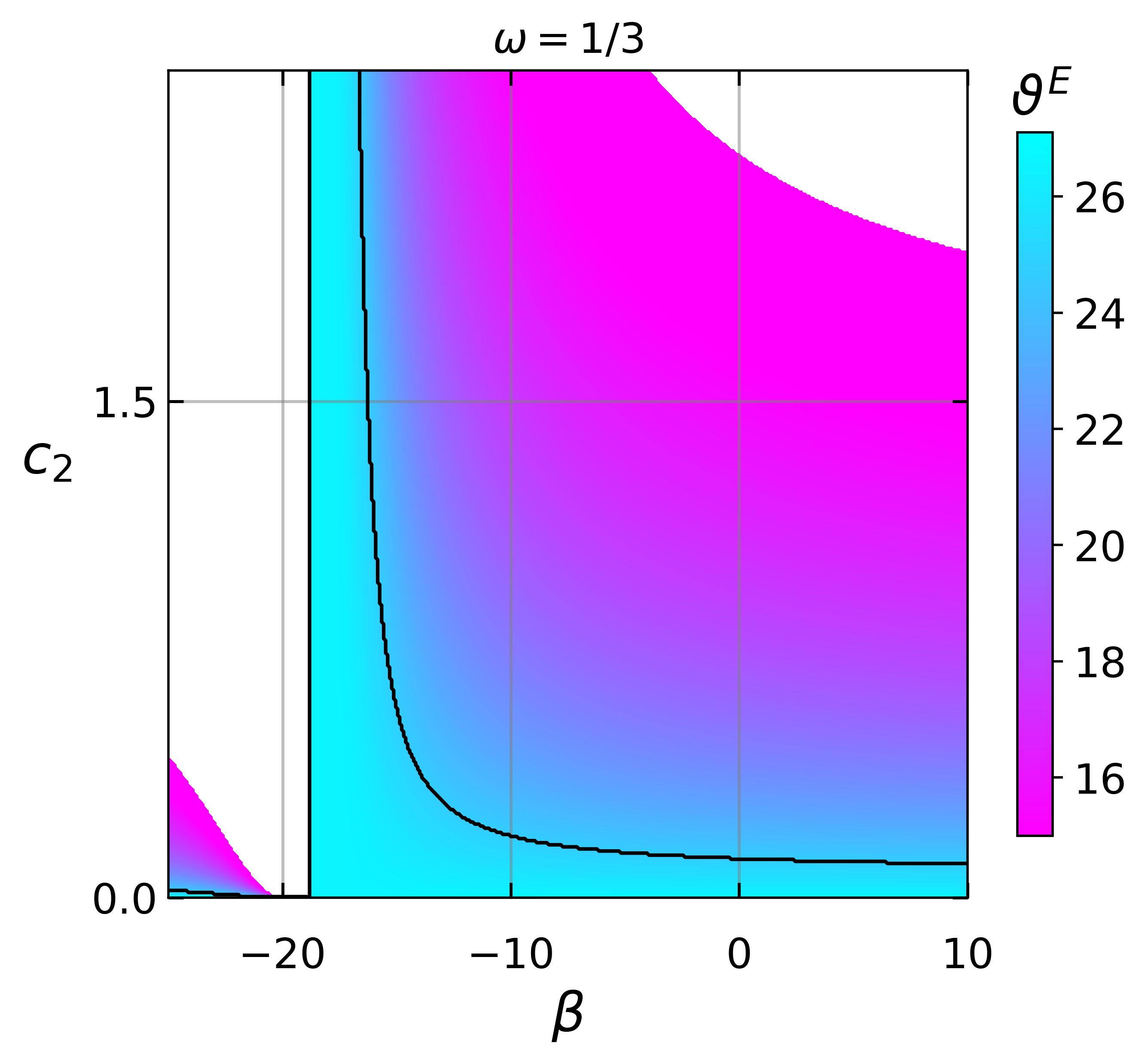}\hspace{.4cm}
	\includegraphics[scale = 0.4]{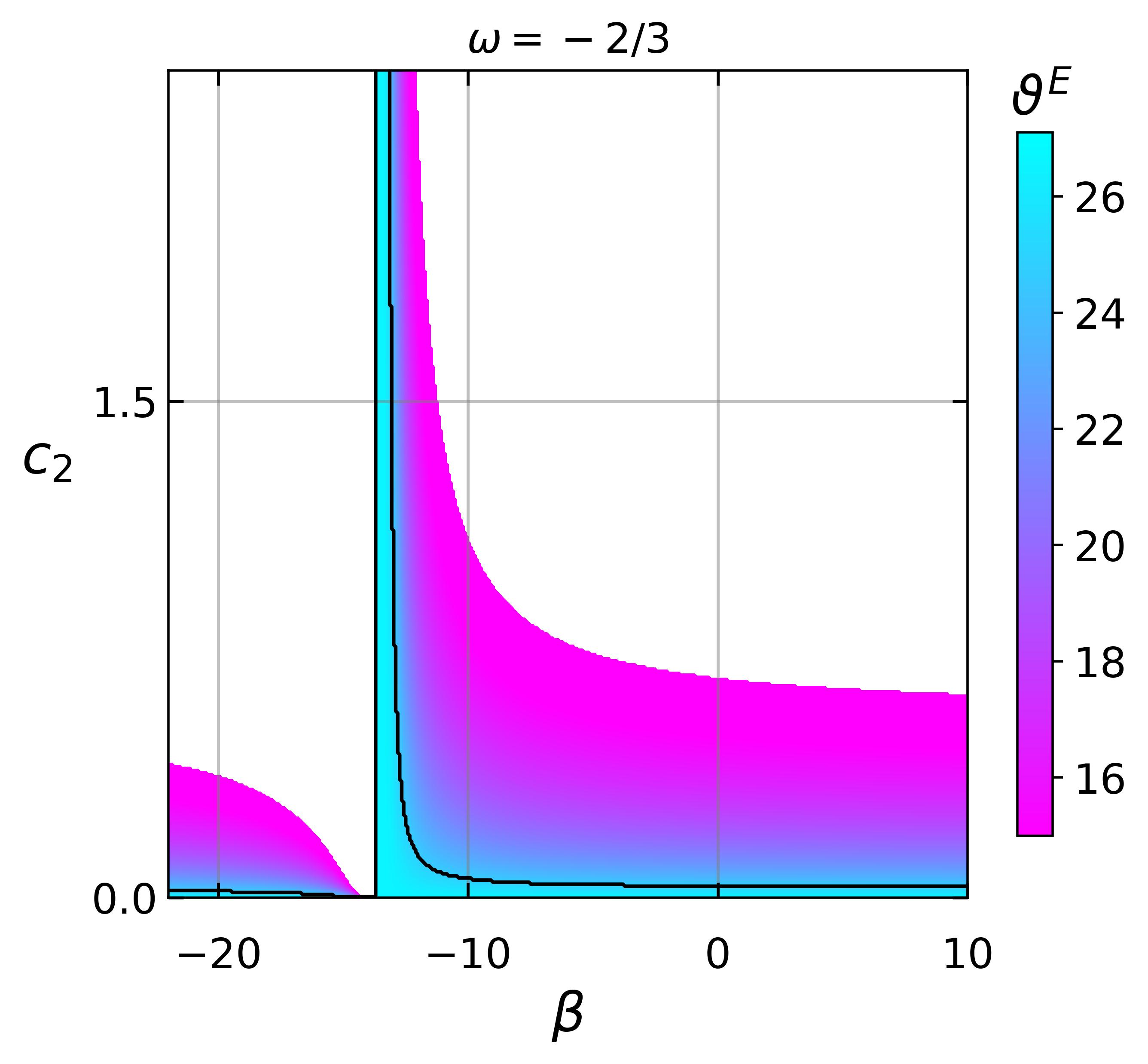}}
\vspace{-0.2cm}
\caption{Density plots for constraining the free model parameters 
$\beta$ and $c_2$ of $f(R,T)$ gravity BHs spacetime from the observed EHT 
Einstein radius $\vartheta^E$ of Sgr A* for three respective surrounding 
fields $\omega =0, 1/3$ and $-2/3$.}
\label{fig8b}
\end{figure}

In addition, Fig~\ref{fig8b} represents the density plot for 
$\vartheta^{E}$ with parameters $\beta$ and $c_2$ in the x and 
y axes respectively. We try to impose constraints on these two parameters 
again from the bounds of $\vartheta^{E}$ of the Sgr A* BH from EHT 
observations using the density plot. The solid black lines in each plot 
represent the lower and 
upper bounds on $\vartheta^{E}$ from the EHT data of Sgr A* BH. Thus the 
regions between the two lines represent the allowed parameter space whose 
coordinates ($\beta,c_2$) are the allowed values of the model parameters. For 
smaller values of $c_2$, the allowed range of $\beta$ extends towards the 
positive axis to a good extent as seen from the figure. For lower values of 
$\beta$, in the region $-20<\beta<-10$, the allowed range of $c_2$ increases to 
a good extent in the positive y axis as seen clearly from the figure. However, 
for further lower values of $\beta<-20$, there exists a narrow allowed 
parameter region that is not very prominent. 
\section{Quasinormal Modes of Oscillations}
\label{sec.4}
As mentioned earlier the QNMs are the characteristic spacetime oscillations of 
a BH that occur when its spacetime is being perturbed. QNMs are a 
reliable tool for diagnosing spacetime near a BH as they provide 
an important means of probing spacetime with extreme gravity. These are 
basically complex numbers, with the real part signifying the frequency while 
the imaginary part signifies the damping of spacetime oscillations. 

\begin{figure}[!h]
\centerline{
        \includegraphics[scale = 0.3]{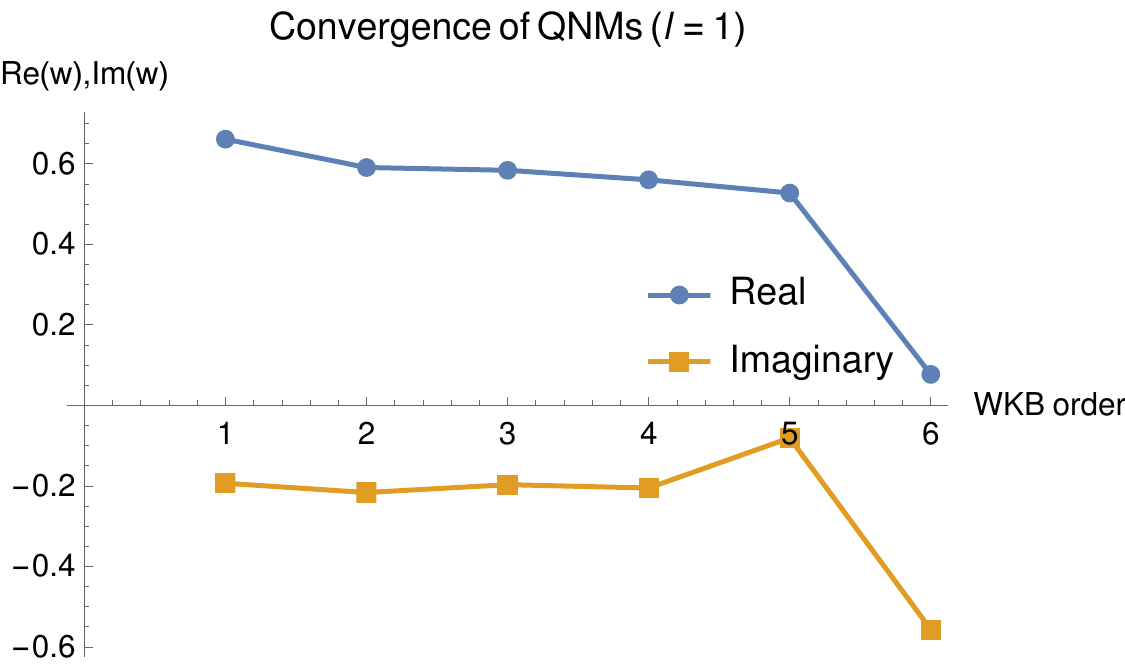}\hspace{1cm}
        \includegraphics[scale = 0.3]{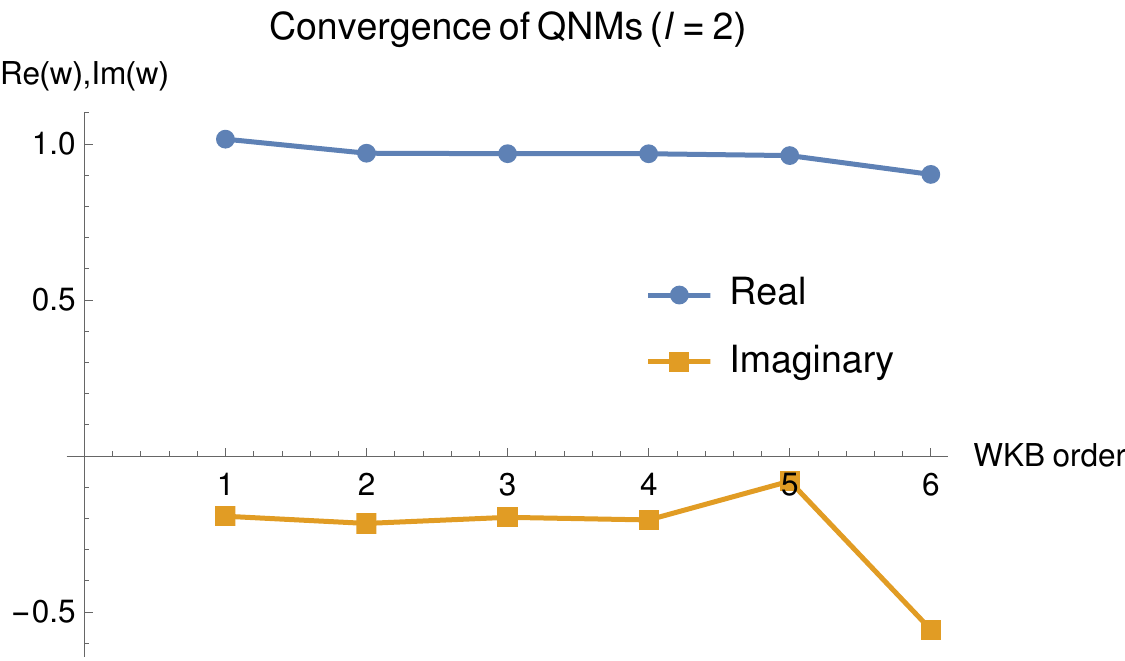}\vspace{0.2cm}}
        \caption{Convergence plots of the different order WKB QNMs of the
minimally coupled $f(R,T)$ gravity BHs for $\beta=0.1$. These are applicable 
for other values of $\beta$ also. \textbf{Here $c_2=0.8$ is considered}.}
        \label{fig9}
\end{figure}

If we perturb a BH spacetime with a scalar field $\Phi$, in response 
to which the BH tries to regain its initial state by radiating any 
deformity in spacetime in the form of gravitational radiation. We describe the 
probe that couples minimally to $\Phi$ by the following equation of 
motion \cite{r30}:
\begin{equation}
	\frac{1}{\sqrt{-g}} \partial_{a}(\sqrt{-g} g^{a b}\partial_b)\Phi=0.
	\label{eq37}
\end{equation}
To obtain the convenient form of this equation, it is essential to express 
the scalar field $\Phi$ in the spherical form \cite{r30}:
\begin{equation}
	\Phi(t, r, \theta, \phi)=\exp^{-i \omega t}\frac{\psi(r)}{r}\,Y_l^m (\theta, \phi), 
	\label{eq38}
\end{equation}
where $\psi$ is the radial part and $Y_{l}^{m}$ is the spherical harmonics. 
$\omega$ represents the frequency of oscillation of the temporal part of the 
wave. Using Eq.~\eqref{eq38} into Eq.~\eqref{eq37} we finally get the 
Schr\"odinger-like wave equation as
\begin{equation}
	\frac{d^2 \psi}{dx^2} +(\omega^2 -V_l(x))\psi=0,
	\label{eq39}
\end{equation}
where we define $x$ as the tortoise coordinate, given by
\begin{equation}
	x=\int\! \frac{dr}{A(r)}.
	\label{eq40}
\end{equation}
The potential $V_l(r)$ in the BH spacetime is obtained as
\begin{equation}
	V_l(r)=A(r)\bigg(\frac{A'(r)}{r}+\frac{l(l+1)}{r^2}\Big),
	\label{eq41}
\end{equation}
where $l$ is the multipole number. For physical consistency, the following 
boundary conditions are imposed on $\psi(x)$ at the horizon and infinity:
\begin{align}
	\psi(x) \rightarrow \Bigg \{ \begin{array}{ll}
	\mathcal{A}\, e^{+i\omega x} & \text{if }x \rightarrow -\infty, \\[3pt]
	\mathcal{B}\, e^{-i\omega x} & \text{if }x \rightarrow +\infty, 
        \end{array}
	\label{eq42}
\end{align} 
where $\mathcal{A}$ and $\mathcal{B}$ denote the constants of integration that 
behave as amplitudes of waves. With these criteria, we are now in a position
to compute the QNMs, for which we follow the Ref.~\cite{r30}. At this stage 
it is to be noted that the QNMs associated with a BH spacetime 
are most often computed using the semi-analytic WKB approximation method as
mentioned earlier. However, here we use the 3rd order WKB approximation only 
due to convergence issues with the higher order as shown in Fig.~\ref{fig9}. 
It is easily observed from the plot that higher order modes are not converging, 
especially for the multipole $l=1$. Thus, in this work, we compute 2nd, 3rd and 
4th order WKB QNMs and then compare the 3rd order results with the time domain 
QNMs for the sake of accuracy of the computation.

\begin{figure}[!h]
	\centerline{
		\includegraphics[scale = 0.25]{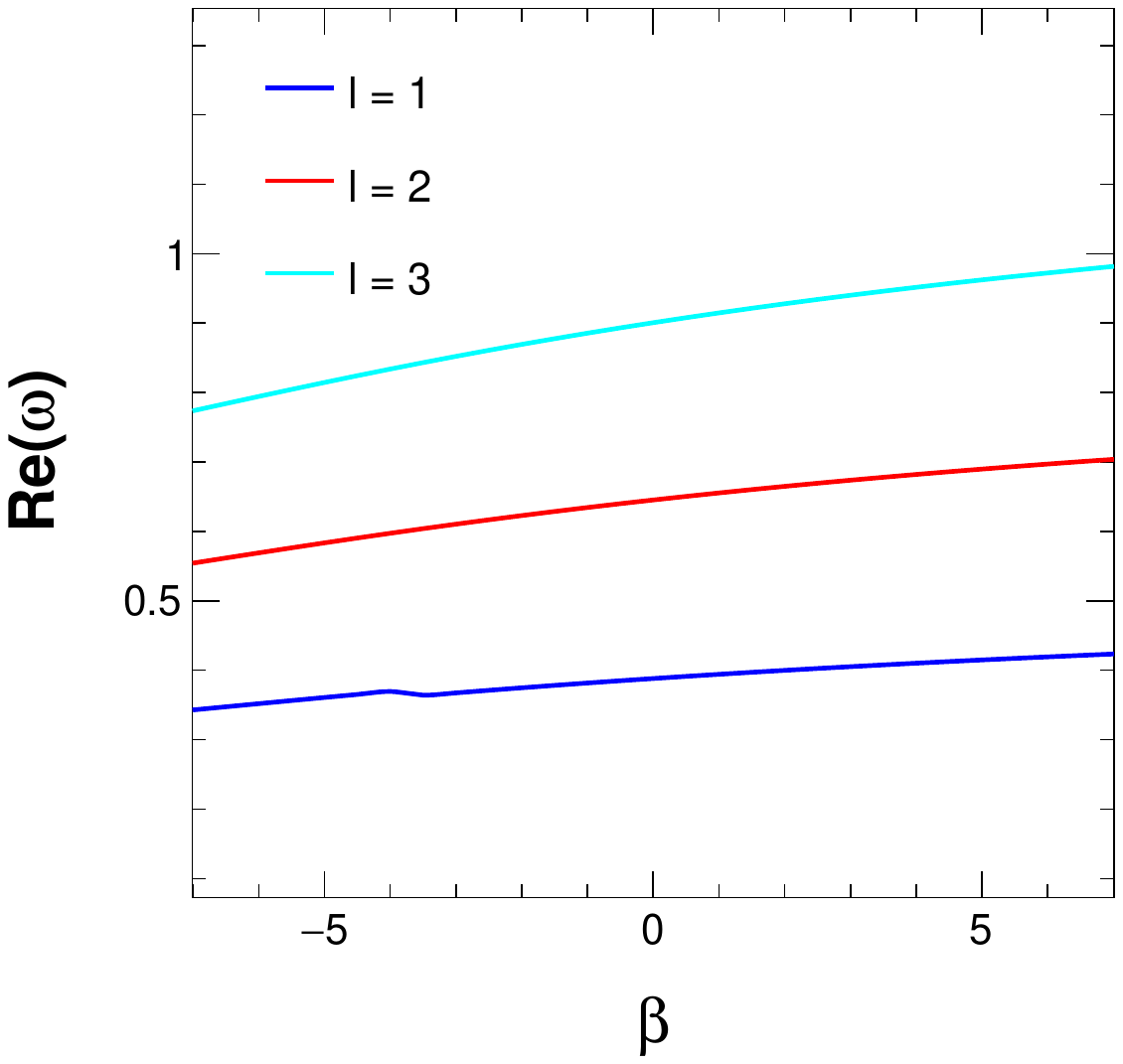}\hspace{0.2cm}
		\includegraphics[scale = 0.25]{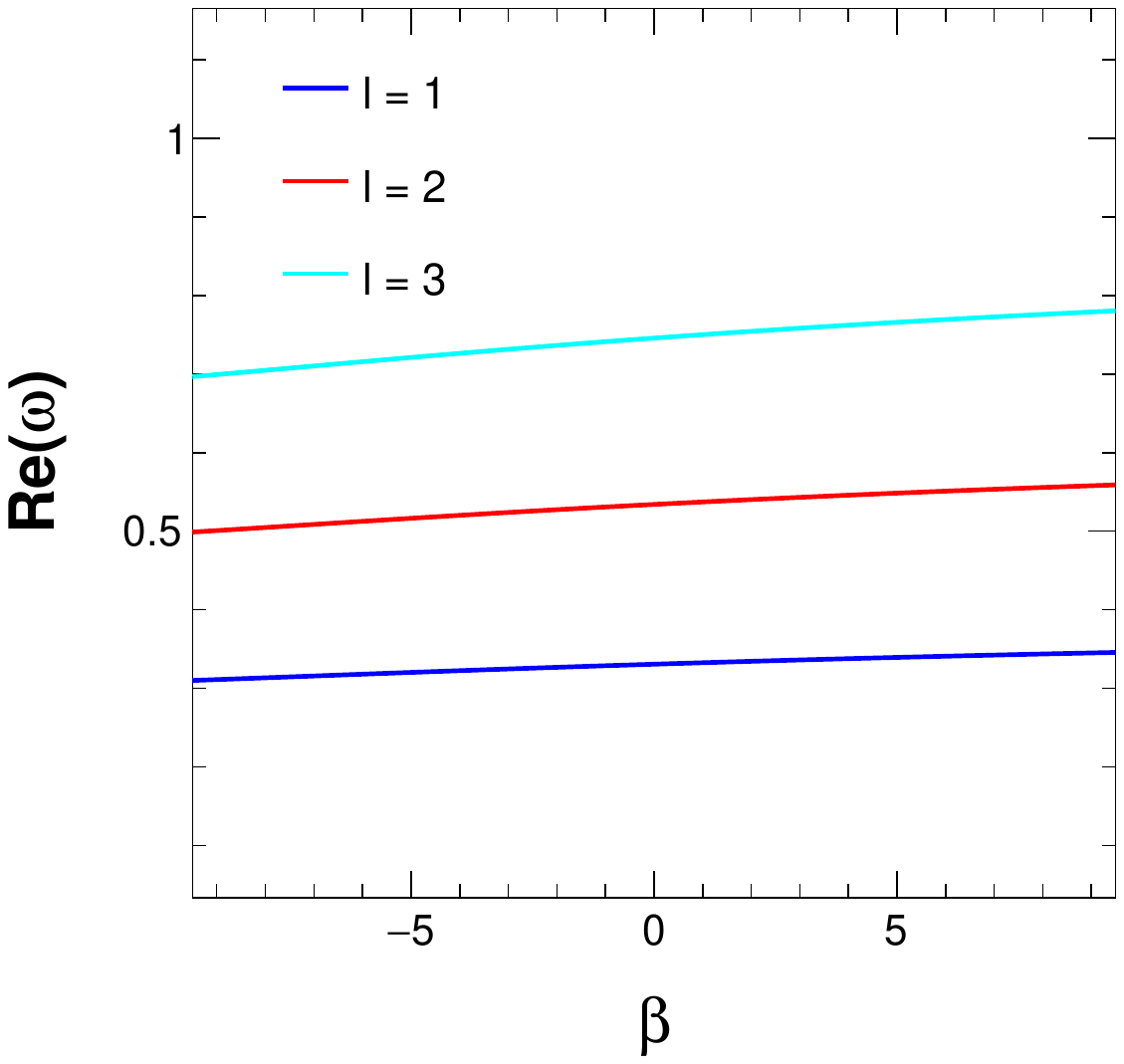}\hspace{0.2cm}
		\includegraphics[scale = 0.25]{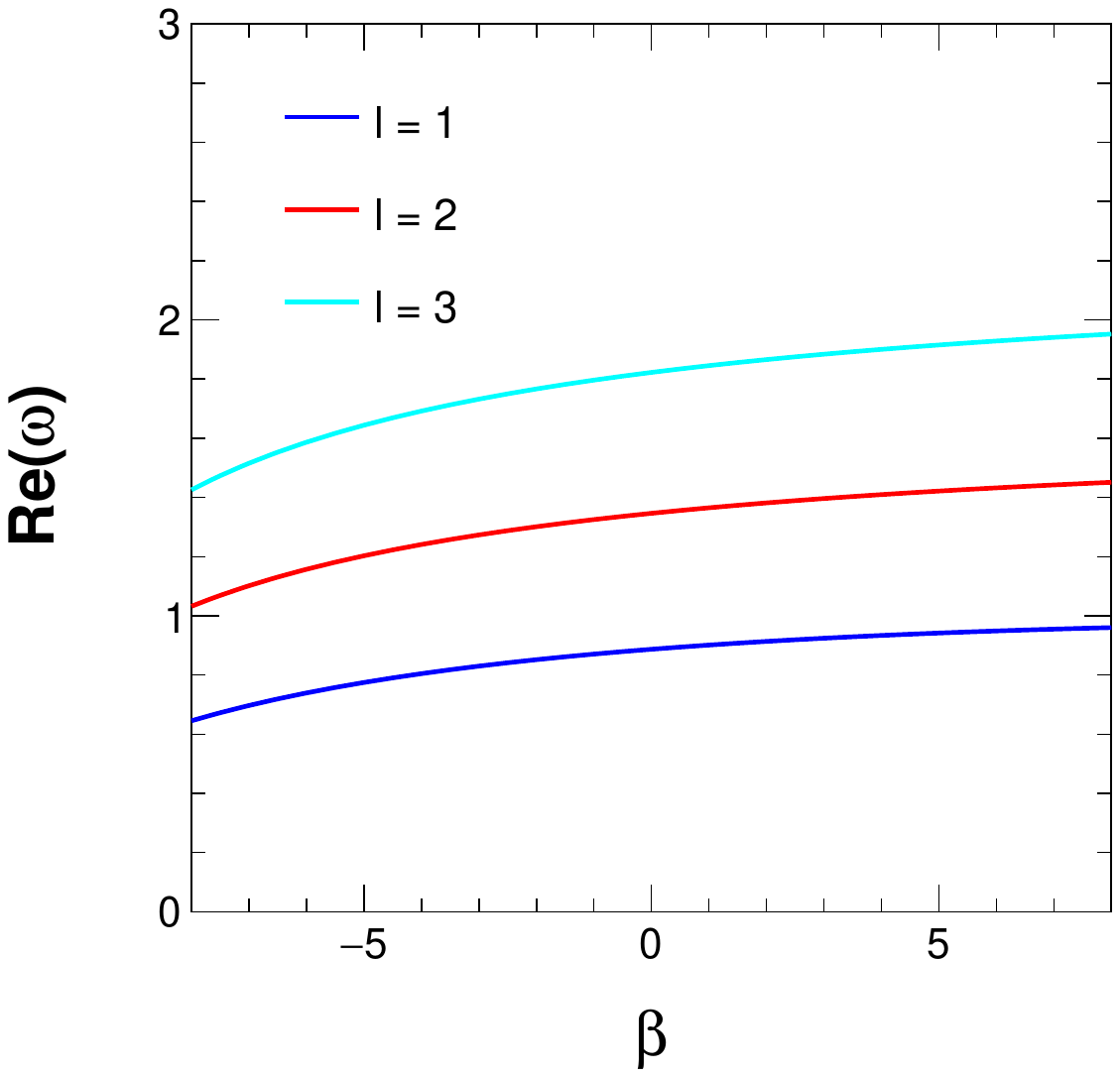}}\vspace{-0.2cm}
\caption{Amplitudes of 3rd order WKB QNMs versus $\beta$ of the minimally 
coupled $f(R,T)$ BHs for $\omega=0$ (left), $\omega=1/3$ (middle) and 
$\omega=-2/3$ (right) cases with three values of multipole $l$.
\textbf{Here $c_2=0.5$ is considered}.}
\label{fig10}
\end{figure}
From Fig.~\ref{fig10} it is seen that for a fixed value of $c_2=0.5$, 
the amplitude shows an increasing trend with an increase in the $\beta$ 
parameter for the case of $\omega=0$. Also, for the radiation case with 
$\omega=1/3$, the trend is almost similar, where amplitude increases as the 
value of parameter $\beta$ increases. A similar situation is observed for the 
$\omega=-2/3$ case where the amplitude increases with $\beta$ but in a 
nonlinear fashion in a lower $\beta$ regime. As usual, for higher multipole, 
we get a larger amplitude in all three cases.
\begin{figure}[!h]
	\centerline{
		\includegraphics[scale = 0.25]{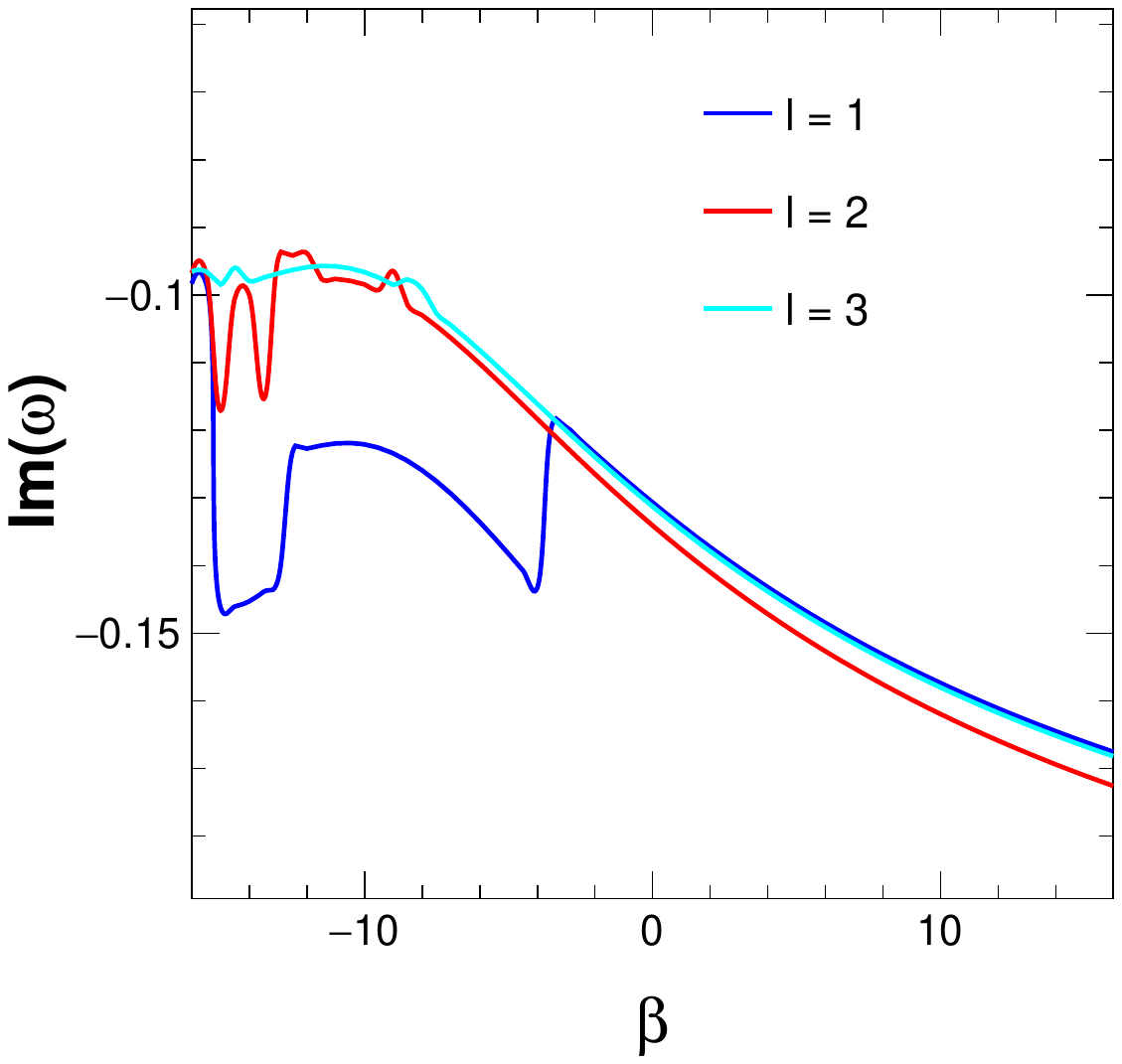}\hspace{0.2cm}
		\includegraphics[scale = 0.25]{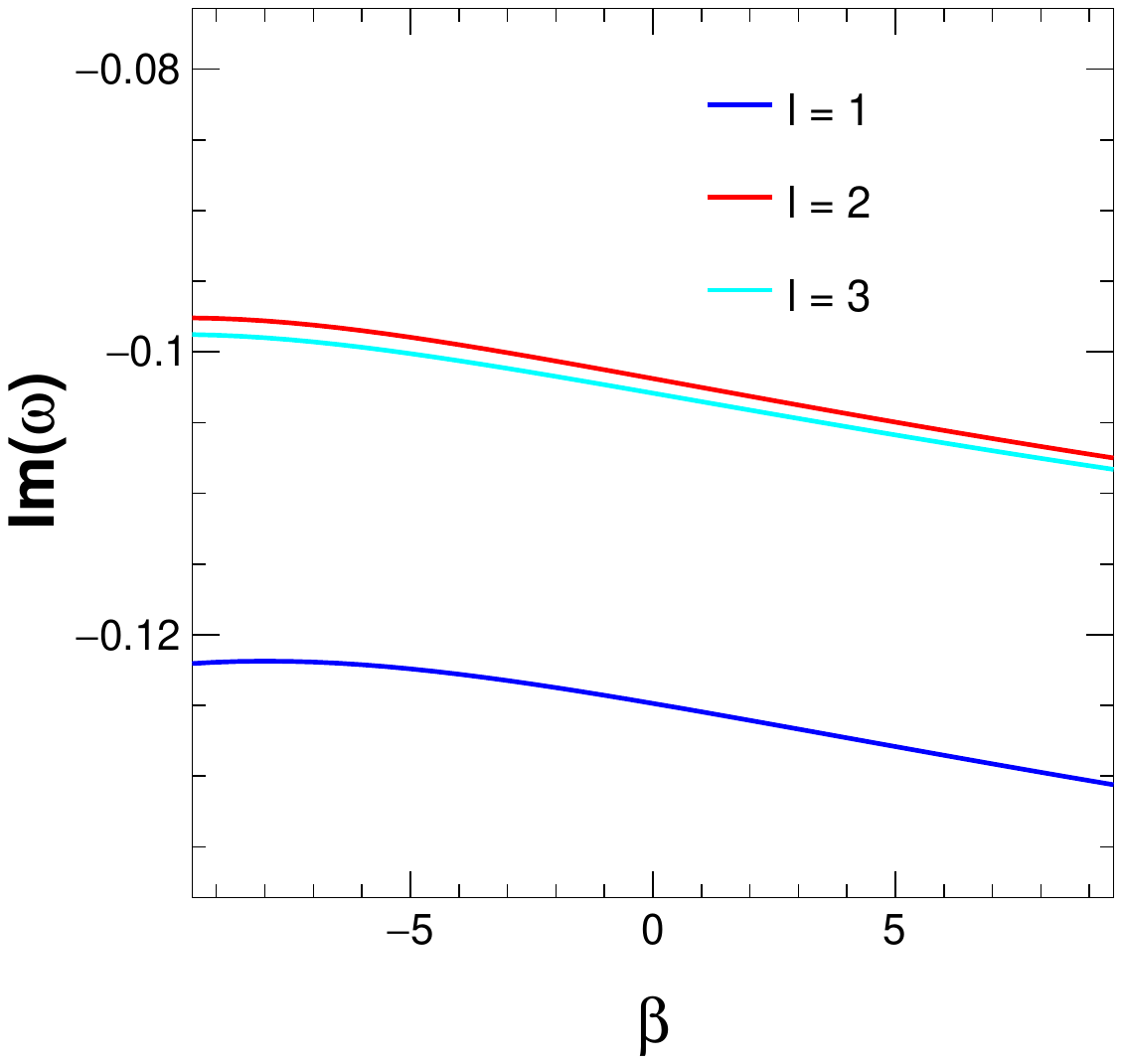}\hspace{0.2cm}
		\includegraphics[scale = 0.25]{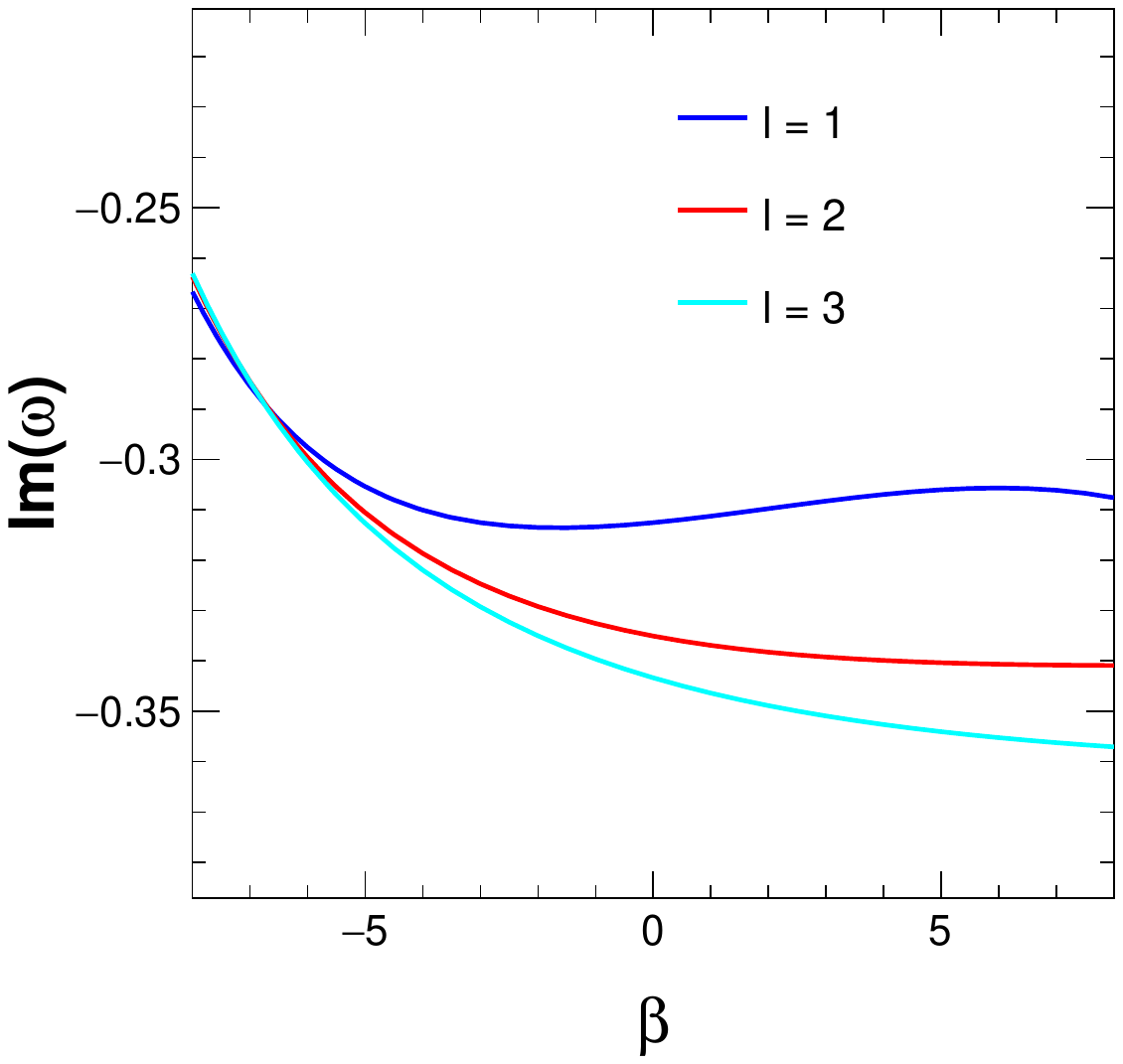}} \vspace{-0.2cm}
	\caption{Dampings of 3rd order WKB QNMs versus $\beta$ of the 
minimally coupled $f(R,T)$ BHs for $\omega=0$ (left), $\omega=1/3$ (middle) 
and $\omega=-2/3$ (right) cases with three values of multipole $l$. 
Here $c_2=0.5$ is considered.}
	\label{fig11}
\end{figure}

Similarly, we plot the damping part of the QNM in Fig.~\ref{fig11} 
with respect to the parameter $\beta$ for a fixed value of $c_2=0.5$. It is 
seen from the figure that for $\omega=0$, the damping is not stable up to 
some negative values of $\beta$ depending on the value of $l$. It decreases a 
little at the beginning but then starts increasing in a nonlinear manner as 
$\beta$ is increased. For higher values of $\beta>-4$, it is seen that the 
damping increases gradually and smoothly which slows down for higher positive 
values of $\beta$. Further, for higher $l$, the increase in the damping is 
less compared to lower $l$ with varying degrees depending on the value of 
$\beta$. For the case of $\omega=1/3$, the damping for the case $l=1$ is 
significantly lower than $l=2$ and $l=3$ case. $l=2$ and $l=3$ cases are close 
to each other. All three cases show an increase in damping with increasing 
$\beta$. For $\omega=-2/3$, there is a larger variation of damping for $l=1$ 
compared to the higher multipoles, where the damping initially increases then 
remains more or less the same with increasing $\beta$ depending on $l$ values. 
$l=2$ and $l=3$ plots show greater damping as $\beta$ is increased, 
compared to $l=1$ case. 

\begin{figure}[!h]
	\centerline{
		\includegraphics[scale = 0.25]{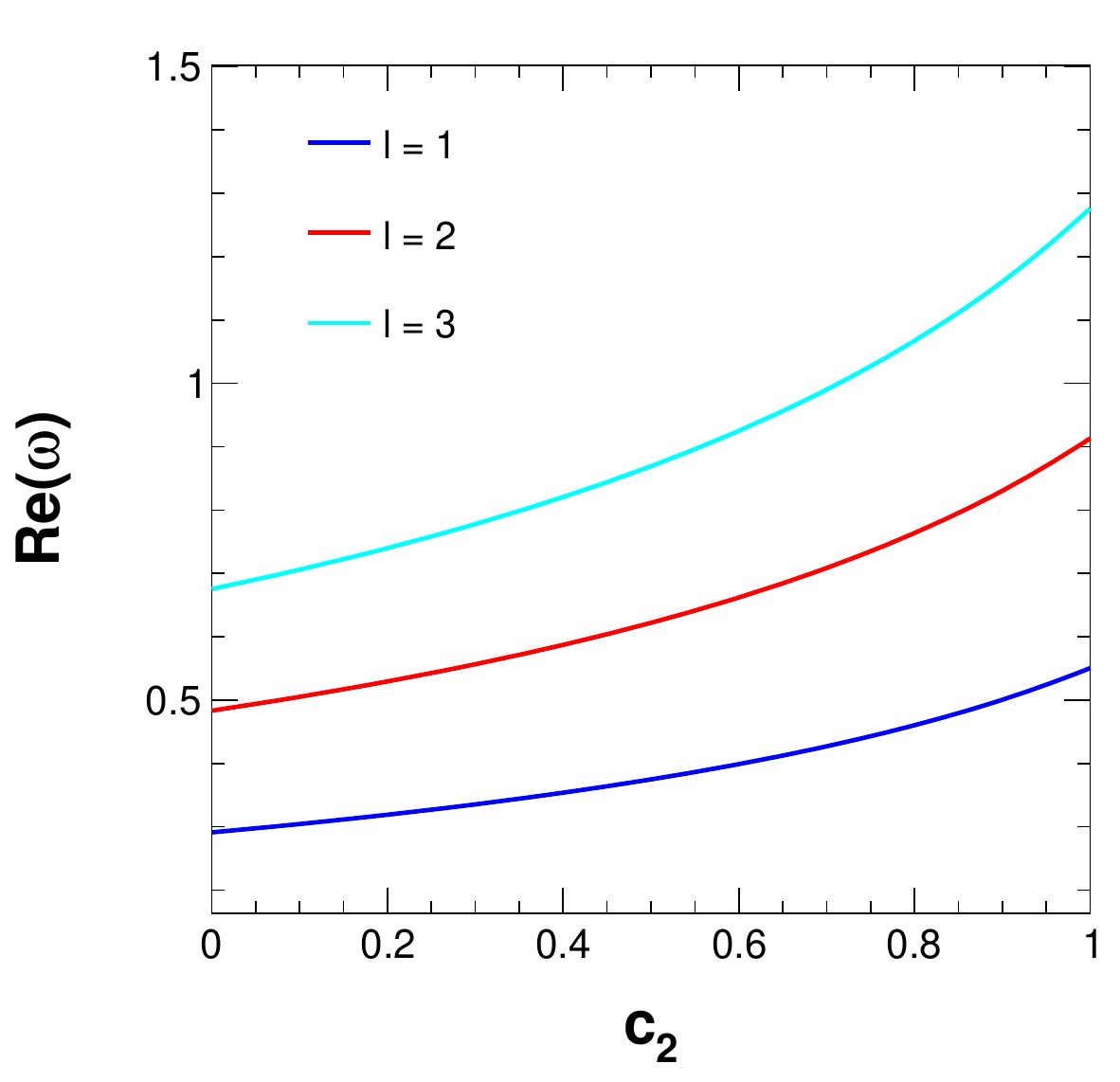}\hspace{0.2cm}
		\includegraphics[scale = 0.25]{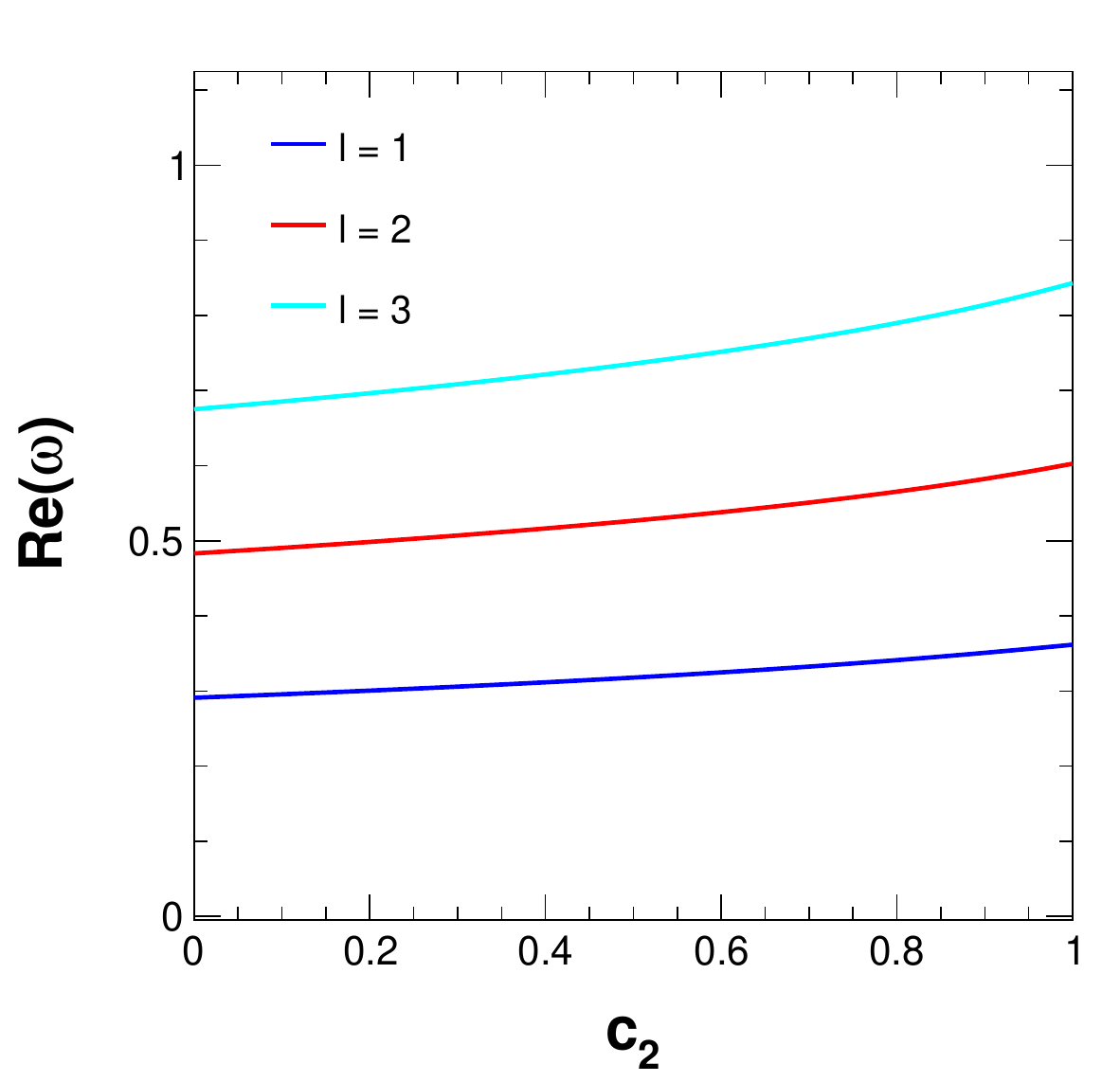}\hspace{0.2cm}
		\includegraphics[scale = 0.25]{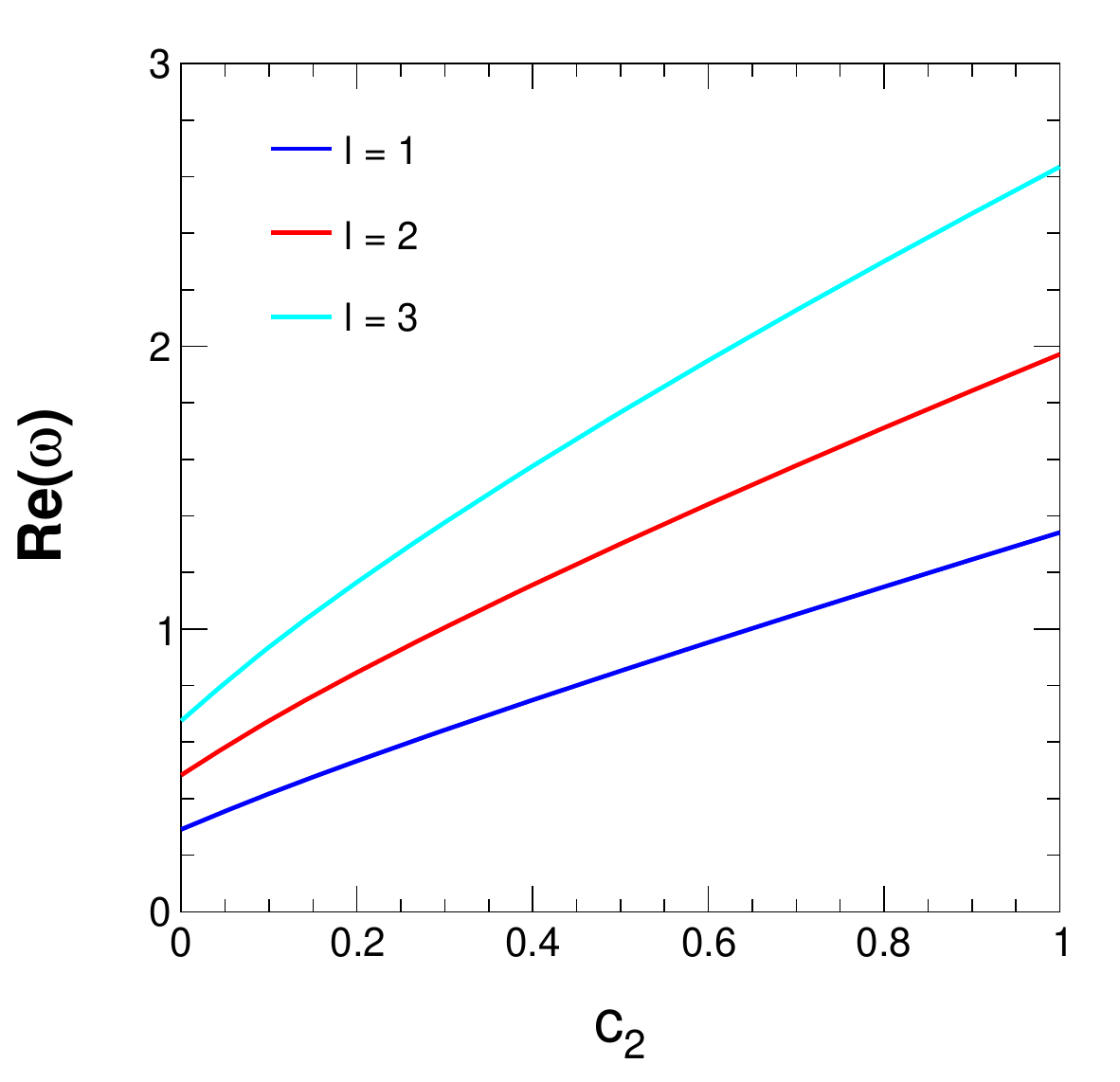}}\vspace{-0.2cm}
\caption{\textbf{Amplitudes of 3rd order WKB QNMs versus $c_2$ of the 
minimally coupled $f(R,T)$ BHs for $\omega=0$ (left), $\omega=1/3$ (middle) 
and $\omega=-2/3$ (right) cases with three values of multipole $l$ and 
$\beta=-2$.}}
\label{fig11-1}
\end{figure}
\begin{figure}[!h]
	\centerline{
		\includegraphics[scale = 0.25]{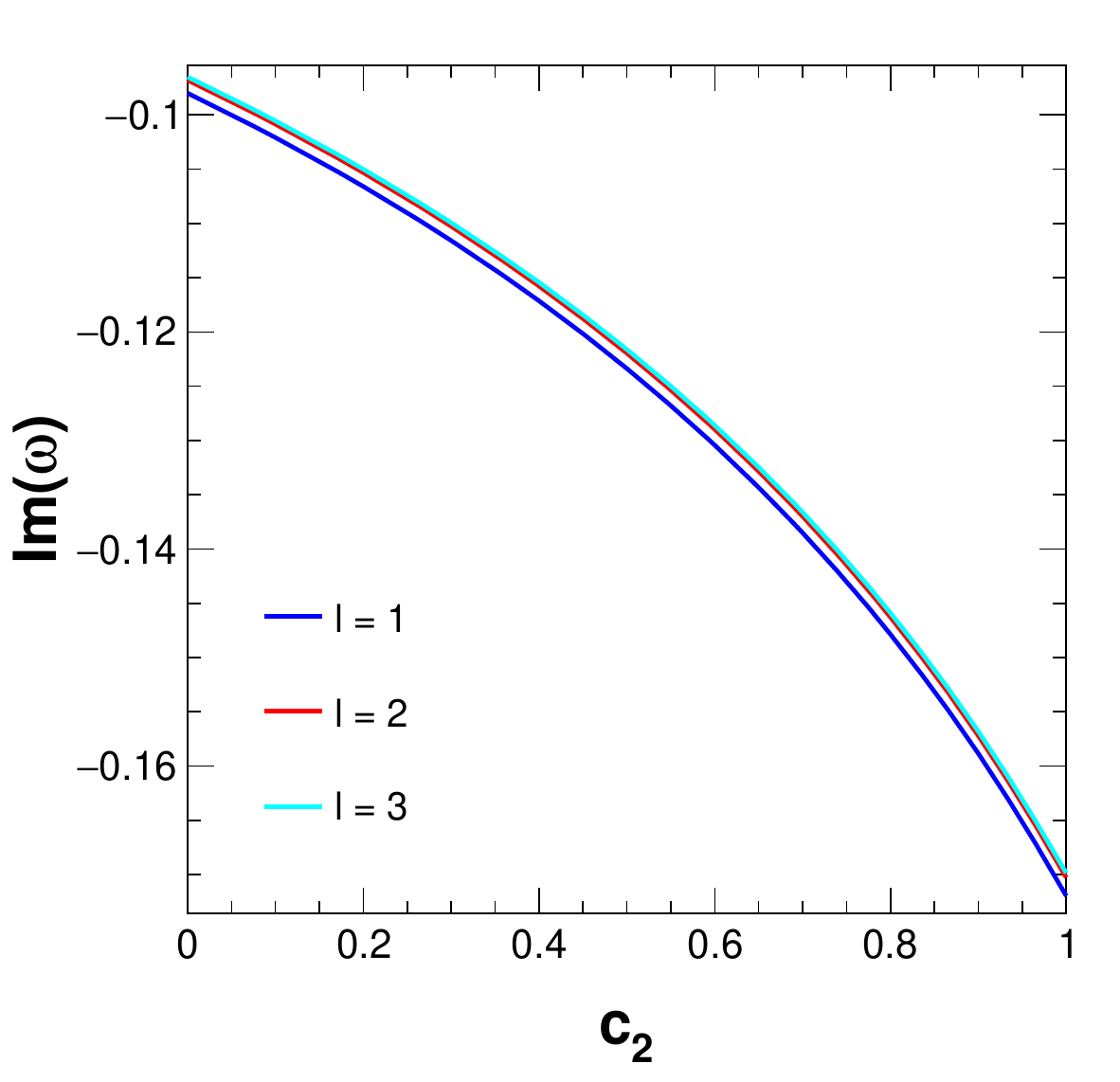}\hspace{0.2cm}
		\includegraphics[scale = 0.25]{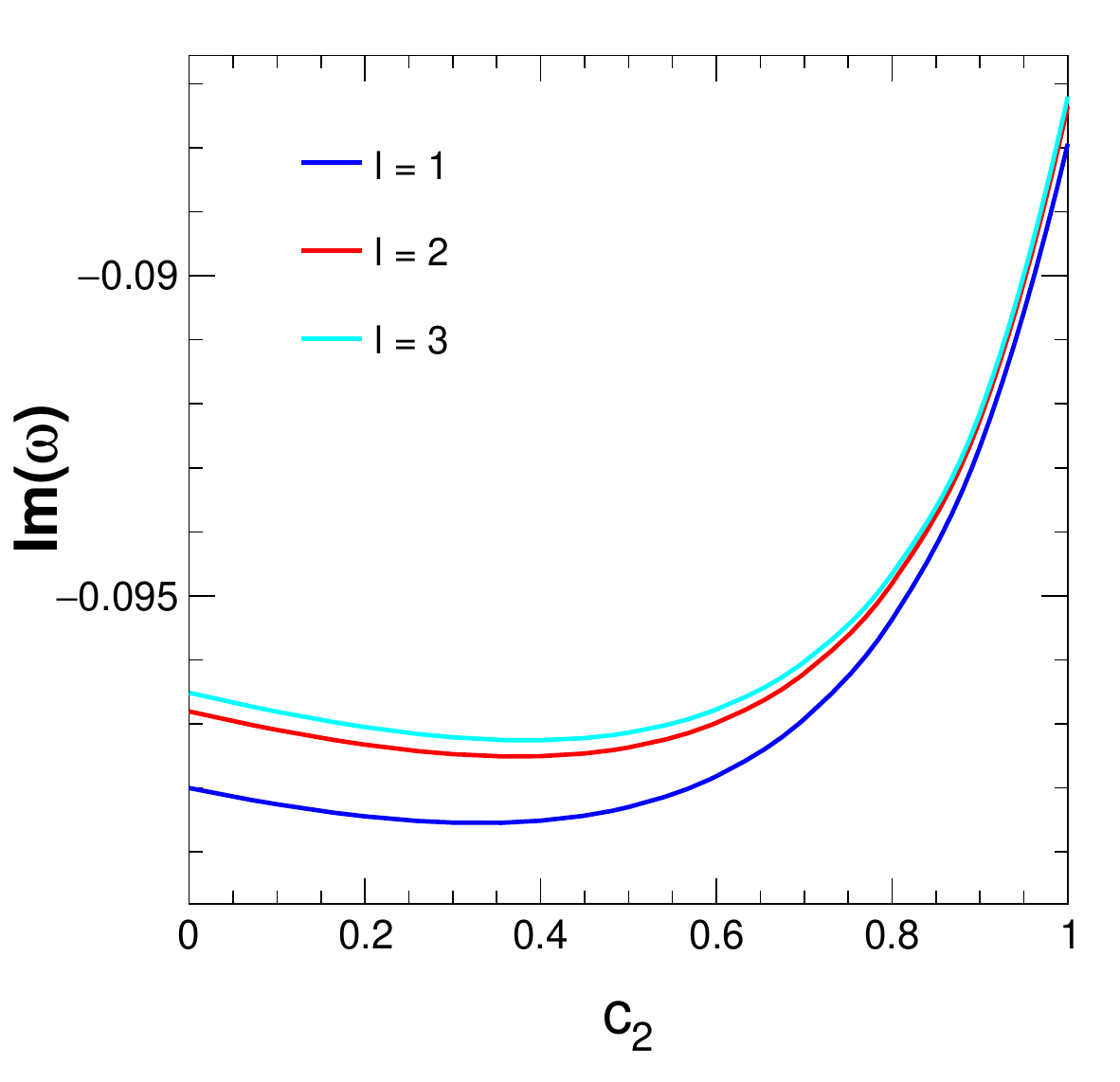}\hspace{0.2cm}
		\includegraphics[scale = 0.25]{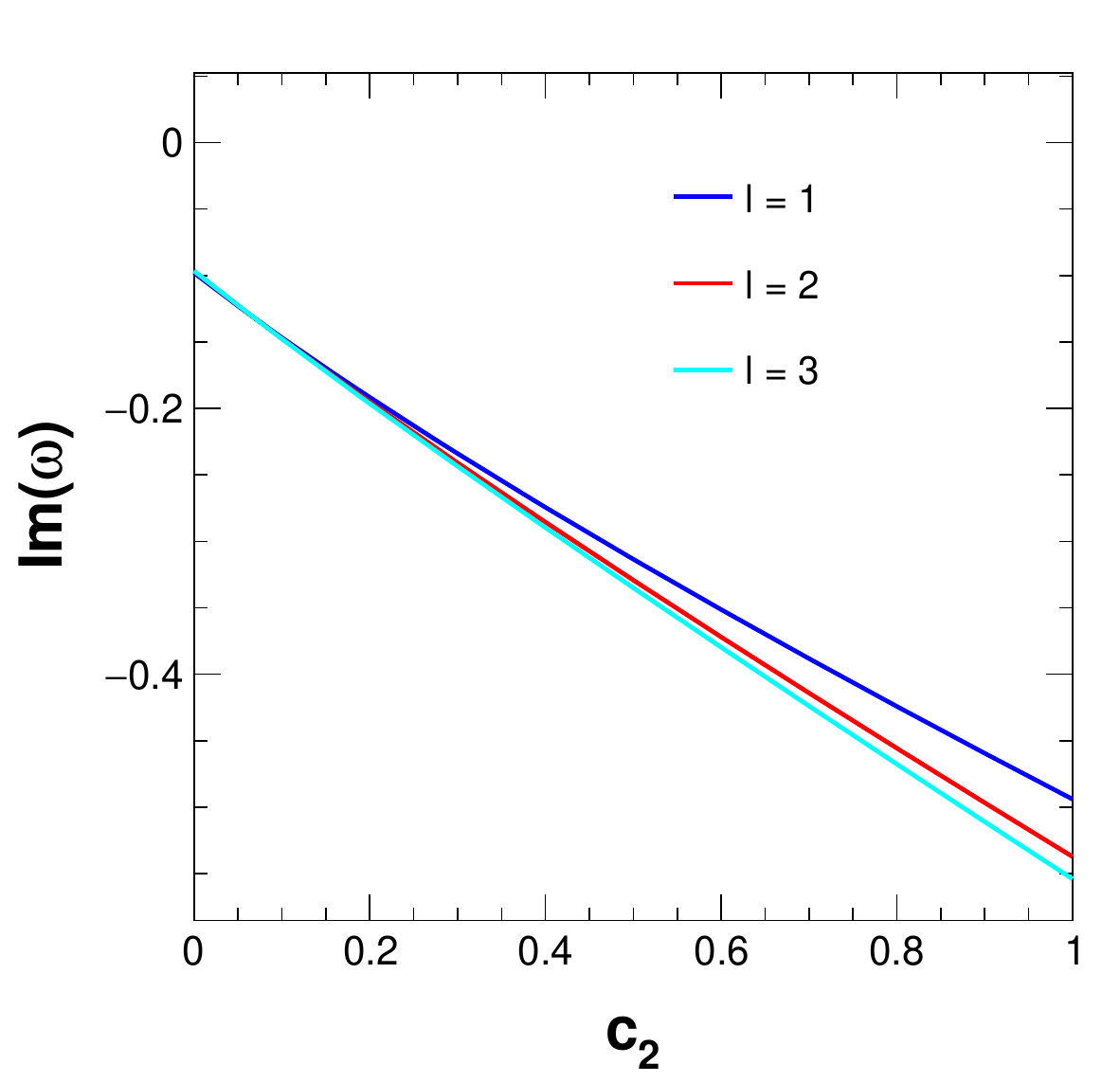}} \vspace{-0.2cm}
	\caption{Dampings of 3rd order WKB QNMs versus $c_2$ of 
the minimally coupled $f(R,T)$ BHs for $\omega=0$ (left), $\omega=1/3$ 
(middle) and $\omega=-2/3$ (right) cases with three values of multipole $l$ 
and $\beta=-2$.}
	\label{fig11-2}
\end{figure}

As seen from Fig.~\ref{fig11-1}, the amplitude increases with $c_2$ 
for the three multipole cases and three different $\omega$ cases. The 
damping (Fig.~\ref{fig11-2}), 
on the other hand, increases for $\omega=0$ and $\omega=-2/3$ cases but 
decreases for $\omega=1/3$. Here the value of parameter $\beta=-2$ is 
considered. It is to be noted that for $c_2$ values higher than 1.1 and 
for $\beta=-11.8$, the QNMs are found to be indeterminate. In Table \ref{table4}, 
we tabulate up to 4th order WKB QNMs for 
the minimally coupled $f(R,T)$ BHs solution \eqref{eq7} along with the 
associated errors.

\begin{table}[!h]
\caption{Up to 4th order WKB QNMs of the BHs solution \eqref{eq7} 
for $l=1,$ $2$, $n=0$, $\beta=-2$ and for different model \textbf{parameters 
$\omega$ and $c_2$}. The approximate errors ($\Delta_3$) associated with the 
WKB QNMs have also been computed.}
\vspace{8pt}
\centering
\scalebox{0.9}{
\begin{tabular}{c@{\hskip 5pt}c@{\hskip 10pt}c@{\hskip 10pt}c@{\hskip 10pt}c@{\hskip 10pt}c@{\hskip 15pt}c@{\hskip 15pt}c@{\hskip 15pt}c@{\hskip 15pt}c}
\hline \hline
\vspace{2mm}
& Multipole & $\omega$ & $\beta$ & $c_2$ & 4th order WKB QNMs & 3rd order WKB QNMs & 2nd order WKB QNMs & $\Delta_{3}$ \\[-3pt]
\hline
& Schwarzschild ($l=1$) & - & - & -              & 0.294510 - 0.107671i & 0.291114 - 0.098001i & 0.292954 - 0.097386i & $2.5762 \times 10^{-3}$ &\\
&\multirow{4}{4em}{$l=1$} & $0$ & $-2.0$ & $0.0$ & 0.292954 - 0.097386i & 0.291114 - 0.098001i & 0.294510 - 0.107671i  & $2.5762 \times 10^{-3}$ &\\ 
&                         & $0$ & $-2.0$ & $0.1$ & 0.306207 - 0.101463i & 0.304313 - 0.102094i & 0.307813 - 0.112096i  & $2.8913 \times 10^{-3}$ &\\
&                         & $0$ & $-5.0$ & $0.1$ & 0.302563 - 0.099759i & 0.300738 - 0.100364i & 0.304121 - 0.110088i & $2.4242 \times 10^{-3}$ &\\
&                         & $0$ & $-5.0$ & $0.5$ & 0.353172 - 0.111042i & 0.351468 - 0.111580i & 0.354682 - 0.121326i & $2.3209 \times 10^{-3}$ &\\

\hline
& Schwarzschild ($l=2$) & - & - & - & 0.483977 - 0.100560i & 0.483211 - 0.096805i & 0.483647 - 0.096718i & $1.1901 \times 10^{-4}$ &\\
&\multirow{4}{4em}{$l=2$} & $0$ & $-2.0$ & $0.0$ & 0.483647 - 0.096718i & 0.483211 - 0.096805i & 0.483977 - 0.100560i  & $1.1901 \times 10^{-4}$ &\\ 
&                         & $0$ & $-2.0$ & $0.1$ & 0.505550 - 0.100770i & 0.505101 - 0.100859i & 0.505891 - 0.104743i  & $1.9871 \times 10^{-3}$ &\\
&                         & $0$ & $-5.0$ & $0.1$ & 0.499527 - 0.099080i & 0.499093 - 0.099166i & 0.499857 - 0.102943i & $5.4445 \times 10^{-4}$ &\\
&                         & $0$ & $-5.0$ & $0.5$ & 0.583196 - 0.110337i & 0.582782 - 0.110416i & 0.583511 - 0.114203i & $5.1997 \times 10^{-4}$ &\\
\hline
&\multirow{4}{4em}{$l=1$} & $1/3$ & $-2.0$ & $0.0$ & 0.292954 - 0.097386i & 0.291114 - 0.098001i & 0.294510 - 0.107671i  & $2.5762 \times 10^{-3}$ &\\ 
&                         & $1/3$ & $-2.0$ & $0.1$ & 0.297469 - 0.097668i & 0.295689 - 0.098256i & 0.298964 - 0.107709i  & $2.6263 \times 10^{-3}$ &\\
&                         & $1/3$ & $-5.0$ & $0.1$ & 0.296439 - 0.097333i & 0.294646 - 0.0979247i & 0.297918 - 0.107368i & $2.3329 \times 10^{-3}$ &\\
&                         & $1/3$ & $-5.0$ & $0.5$ & 0.312994 - 0.096019i & 0.311265 - 0.0965518i & 0.314002 - 0.105040i & $1.8571 \times 10^{-3}$ &\\
\hline
&\multirow{4}{4em}{$l=2$} & $1/3$ & $-2.0$ & $0.0$ & 0.483647 - 0.096718i & 0.483211 - 0.096805i & 0.483977 - 0.100560i  & $1.1901 \times 10^{-4}$ &\\ 
&                         & $1/3$ & $-2.0$ & $0.1$ & 0.491023 - 0.097009i & 0.490596 - 0.097094i & 0.491337 - 0.100769i  & $1.8799 \times 10^{-3}$ &\\
&                         & $1/3$ & $-5.0$ & $0.1$ & 0.489297 - 0.096684i & 0.488866 - 0.096769i & 0.489607 - 0.100441i & $5.2277 \times 10^{-4}$ &\\
&                         & $1/3$ & $-5.0$ & $0.5$ & 0.516199 - 0.095577i & 0.515768 - 0.095657i & 0.516391 - 0.098960i & $4.0751 \times 10^{-4}$ &\\
\hline
&\multirow{4}{4em}{$l=1$} & $-2/3$ & $-2.0$ & $0.0$ & 0.292954 - 0.097386i & 0.291114 - 0.098001i & 0.294510 - 0.107671i  & $2.5762 \times 10^{-3}$ &\\ 
&                         & $-2/3$ & $-2.0$ & $0.1$ & 0.421151 - 0.145533i & 0.417478 - 0.146813i & 0.423742 - 0.163778i  & $5.5869 \times 10^{-3}$ &\\
&                         & $-2/3$ & $-5.0$ & $0.1$ & 0.398722 - 0.139841i & 0.395415 - 0.141010i & 0.401333 - 0.156839i & $4.1784 \times 10^{-3}$ &\\
&                         & $-2/3$ & $-5.0$ & $0.5$ & 0.786104 - 0.300951i & 0.774726 - 0.305371i & 0.793324 - 0.349878i & $1.2654 \times 10^{-2}$ &\\
\hline
&\multirow{4}{4em}{$l=2$} & $-2/3$ & $-2.0$ & $0.0$ & 0.483647 - 0.096718i & 0.483211 - 0.096805i & 0.483977 - 0.100560i  & $1.1901 \times 10^{-4}$ &\\ 
&                         & $-2/3$ & $-2.0$ & $0.1$ & 0.676258 - 0.147199i & 0.675357 - 0.147396i & 0.676897 - 0.154297i  & $3.5725 \times 10^{-3}$ &\\
&                         & $-2/3$ & $-5.0$ & $0.1$ & 0.644813 - 0.140203i & 0.643994 - 0.140381i & 0.645424 - 0.146803i & $1.0148 \times 10^{-3}$ &\\
&                         & $-2/3$ & $-5.0$ & $0.5$ & 1.206000 - 0.309746i & 1.202880 - 0.310548i & 1.208190 - 0.330497i & $3.7180 \times 10^{-3}$ &\\
\hline \hline \vspace{4mm}
\end{tabular}}
\label{table4}
\end{table}

As seen from this Table \ref{table4}, the QNMs are tabulated for the cases of 
$\omega=0$, i.e.~the dust matter case, $\omega=1/3$, i.e.~the radiation case 
and $\omega=-2/3$, i.e.~the quintessence case, considering $c_2$ and 
$\beta$ as free parameters. We have also presented the Schwarzschild case 
for multipole $l=1$ and $l=2$ for the purpose of comparison. 
It is seen that the QNMs have significant deviations from Schwarzschild case 
when we introduce a MTG model. As already seen from
Fig.~\ref{fig10}, a common observation is that with increasing multipole, 
the amplitude increases. Similar is the trend for the parameter $\beta$ where 
the amplitude shows an increasing trend. The damping part also shows the same 
trend with respect to $\beta$ as discussed from Fig.~\ref{fig10} analysis. The 
associated errors have been computed and come out to be in the range of 
$10^{-5}$ to $10^{-1}$ depending on the value of $\beta$, $l$ and $\omega$. 
In general, a specific lower value of $\beta=-16$ and higher $l$ values give 
lower errors. Whereas for $\omega=-2/3$, the errors are maximum. It needs to be
mentioned that the error $\Delta_3$ is calculated as \cite{r19}
\begin{equation}
\Delta_3 = \frac{W\!K\!B_4-W\!K\!B_2}{2},
\end{equation}
where $W\!K\!B_2$ and $W\!K\!B_4$ are QNMs calculated respectively from the
2nd and 4th order WKB approximation method.  

For the sake of the accuracy of our calculations, we compare the results of
QNMs obtained by the WKB method with that of time domain analysis for which we 
plot the time profile of the wave. The next section deals with the time 
profile and QNMs computed by a wave fitting algorithm.

\section{Evolution of a scalar perturbation around the black holes}
\label{sec.5}
In this section, we study the evolution of a scalar perturbation around our 
considered BHs' spacetime for three different cases: $\omega=0$, $\omega=1/3$ 
and $\omega=-2/3$ for different values of $\beta$ and $c_2$. We use 
the time domain integration method to study this 
evolution. For this, we follow the method described in Refs.~\cite{Dhruba1,
Gundlach}. Thus, considering $\psi(x,t) = \psi(i{\Delta}x,j{\Delta}t)=
\psi_{i,j}$ and $V(r(x)) = V(x,t)=V_{i,j}$, the equation \eqref{eq37} can be 
expressed in the following form: 
\begin{equation}
\frac{\psi_{i+1,j}-2\psi_{i,j}+\psi_{i-1,j}}{{\Delta}x^2}-\frac{\psi_{i,j+1}-2\psi_{i,j}+\psi_{i,j-1}}{{\Delta}t^2}-V_{i}\psi_{i,j}=0.
\end{equation} 
This equation can be rewritten as
\begin{equation}
\psi_{i,j+1} = -\,\psi_{i,j-1}+\left(\frac{{\Delta}t}{{\Delta}x}\right)^{\!2}\!\!(\psi_{i+1,j}+\psi_{i-1,j})+\left(2-2\,\left(\frac{{\Delta}t}{{\Delta}x}\right)^{\!2}-V_{i}\Delta{t}^2\right)\!\psi_{i,j}.
\end{equation}
Here, we use initial conditions as 
$\psi(x,t) = \exp\left[-(x-\mu)^2/2{\sigma}^2\right]$ and 
$\left.\psi(x,t)\right|_{t<0} = 0$, where $\mu$ is the median and $\sigma$ is 
the width of the initial wave packet. Further, we use the Von Neumann 
stability condition $\frac{{\Delta}t}{{\Delta}x}\!<\!1$ for stability of 
results and calculate the time profiles.
\begin{figure}[!h]
    \centerline{
     \includegraphics[scale=0.300]{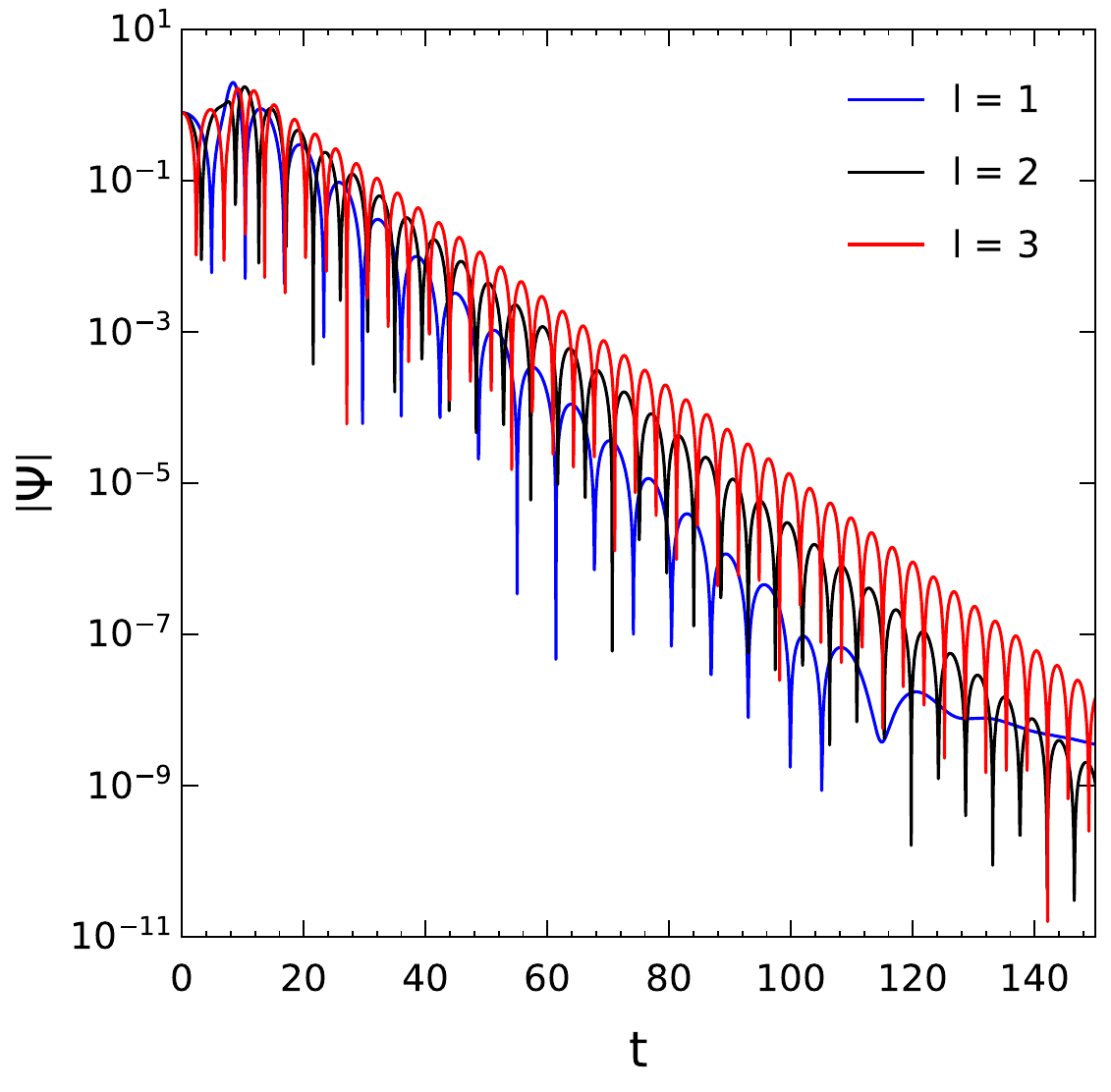}\hspace{0.3cm}
     \includegraphics[scale=0.300]{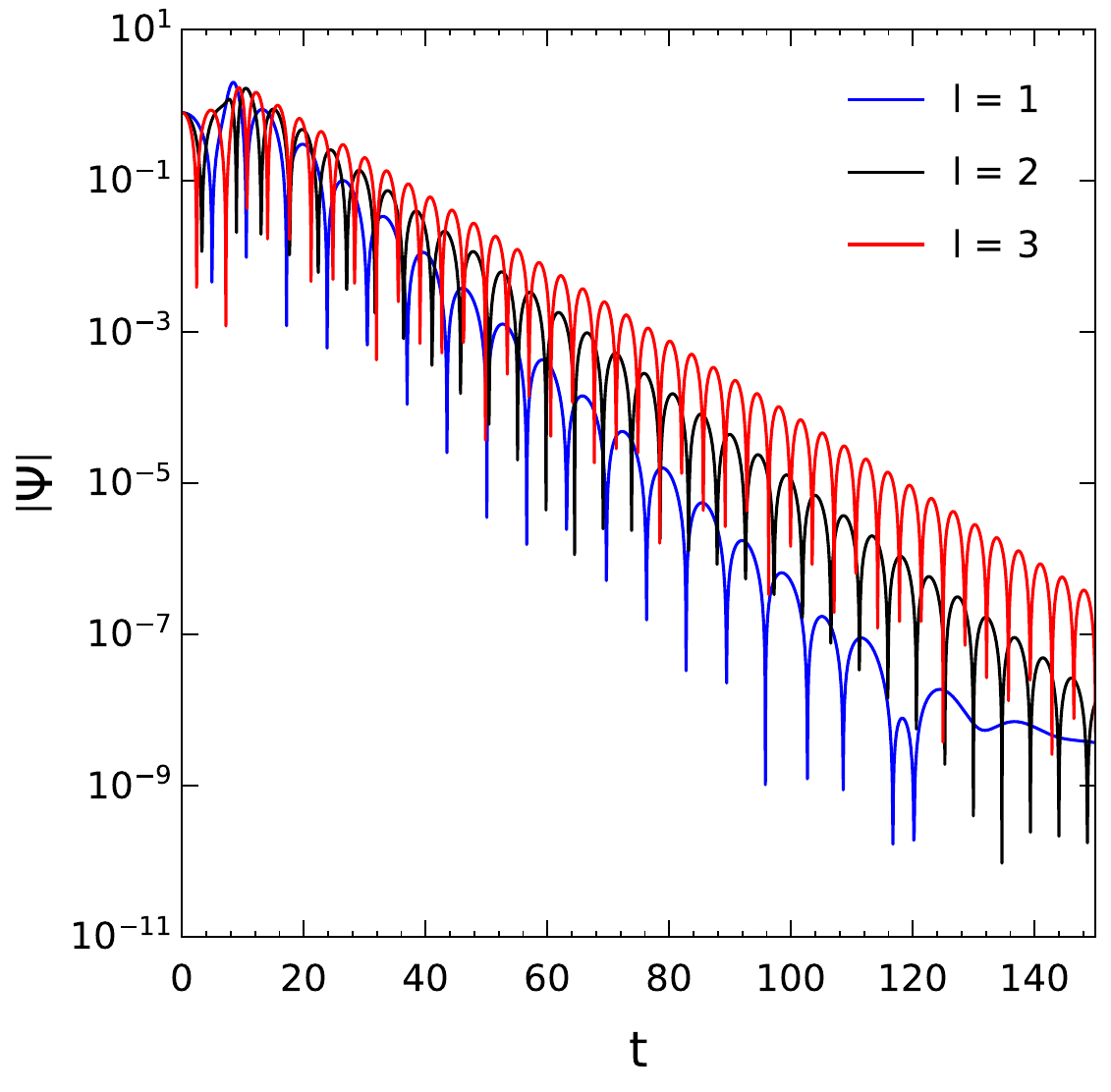}\hspace{0.3cm}
     \includegraphics[scale=0.3]{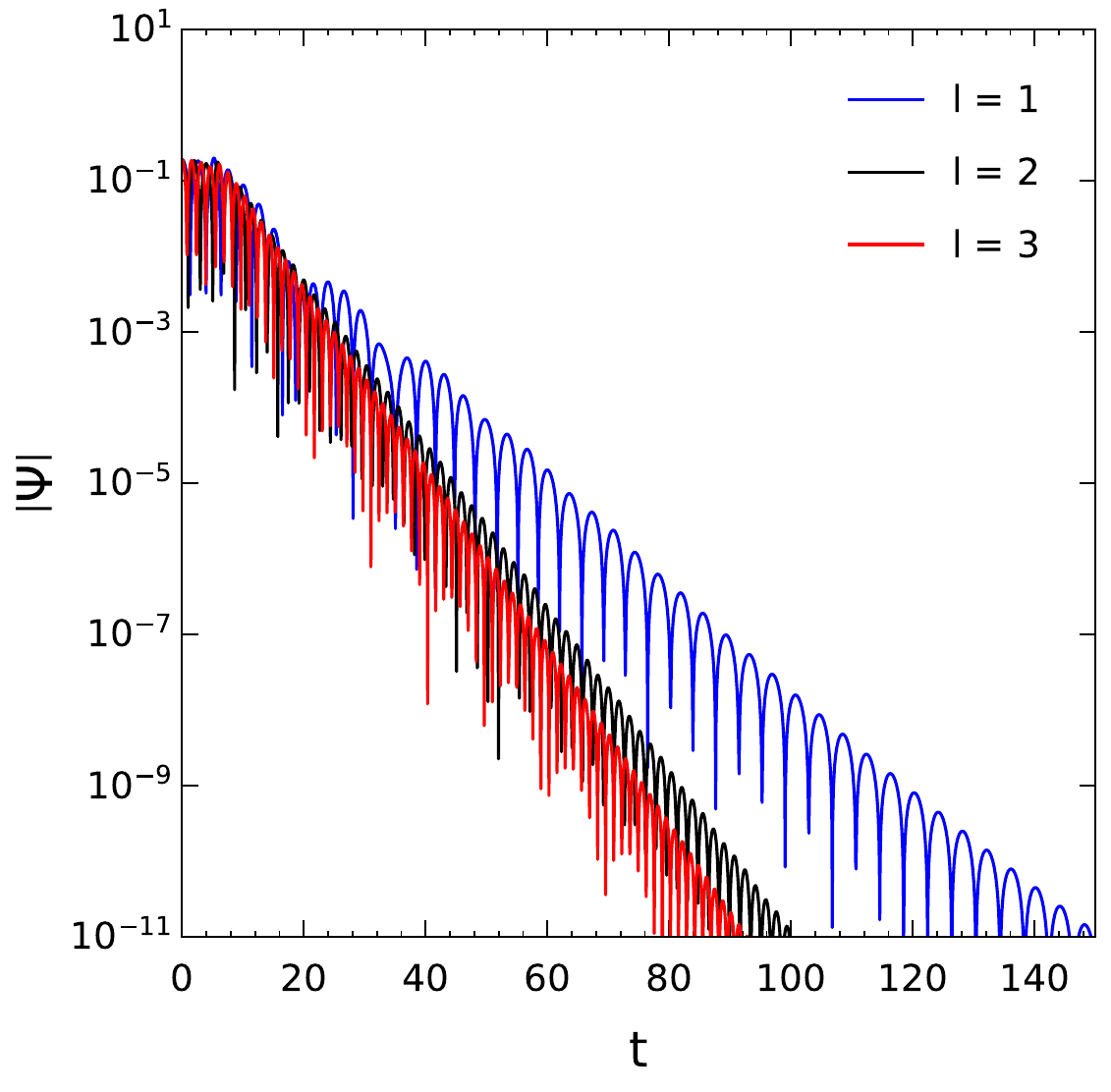}}
     \centerline{
     \includegraphics[scale=0.300]{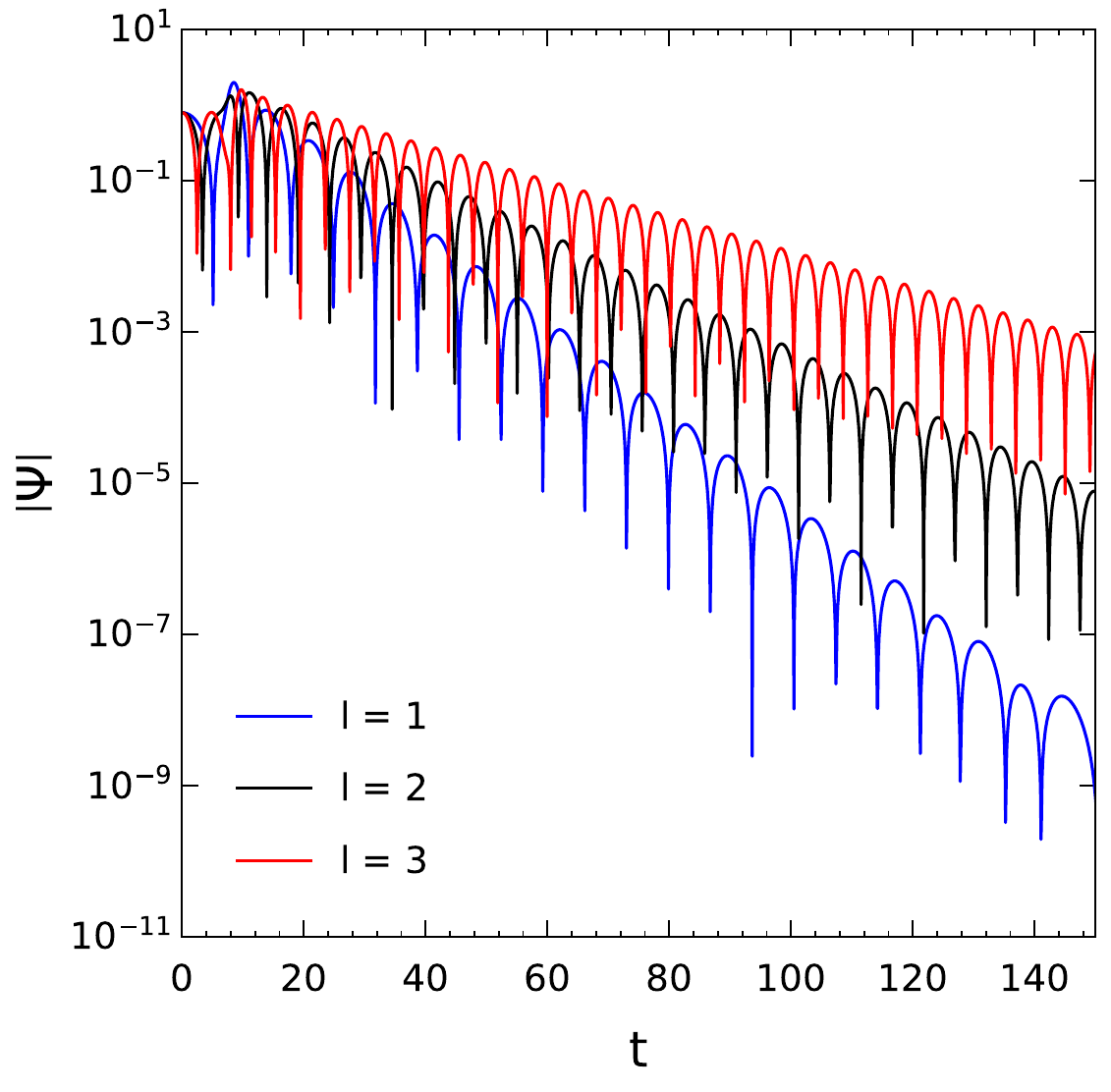}\hspace{0.3cm}
     \includegraphics[scale=0.300]{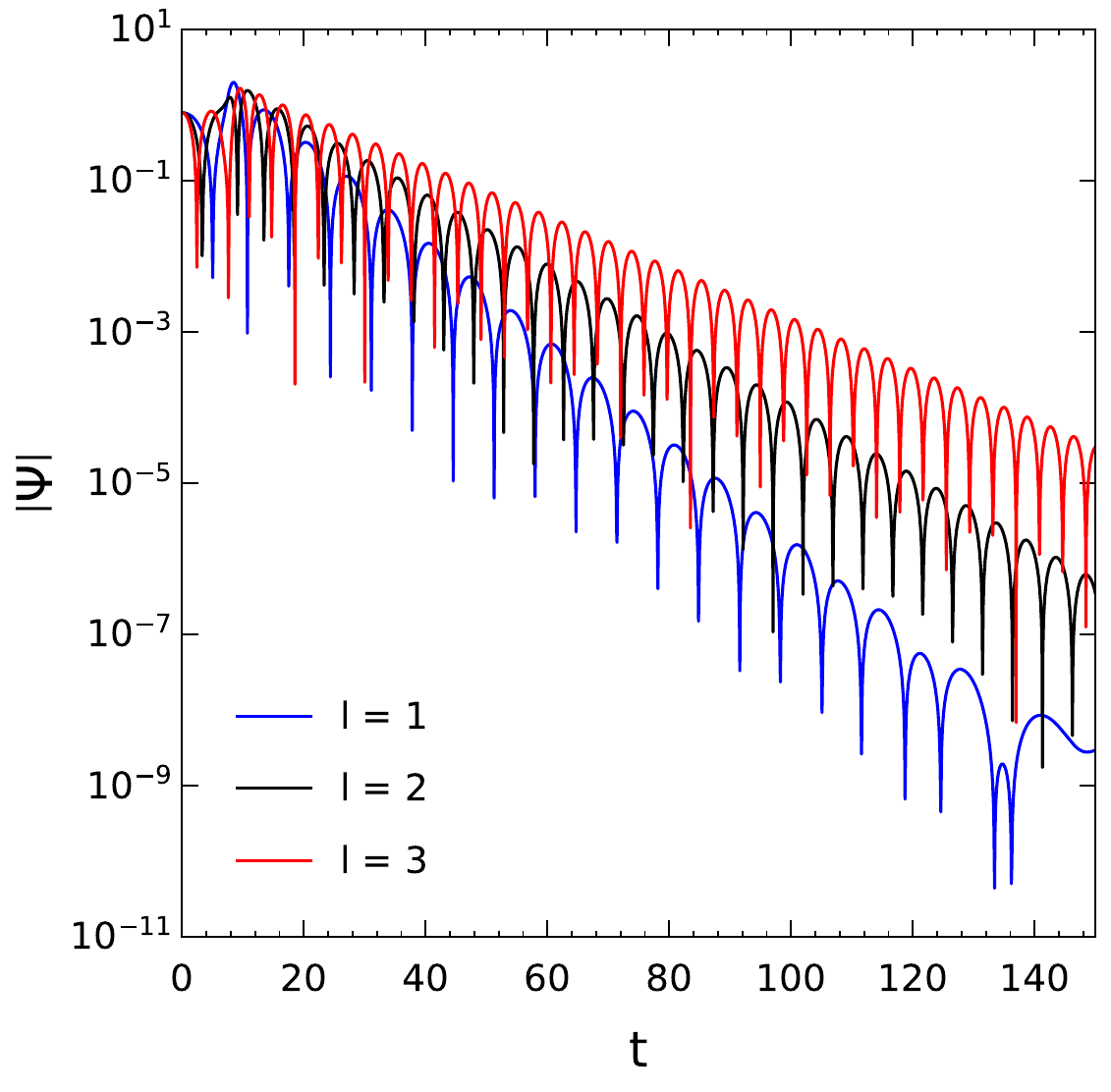}
\hspace{0.3cm}
     \includegraphics[scale=0.300]{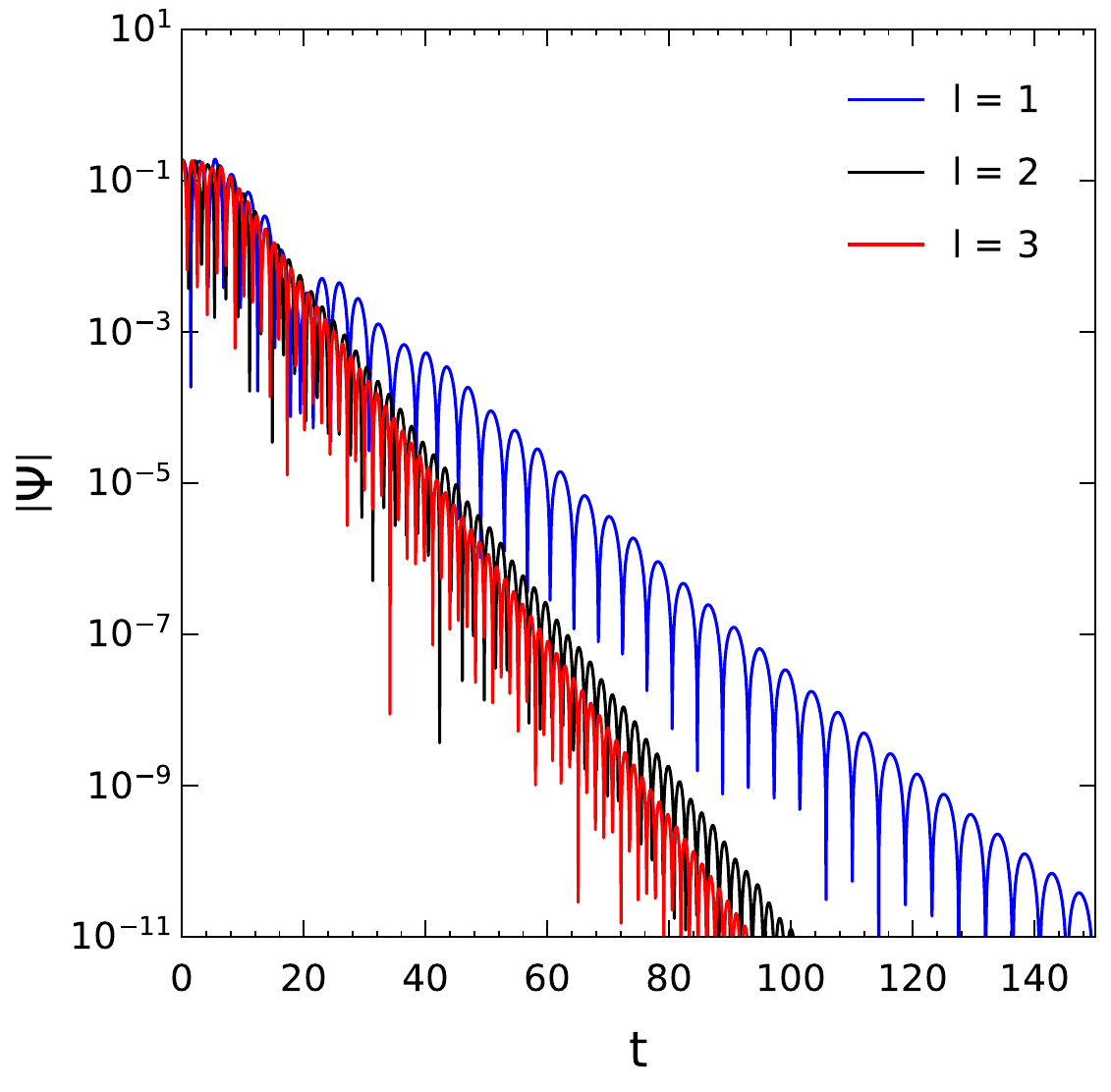}}
\vspace{-0.2cm}
    \caption{Time domain profile for the scalar perturbation in the 
spacetime of the minimally coupled $f(R,T)$ BHs. In the upper row, the 
left plot is for $\omega=0$, $\beta=-8.8$ and $c_2 = 1$, the middle plot is 
for $\omega=1/3$, $\beta=-6$ and $c_2 = 1$, and the right plot is for 
$\omega=-2/3$, $\beta=-2$ and $c_2 = 1$. In the lower row, the left plot is 
for $\omega = 0$, $\beta = -8.8$ and $c_2 = 0.1$, the middle plot is for 
$\omega = 1/3$, $\beta = -5.5$ and $c_2 = 0.5$, and the right plot is for 
$\omega = -2/3$, $\beta = -2$ and $c_2 = 0.8$.}
\label{fig:12}
\end{figure}

In Fig.~\ref{fig:12} we plot the time domain profiles for $\omega=0$, 
$\omega=1/3$ and $\omega=-2/3$ in the unit of $M$. It can be observed from 
the figure that the pattern of variation of the time domain profile with the 
multipole $l$ is in agreement with the results obtained from the 3rd order WKB 
approximation method. 
\begin{figure}[!h]
\centerline{
\includegraphics[scale=0.300]{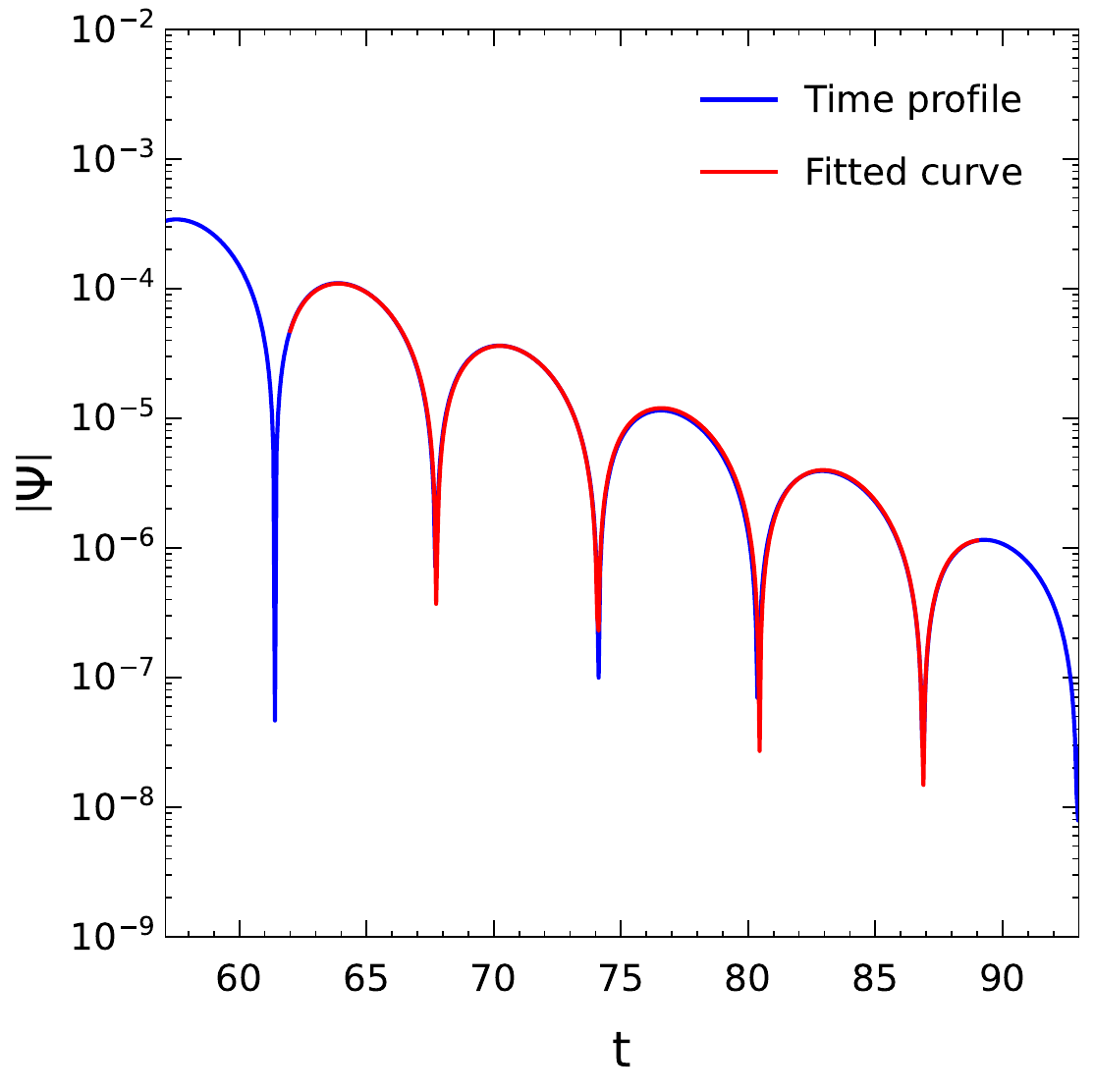}\hspace{0.3cm}
\includegraphics[scale=0.300]{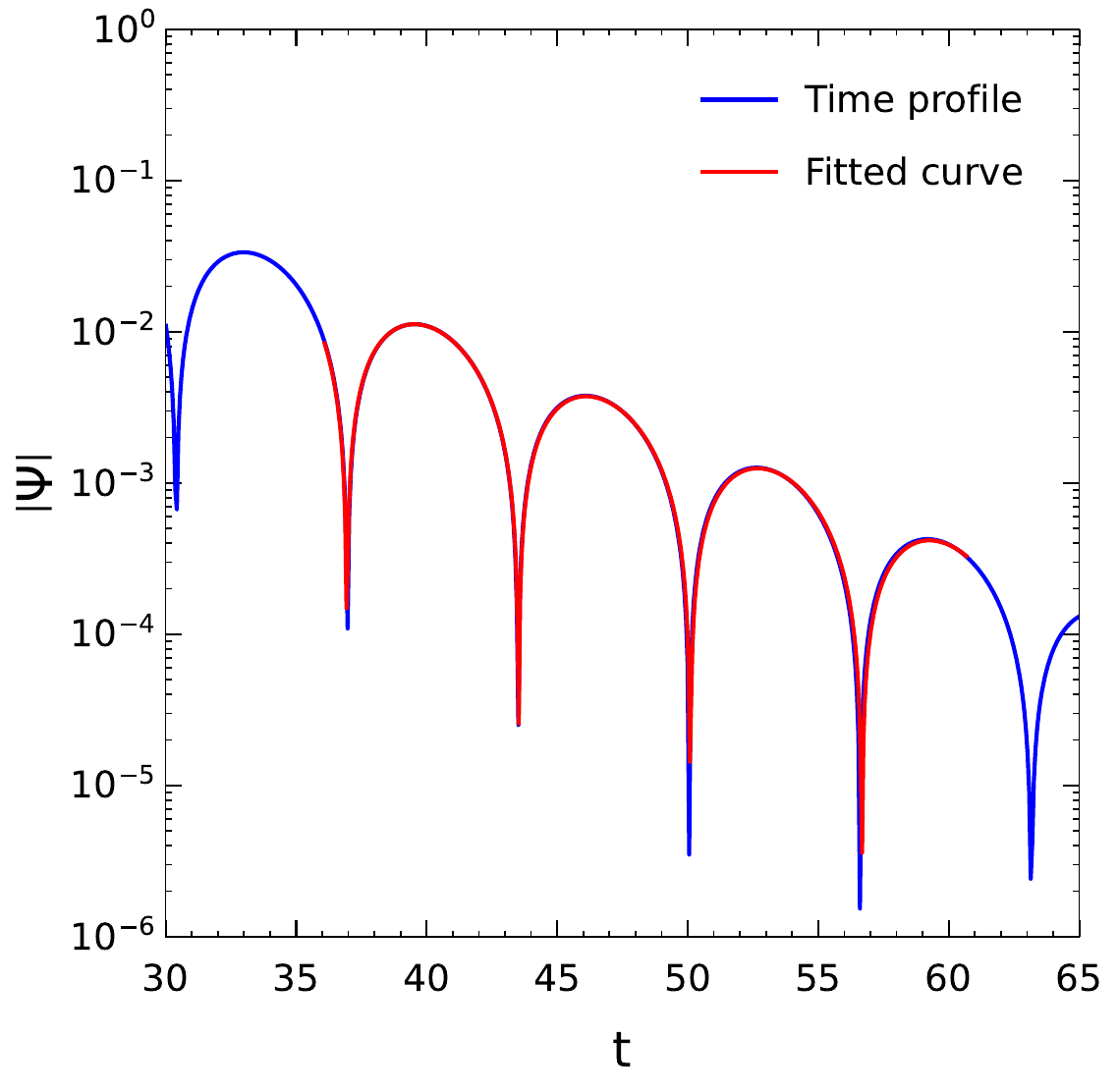}\hspace{0.3cm}
\includegraphics[scale=0.300]{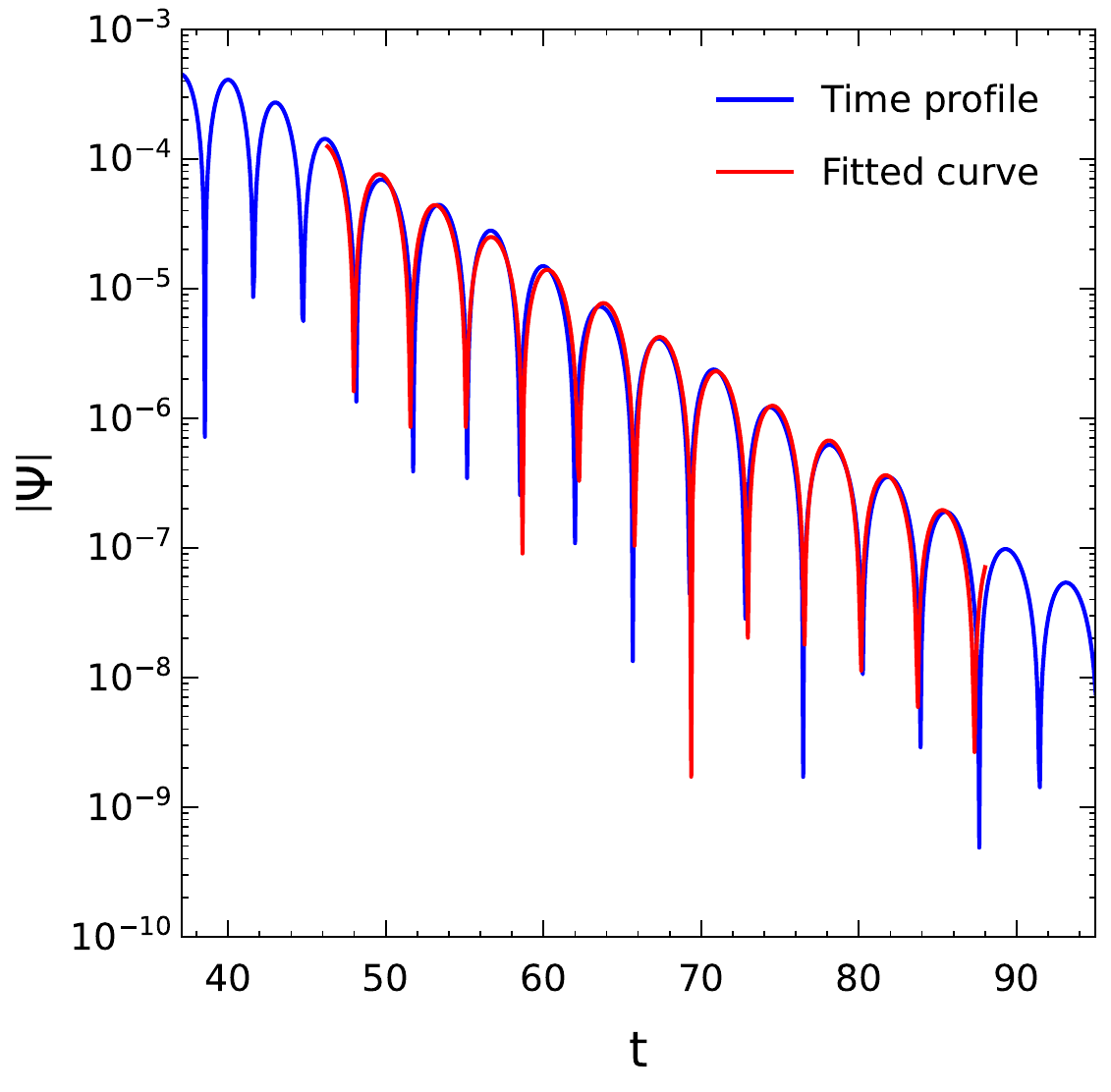}}
\centerline{
\includegraphics[scale=0.300]{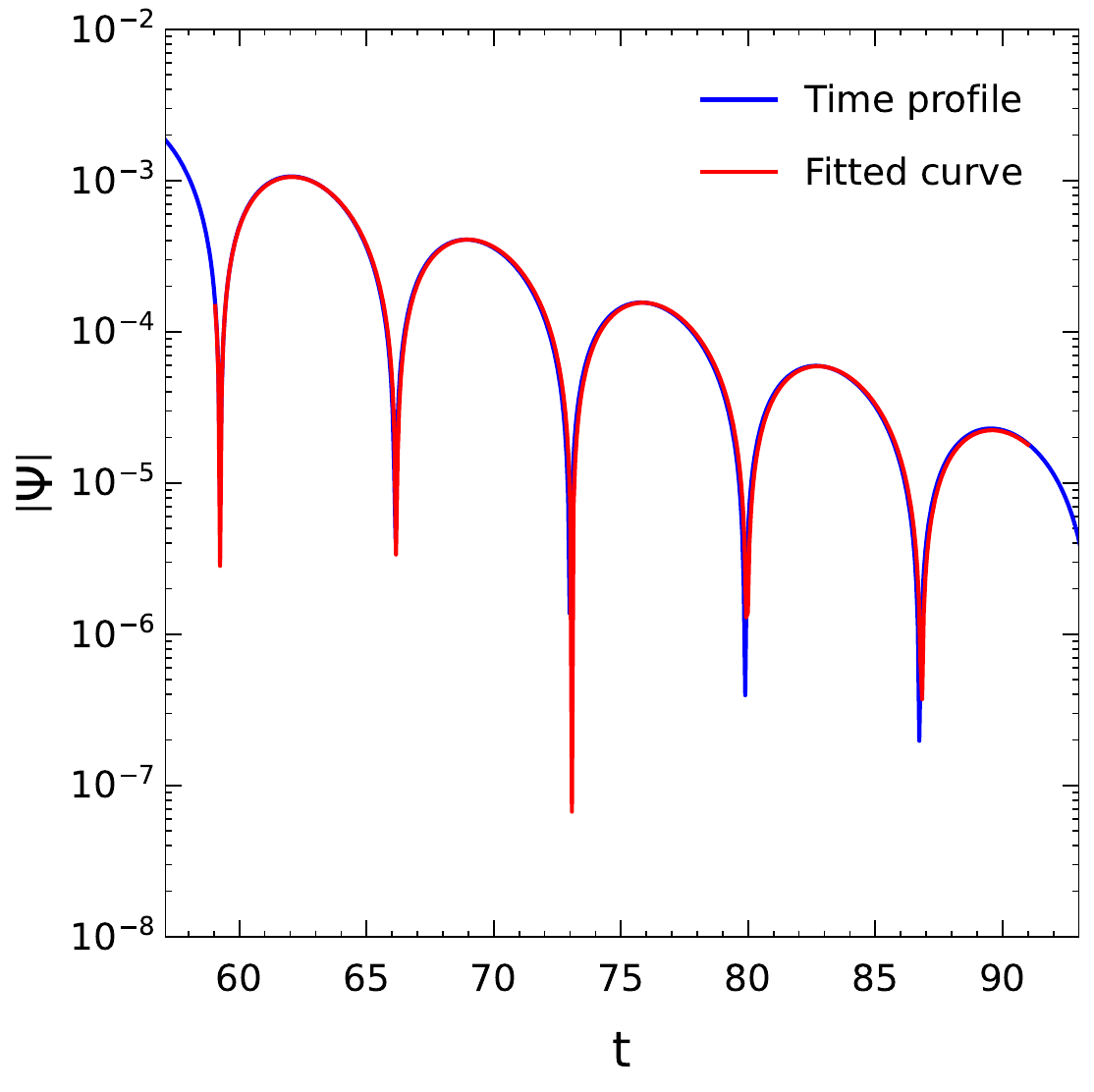} \hspace{0.3cm}
\includegraphics[scale=0.300]{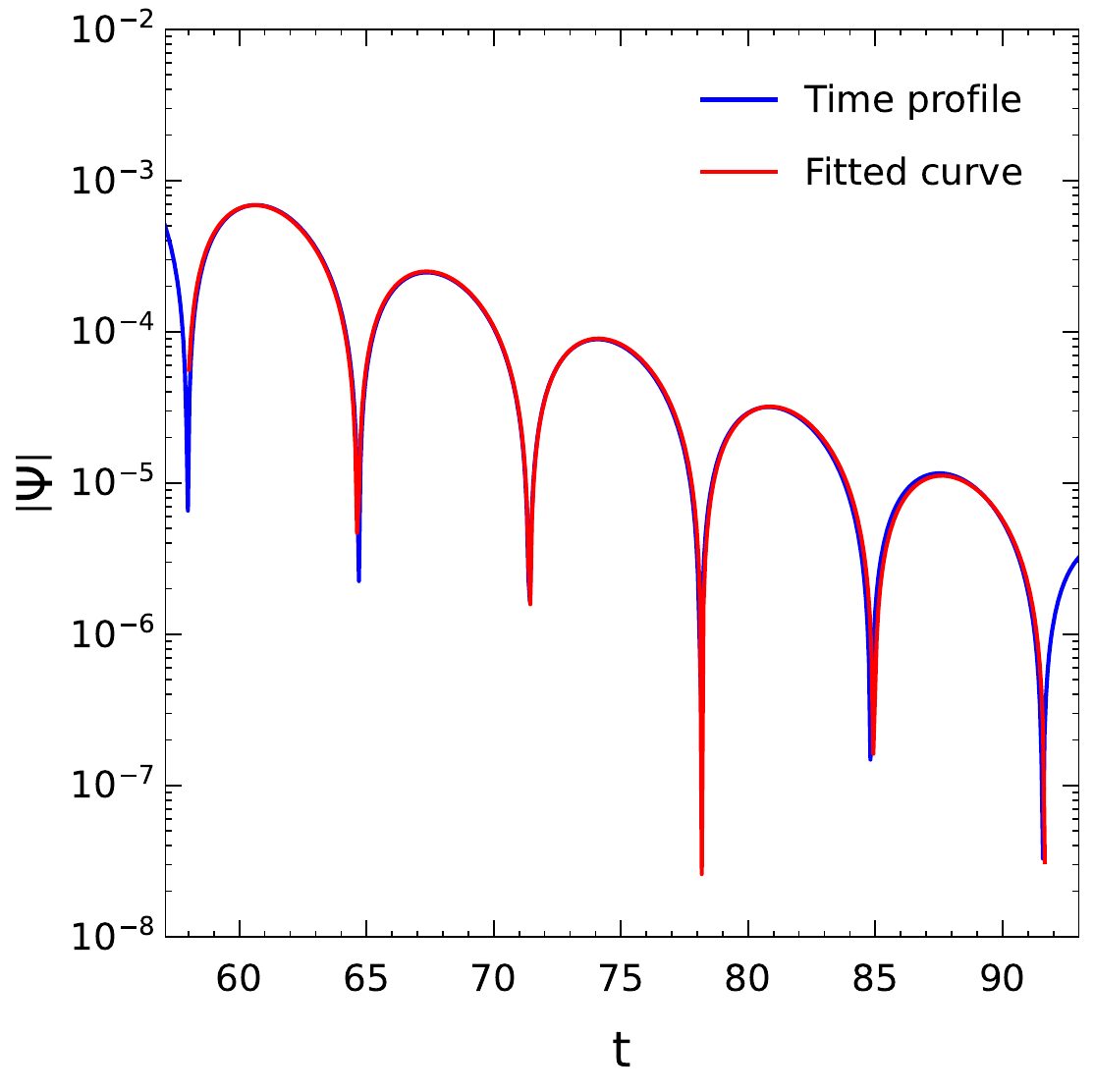}\hspace{0.3cm}
\includegraphics[scale=0.300]{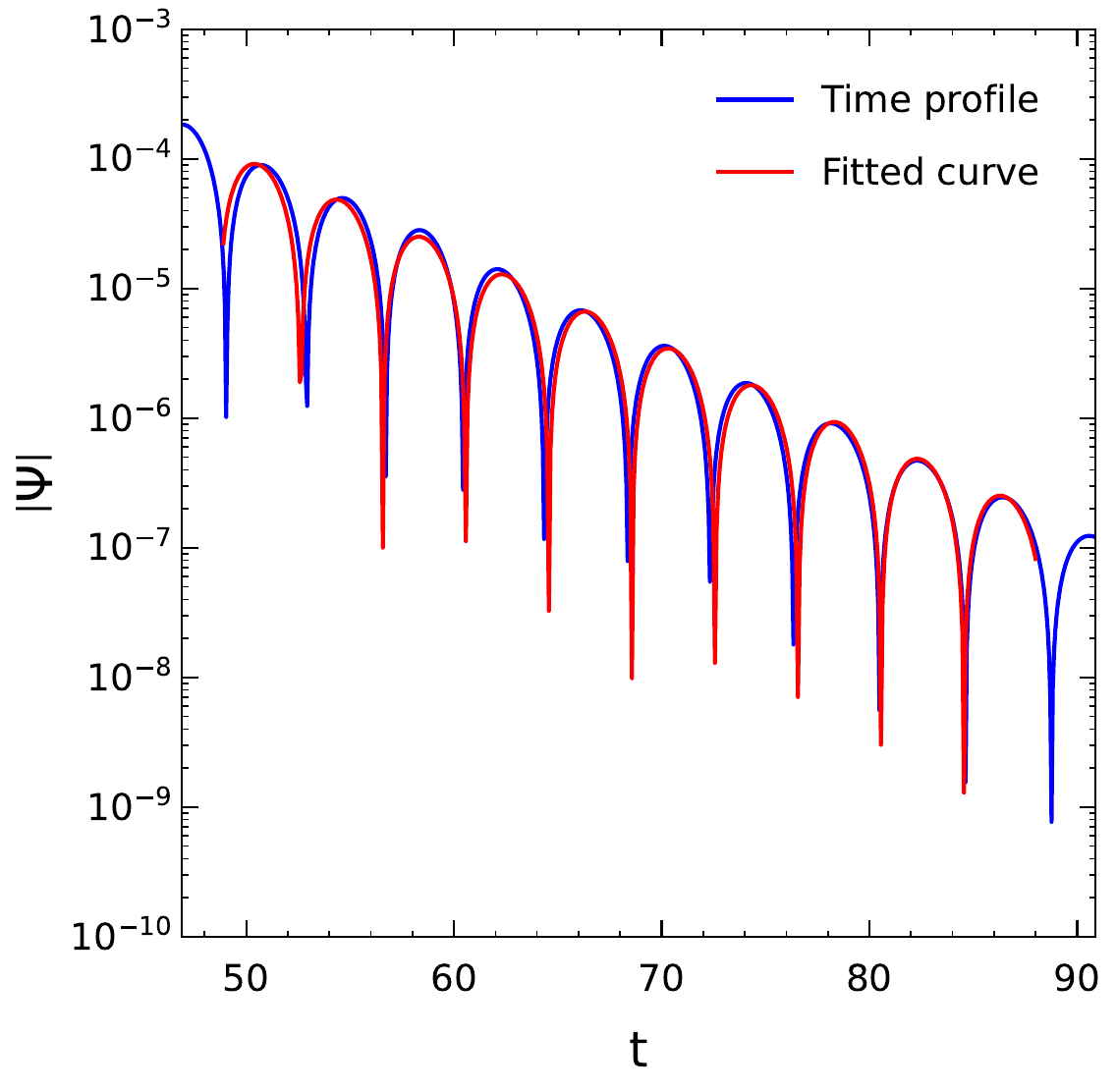}}
\vspace{-0.2cm} 
\caption{Fitting of the time domain profile to estimate the QNMs as an
example for $l=1$. In the upper row, the left plot is for $\omega=0$, 
$\beta=-8.8$ and $c_2 = 1$, the middle plot is for $\omega=1/3$, $\beta=-6$ 
and $c_2 = 1$, and the right plot is for $\omega=-2/3$, $\beta=-2$ and 
$c_2 = 1$. In the lower row, the left plot is for $\omega=0$, $\beta=-8.8$ 
and $c_2 = 1.1$, the middle plot is for $\omega=1/3$, $\beta=-5.5$ and 
$c_2 = 0.5$, and the right plot is for $\omega=-2/3$, $\beta=-2$ and 
$c_2 = 0.8$.}
\label{fig:13}
\end{figure} 
Fig.~\ref{fig:13} shows the fitting of the time domain profiles using the 
Levenberg Marquardt algorithm \cite{Levenberg1,Levenberg2,Levenberg3}. From 
the fitting, we calculate the QNMs of the BHs.  Further, we calculated the 
difference in the magnitudes of the QNMs obtained from the 3rd order WKB 
approximation method and those obtained from the time domain analysis method 
as follows: 
\begin{equation}
\Delta_{Q\!N\!M} = \frac{\left| QNM_{\text{WKB}} - QNM_{\text{Time domain}} \right|}{2}.
\label{eq48}
\end{equation} 
Table \ref{tab:1} shows the calculated QNMs of the BHs for 
$\omega=0$, $\omega = 1/3$ and $\omega = -2/3$ with different values of 
$\beta$ and $c_2$.
\begin{table}[!h]
\caption{QNMs of the minimally coupled $f(R,T)$ gravity BHs computed from the
fitting of the time domain profiles using the Levenberg Marquardt algorithm 
\cite{Levenberg1,Levenberg2,Levenberg3} in comparison with that obtained from
3rd order WKB approximation method for different values of $\omega$, $\beta$, and $c_2$.}
\vspace{0.2cm}
\centering
    \scalebox{0.92}{
    \begin{tabular}{c@{\hskip 5pt}c@{\hskip 5pt}c@{\hskip 5pt}c@{\hskip 10pt}c@{\hskip 15pt}c@{\hskip 15pt}c@{\hskip 15pt}c@{\hskip 15pt}c@{\hskip 15pt}c}
     \hline \hline \vspace{2pt}
      & $\omega$ & $\beta$ & $c_2$ & Multipole $l$ & 3rd order WKB QMNs & Time domain QNMs & $R^2$ & $\Delta_{Q\!N\!M}$ & \\
        \hline
&\multirow{3}{5em}{$\omega = 0$} & \multirow{3}{5em}{$\beta = -8.8$} & \multirow{3}{5em}{$c_2 = 1$} & $1$ & $0.371088 - 0.116790i$ & $0.371218 - 0.118280i$ & 0.995280 & $2.818530 \times 10^{-4}$&\\
&   &    &   &$ 2$ & $0.606594 - 0.098307i$ & $0.602417 - 0.097901i$ & 0.985772 & $2.093900 \times 10^{-3}$&\\
&    &    &   &$3$ & $0.846850 - 0.092873i$ & $0.842375 - 0.091867i$ & 0.988405 & $2.278800 \times 10^{-3}$&\\
        \hline
&\multirow{3}{5em}{$\omega = 1/3$} & \multirow{3}{5em}{$\beta = -6$} &\multirow{3}{5em}{$c_2 = 1$}& $1$ & $0.333450 - 0.087057i$ & $0.324891 - 0.149260i$ & 0.981999 & $6.455010 \times 10^{-3}$&\\
&   &    &   &$2$ & $0.554941 - 0.086645i$ & $0.522741 - 0.095422i$ & 0.987519 & $1.514250 \times 10^{-2}$ & \\
&   &    &   &$ 3$ & $0.775963 - 0.086548i$ & $0.723183 - 0.091149i$ & 0.989405 & $2.593500 \times 10^{-2}$ & \\
        \hline
&\multirow{3}{5em}{$\omega = -2/3$} & \multirow{3}{5em}{$\beta = -2$} &\multirow{3}{5em}{$c_2 = 1$}& $1$ & $1.341550 - 0.494087i$ & $1.343261 - 0.472752i$ & 0.971959 & $2.808810 \times 10^{-3}$&\\
&   &    &   &$2$ & $1.972000 - 0.537327i$ & $1.971946 - 0.516896i$ & 0.989519 & $2.664080 \times 10^{-3}$&\\
&   &    &   &$3$ & $2.635360 - 0.554069i$ & $2.635490 -0.554137i$ & 0.989134 & $7.061500 \times 10^{-5}$&\\
        \hline
& \multirow{3}{5em}{$\omega = 0$}& \multirow{3}{5em}{$\beta = -8.8$}& \multirow{3}{5em}{$c_2 = 0.1$}& $1$ & $0.309761 - 0.123334i$ & $0.309543 - 0.122674i$& 0.991752& $3.818000 \times 10^{-4}$&\\
&  &  &  & $2$ & $0.492441 - 0.105570i$& $0.493673 - 0.106510i$& 0.997334 & $0.775000 \times 10^{-4}$&\\
&  &  &  & $3$ & $0.686222 - 0.101014i$ & $0.685122 - 0.102673i$ & 0.985176 & $0.997300 \times 10^{-3}$&\\
\hline
& \multirow{3}{5em}{$\omega = 1/3$}& \multirow{3}{5em}{$\beta = -5.5$}& \multirow{3}{5em}{$c_2 = 0.5$}& $1$ & $0.319104 -0.122256i$ & $0.311172 - 0.122400i$& 0.993911& $3.967500 \times 10^{-3}$&\\
&  &  &  & $2$ & $0.515656 -0.104737i$& $0.515723 - 0.103974i$& 0.99632 & $4.283255 \times 10^{-5}$&\\
&  &  &  & $3$ & $0.718415 -0.0998847i$ & $0.718243 - 0.093451i$ & 0.985176 & $3.217500 \times 10^{-3}$&\\
\hline
\hline
& \multirow{3}{5em}{$\omega = -2/3$}& \multirow{3}{5em}{$\beta = -3$}& \multirow{3}{5em}{$c_2 = 0.8$}& $1$ & $1.149490 -0.424259i$ & $1.127632 - 0.423921i$& 0.981517& $1.093100 \times 10^{-2}$&\\
&  &  &  & $2$ & $1.711980 -0.455701i$& $1.709850 - 0.454871i$& 0.98532 & $1.343781 \times 10^{-3}$&\\
&  &  &  & $3$ & $2.300760 -0.467673i$ & $2.354610 - 0.461734i$ & 0.985176 & $2.708001 \times 10^{-2}$&\\
\hline
\end{tabular}}
\label{tab:1}
\end{table}
From this time domain analysis we have seen that the QNMs obtained from this 
analysis are in good agreement with the results obtained from the WKB 
approximation method. As the $R^2$s for the fits are also found to be 
satisfactory, the accuracy of the fittings is ensured.

\section{Conclusion}
\label{sec.6}  
Gravitational lensing is a versatile technique for exploring the properties of 
spacetime and for testing the predictions of different theories of gravity. In 
the first part of this study, we have analyzed the strong gravitational 
lensing properties of BHs described by the $f(R,T)$ gravity theory, focusing 
on a minimally coupled model of the theory. The analysis has relied on a 
strong field limit universal method developed by V.~Bozza, applicable to all 
spacetime in any gravity theory where the photons follow the standard geodesic 
equations \cite{2002_Bozza}. We have conducted the analysis considering three 
types of surrounding fields around the BHs, namely the dust field, the 
radiation field and the quintessence field. Here, we have concentrated on the 
investigation of the characteristics of the region just outside the event 
horizon of the minimally coupled $f(R,T)$ gravity BHs for these fields 
separately through the analysis of photon deflection extremely close to the 
photon sphere of the BHs. For this, the influences of the model 
parameters $\beta$ and $c_2$ on different photon trajectories are examined. 
Our analysis reveals that the lensing coefficients $\bar{a}$ and $\bar{b}$ 
exhibit distinct and contrasting behaviors with $\beta$ depending on the 
values of $\omega$, and result in nearly identical trends for both 
$\omega = 0$ and $1/3$, however, show contrasting behavior for $\omega = -2/3$ 
with $c_2$. These variations directly affect the deflection angle 
$\hat{\alpha}$ which decreases 
with increasing impact parameter $\zeta$ and becomes negative for larger 
values of $\zeta$, highlighting repulsive interactions by the gravitational 
field of the BHs. Furthermore, by considering the supermassive BH SgrA* as a 
minimally coupled $f(R,T)$ gravity BH, we have computed the numerical values 
of the lensing observables. The computed lensing observables: the angular 
position of relativistic images $\vartheta_\infty$, angular separation $s$, 
and relative magnification $r_{\text{mag}}$ demonstrate a consistent decline 
with increasing $\beta$ and $c_2$ in scenarios with $\omega = 0 $ and 
$\omega = 1/3$. Remarkably, when comparing the observables of the first two 
classes of BHs, the respective observables are found smaller in a dust 
dominated environment than in the presence of the radiation field. 
Eventually, the variation as a function of $\beta$ shows a sudden 
growth of dust field observables beyond $\beta \leq -12.23$. This sudden 
increase of observables could be due to the dominance of the physical 
parameter linked to the dust field over those of the radiation field 
beyond this value. In addition, the analysis for $\omega = -2/3$ reveals a 
distinct deviation in the behavior of the observables beyond 
$\beta \approx -9.8$ in contrast 
to the cases of $\omega = 0 $ and $\omega = 1/3$. The sudden decline 
of $\vartheta_{\infty}$ 
and the steep growth of $s$ and $r_{mag}$ beyond $\beta =-9.8$ indicate 
a significant dependence of the observables on the parameter $\beta$. These 
results highlight that reliable predictions for the observables are confined
within this specific range of $\beta$, emphasizing the intricate relationship 
between $\beta$ and the underlying physical properties of the system. 
However, our findings underscore the significant 
role of $\beta$ and the type of surrounding field in shaping the strong lensing behavior of the considered BHs spacetime with deviation from the GR, greater 
in case of $\omega = 0$, and lesser in $\omega = - 2/3$ situation respectively 
for any positive value of $c_2$. Additionally, we have constrained parameters 
$\beta$ and $c_2$ from the EHT observational data of SgrA* to test the 
viability and observational consistency of the model. Indeed, the constraining 
of the parameters enables us to confirm a limit to $\beta$ and $c_2$ in such a 
way that the upper bound of $\beta$ increases with the decrease of $c_2$ 
providing a potential avenue for examining analyses through strong lensing 
observations.
 
The QNMs associated with the BHs' spacetime have also been studied. For three 
values of equation of state parameter $\omega$, we calculated the QNMs of
BHs using the 3rd order WKB approximation method as the higher order 
($>$ 5th order) values of QNMs of the BHs obtained by this method are not 
converging. We plot the real part of calculated QNMs denoting the amplitude of 
frequency of the mode against the model parameter $\beta$. It is seen that 
the amplitude increases with increasing $\beta$ for all three cases. For the 
imaginary part of the QNMs denoting damping rate, we found that for 
$\omega=-2/3$, the highest damping has been observed compared to the other two 
cases. For $\omega=1/3$, the plots converse near $\beta=0$ and then show a 
steady increase in damping as $\beta$ increases. The analysis is done 
again for varying $c_2$ parameter. It is seen that the parameter has profound 
impact on the amplitude as well as damping of QNMs, with its influence 
increasing for higher multipole $l$. Time evolution of the QNMs 
of oscillations has also been studied and the decaying modes are clearly visible
from their profiles. For the sake of accuracy, we have compared the results 
obtained by the WKB method with that obtained from the time profile 
fitting method based on the Levenberg Marquardt algorithm and found a good 
agreement in results as shown in Table \ref{tab:1}. The deviation between 
the results of these two methods has been represented via a quantity parameter 
$\Delta_{Q\!N\!M}$ defined in equation \eqref{eq48}, which comes out to be of 
the order of about $10^{-3}$, thus signifying a good match between the two 
methods.
 
We hope that this work will contribute towards a better understanding of the 
nature of GWs and the lensing phenomena in MTGs. This study enriches our 
understanding of BH spacetimes in MTGs and offers valuable insights for 
future observational and theoretical investigations in astrophysics.
   
\section*{Acknowledgements}
UDG is thankful to the Inter-University Centre for Astronomy and Astrophysics
(IUCAA), Pune, India for awarding the Visiting Associateship of the institute.

\bibliographystyle{apsrev}

\end{document}